\RequirePackage{fix-cm}
\documentclass[twocolumn]{svjour3}          %
\smartqed  %
\usepackage{graphicx}
\usepackage[normalem]{ulem}
\usepackage{bbm}
\usepackage{bm}
\usepackage{xspace}
\usepackage{multirow}
\usepackage{algorithm}
\usepackage{algpseudocode}
\usepackage{amsmath,amssymb,amsfonts}
\usepackage{xcolor}

\usepackage{amsmath}
\DeclareMathOperator{\Tr}{tr}

\newcommand{\malinga}[1]{\textcolor{black}{#1}}

\newcommand{\rproof}[1]{\textcolor{black}{#1}}
\newcommand{\makeorange}[0]{\color{black}}
\newcommand{\maketeal}[0]{\color{black}}

\newcommand{\eat}[1]{}

\newcommand{\ie}[0]{i.e.,\xspace}
\newcommand{\eg}[0]{e.g.,\xspace}
\newcommand{\etal}[0]{et al.\xspace}

\begin{document}

\title{\malinga{No DBA? No regret! Multi-armed bandits for index tuning of analytical and HTAP workloads with provable guarantees}}
\titlerunning{No DBA? No regret!}

\author{R. Malinga Perera \and Bastian Oetomo \and Benjamin I. P. Rubinstein \and  Renata Borovica-Gajic %
}

\institute{R. Malinga Perera \at
              School of Computing and Information Systems, \\
              University of Melbourne,\\
              Australia.\\
              \email{malinga.perera@student.unimelb.edu.au}
          \and
          Bastian Oetomo \at
              \email{b.oetomo@student.unimelb.edu.au} 
            \and
          Benjamin I. P. Rubinstein \at
              \email{brubinstein@unimelb.edu.au} 
            \and
          Renata Borovica-Gajic \at
              \email{renata.borovica@unimelb.edu.au} 
}

\date{Received: date / Accepted: date}

\maketitle

\begin{abstract}
Automating physical database design has remained a long-term interest in database research due to substantial performance gains afforded by optimised structures. Despite significant progress, a majority of today's commercial solutions are highly manual, requiring offline invocation by database administrators \linebreak[4] (DBAs) who are expected to identify and supply representative training workloads. Even the latest advancements like query stores provide only limited support for dynamic environments. This status quo is untenable: identifying \emph{representative} static workloads is no longer realistic; and physical design tools remain susceptible to the query optimiser's cost misestimates. \malinga{Furthermore, modern application environments such as hybrid transactional and analytical processing (HTAP) systems  render analytical modelling next to impossible.}

We propose a self-driving approach to online index selection that eschews the DBA and query optimiser, and instead \emph{learns} the benefits of viable structures through strategic exploration and direct performance observation. We view the problem as one of sequential decision making under uncertainty, specifically within the bandit learning setting. 
Multi-armed bandits balance exploration and exploitation to \emph{provably} guarantee average performance that converges to policies that are optimal with perfect hindsight.   \malinga{Our comprehensive empirical evaluation against a state-of-the-art commercial tuning tool demonstrates up to 75\% speed-up on shifting and ad-hoc workloads and up to 28\% speed-up on static workloads in analytical processing environments. In HTAP environments, our solution provides up to 59\% speed-up on shifting and 51\% speed-up on static workloads. Furthermore, 
our bandit framework outperforms deep reinforcement learning (RL) in terms of convergence speed and performance volatility (providing up to 58\% speed-up).}
\end{abstract}

\section{Introduction}
\label{sec:intro}
With the growing complexity and variability of database applications and their hosting platforms (\eg multi-tenant cloud environments), automated physical design tuning, particularly automated index selection, has re-emerged as a contemporary challenge for database management systems. 
Most database vendors offer automated tools for physical design tuning within their product suites~\cite{DTA2005,DB2_IntegratedApproach,OracleAdvisor}. 
Such tools 
form an integral part of broader efforts toward fully automated database management systems which aim to:
a) decrease database administration costs and thus total costs of ownership~\cite{TCO,pavlo};
b) help non-experts use database systems; and
c) facilitate hosting of databases 
on dynamic environments such as cloud-based services~\cite{CloudDBAsAService,sqlvm,das2019automatically,AIMeetsAI}. 
Most physical design tools take an \emph{off-line} approach, where the representative training workload is provided by the data\-base administrator (DBA)~\cite{SurajitDecade}.
Where \emph{online} solutions are provided~\cite{COLT2,QUIET,OnlineApproachAutoAdmin,ToTuneOrNot,das2019automatically,forecastingPavlo}, questions remain: How often should the tools be invoked? 
And importantly, is the quality of proposed designs in any way guaranteed? How can tools generalise beyond queries seen 
to dynamic ad-hoc workloads, where queries are unpredictable and non-stationary?

Modern analytics workloads are dynamic in nature with ad-hoc queries 
common~\cite{ResearcherGuide}, \eg data exploration 
workloads adapt to past query responses~\cite{noDBSIGMOD2012}. 
Such ad-hoc workloads hinder automated tuning since:
a) inputting representative information to design tools is infeasible under time-evolving workloads; and
b) reacting too quickly to changes may result in  performance variability, where indices are continuously dropped and created. 
Any robust automated physical design solution must address such challenges~\cite{COLT2}. 

\malinga{The situation is further aggravated in HTAP environments, that consist of online transaction processing (OLTP) and online analytical processing (OLAP) workloads.
While indices provide (primarily) positive benefits to OLAP queries, they hinder the OLTP query performance due to the additional index maintenance overhead. Furthermore, in dynamic settings, workload composition (i.e., analytical to transaction ratio) can vary over time, making it even more challenging to identify useful indices that boost overall workload performance.}

To compare alternative physical design structures, automated design tools use a cost model employed by the query optimiser, typically exposed through a ``what-if" interface~\cite{ChaudhuriWhatIf}, as the sole source of truth. 
However such cost models 
make inappropriate assumptions about data characteristics~\cite{ImplicationsOfAssumptions,HowGoodQO}: 
commercial DBMSs often assume attribute value independence and uniform data distributions when sufficient statistics are  unavailable \cite{SThistograms,HowGoodQO,vldbjborovica}. 
As a result, estimated benefits of proposed designs may diverge significantly from actual workload performance~\cite{DBTest2012PhysicalDesigners,GebalyRobustness,vldbjborovica,icde15Borovica,AIMeetsAI,das2019automatically}. 
Even with more complex data distribution statistics such as single- and multi-column histograms, the issue remains for complex workloads~\cite{vldbjborovica}. \malinga{Moreover, data additions and updates in HTAP environments continuously invalidate statistics, adding on to the optimiser misestimates. Keeping statistics up-to-date in such a setting requires extra effort.}

\begin{figure*}[t]
    \centering
    \includegraphics[width = \textwidth]{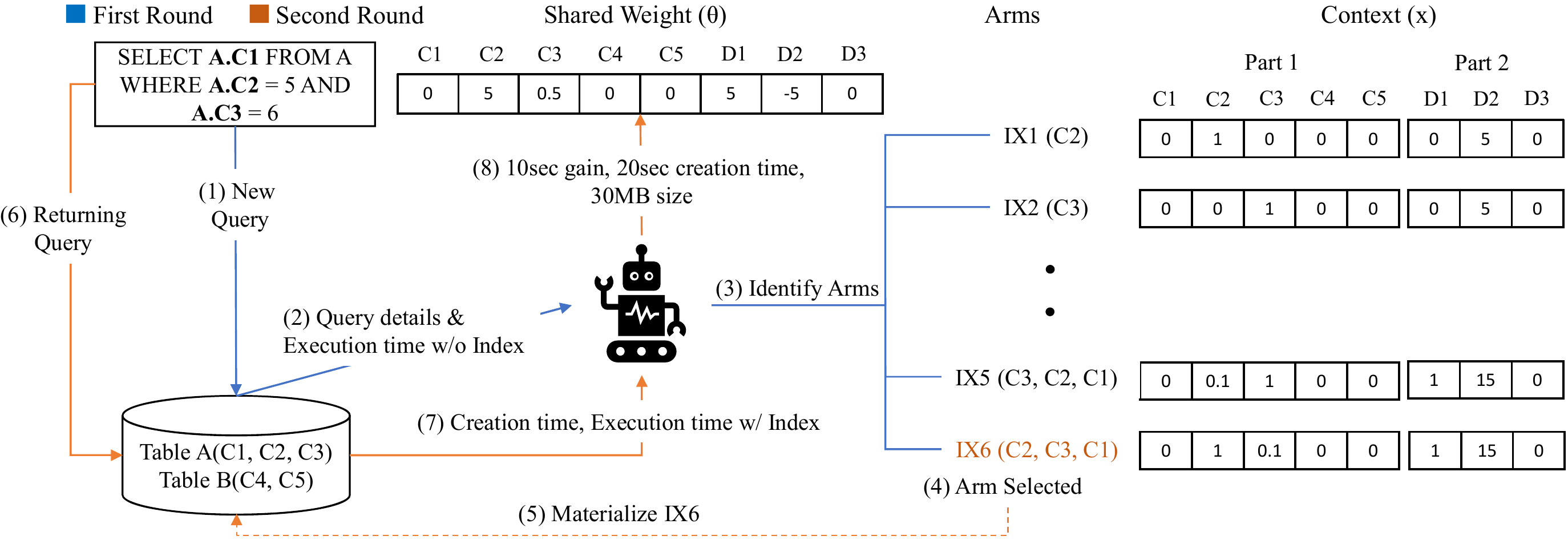}
\caption{An abstract view of the proposed bandit learning-based online index selection.}
    \label{fig:bandit_overall}
\end{figure*}

In this paper, we demonstrate that even in ad-hoc environments where queries are unpredictable, there are opportunities for index optimisation. We argue that the problem of online index selection under ad-hoc, analytical \malinga{and HTAP} workloads can be efficiently formulated within the 
multi-armed bandit (MAB) learning setting---a tractable form of Markov decision process.
MABs take arms or actions (selecting indices) to maximise cumulative rewards, trading off exploration of untried actions with exploitation of actions that maximise rewards observed so far (see Figure~\ref{fig:bandit_overall}). MABs permit learning from observations of actual performance, and need not rely on potentially misspecified cost models. 
Unlike initial efforts with applying learning for physical design, \eg more general forms of reinforcement learning~\cite{no_dba}, bandits offer regret bounds that \emph{guarantee} the fitness of dynamically-proposed indices~\cite{c2ucb}.

The key contributions of the paper are summarised next:
\begin{itemize}
    \item We model index tuning as a multi-armed bandit, proposing design choices that lead to a practical, competitive solution.
    \item Our proposed design achieves a worst-case safety guarantee
    against any optimal fixed policy, as a consequence of a corrected regret analysis of the C$^2$UCB bandit.
    \item We introduce a new bandit flavour that extends the existing contextual and combinatorial bandit where structured rewards are observed for each arm, providing additional feedback for the learned weight. This bandit variation enjoys a better regret bound compared to the C$^2$UCB bandit.
    \item Our comprehensive experiments demonstrate MAB's superiority over a state-of-the-art commercial physical design tool and deep reinforcement learning agent, with up to 75\% speed-up in the former and 58\% speed-up in the latter case, under dynamic, analytical workloads.
    \malinga{\item We showcase MAB's ability to perform in complex HTAP environments, which are notoriously challenging for index tuning, delivering 59\% speed-up over the state-of-the-art commercial design tool.}
\end{itemize}
\section{Problem Formulation}
\label{sec:problem_formulation}

The goal of the \emph{online database index selection problem} is to choose a set of indices (referred to as a \emph{configuration}) that minimises the total running time of a workload sequence within a given memory budget. Neither the workload sequence, nor system run times, are known in advance.

We adopt the problem definition of~\cite{OnlineApproachAutoAdmin}. Let the \emph{workload} $W = (w_{1},w_{2},\ldots,w_{T})$ be a sequence of \emph{mini-workloads} (\eg a sequence of single queries), $I$ the set of \emph{secondary indices}, 
$C_{mem}(s)$ represent the memory space required to materialise a configuration $s\subseteq I$,
and $\mathcal{S}=\left\{s\subseteq I \left| \, C_{mem}(s) \leq M\right\} \right. \subseteq 2^I$ be the class of \emph{index configurations} feasible within our total memory allowance $M$. Our goal is to propose a configuration sequence $S = (s_{0},s_{1},\ldots,s_{T})$, with $s_t \in \mathcal{S}$ as the configuration in round $t$ and $s_{0}=\emptyset$ as the starting configuration, which minimises the \emph{total workload time $C_{tot}(W,S)$} defined as:
$$C_{tot}(W,S) = \sum_{t=1}^T C_{rec}(t) + C_{cre}(s_{t-1},s_t) + C_{exc}(w_t,s_t)\enspace.$$
Here $C_{rec}(t)$ refers to the \emph{recommendation time} in round $t$ (defined as running time of the recommendation tool) and $C_{cre}(s_{t-1},s_t)$ refers to the incremental index creation time in transitioning from configuration $s_{t-1}$ to $s_{t}$. Finally, $C_{exc}(w_t,s_t)$ denotes the execution time of mini-workload $w_{t}$ under the configuration $s_{t}$, namely the sum of response times of individual queries.

At round $t$, the system:
\begin{enumerate}
    \item Chooses a set of indices $s_{t} \in \mathcal{S}$ in preparation for upcoming workload $w_t$, without  direct access to $w_t$. 
    
    $s_t$
    only depends on observation of historical workloads ($w_1, \ldots, w_{t-1}$), corresponding sets of chosen indices, and resulting performance;
    
    \item Materialises the indices in $s_t$ which do not exist yet, that is, all indices in the set difference
    $s_t \backslash s_{t-1}$; and
    \item Receives workload $w_t$, executes all the queries \linebreak[4] therein, and measures elapsed time of each individual query and each operator in the corresponding query plan.
\end{enumerate}

\section{Contextual Combinatorial Bandits}
\label{sec:background}
In this paper, we argue that online index selection can be successfully addressed using  multi-armed bandits \linebreak[4] (MABs) from statistical machine learning, where different arms correspond to chosen indices. 
We first present necessary background on MABs, outlining the \textbf{essential properties} that we exploit in our work (i.e., bandit context and combinatorial arms) to converge to \linebreak[4] highly performant index configurations.

We use the following notation.
We denote non-scalar values with boldface: lowercase for (by default column) vectors and uppercase for matrices. 
We also write $[k]=\{1,2,\ldots,k\}$ for $k\in\mathbb{N}$, and denote the transpose of a matrix or a vector with a prime.

The contextual combinatorial bandit setting under  \emph{semi-bandit} feedback involves repeated selections from $k$ possible actions, over rounds $t=1,2,\ldots$, in which the MAB:
\begin{enumerate}
    \item Observes a \emph{context} feature vector (possibly random or adversarially chosen) of each action or \emph{arm} $i \in [k]$, denoted as $\bm{X}_t = \{ \bm{x}_t(i) \}_{i \in [k]}$, for $\bm{x}_t(i) \in \mathbb{R}^d$, along with their costs, $c_i$;
    \item Selects or \emph{pulls}\ a set of arms (referred to as \emph{super arm}) $s_t\in \mathcal{S}_t$, where we restrict the class of possible super arms $\mathcal{S}_t \subseteq \mathcal{S}^\prime_t = \left\{s \subseteq [k] \left| \, \sum_{i\in s}c_i \leq M \right. \right\} \subseteq 2^{[k]}$; and 
    \item For each $i_t \in s_t$, observes random \emph{scores} $r_t(i_t)$ drawn from fixed but unknown arm distribution which depends solely on the arm $i_t$ and its context $\bm{x}_t(i_t)$, whose true expected values are contained in the unknown variable $\bm{r}^\star_t = \{\mathbb{E}[r_{t}(i)]\}_{i \in [k]}$.
\end{enumerate}

\malinga{\begin{remark}
    \label{remark:mabvsrl}
    The contextual combinatorial bandit setting is a special case of a Markov decision process, which is solved in general by  reinforcement learning (RL). The key difference is that in bandits, state transition is not affected by MAB actions, only rewards are. States (observed via contexts) arrive arbitrarily. This simplicity admits theoretical guarantees for practical MAB learners, where state-of-the-art RL agents regularly have \linebreak[4]none. When playing in a bandit setting, in practice MAB learners may converge faster than their (typically over parametrised) RL cousins. %
\end{remark}}

A MAB's goal is to maximise the cumulative expected reward $\sum_t \mathbb{E}[R_t(s_t)] = \sum_t g(s_t, \bm{r}^\star_t, \bm{X}_t)$ for a known function $g$. This function $g$ need not be a simple summation of all the scores\malinga{, but is typically assumed to be monotonic and Lipschitz smooth in the arm scores.}

\maketeal
\begin{definition} A \emph{monotonic} function $g(s, \mathbf{r}, \mathbf{X})$ is non-decreasing in $\mathbf{r}$: 
for all $s, \mathbf{X}$, if $\mathbf{r} \preceq \mathbf{r}^\prime$
then $g(s, \mathbf{r}, \mathbf{X}) \leq g(s, \mathbf{r}^\prime, \mathbf{X})$.
\end{definition}

\begin{definition}
Function $g(s, \mathbf{r}, \mathbf{X})$ is \emph{$C$-Lipschitz (uniformly) in $\mathbf{r}$}, if 
    $\left|g(s, \mathbf{r}, \mathbf{X}) - g(s, \mathbf{r}^\prime, \mathbf{X})\right| \leq C \cdot \left\| \mathbf{r} - \mathbf{r}^\prime\right\|_2$, 
    for all
$\mathbf{r}, \mathbf{r}^\prime$, $\mathbf{X}, s$.
\end{definition}

\color{black}
The core challenge in this problem is that the expected scores for all arms $i\in[k]$ are unknown. Refinement of a bandit learner's approximation for arm $i$ is \emph{generally} only possible by including arm $i$ in the super arm, as the score for arm $i$ is not observable when $i$ is not played. This suggests solutions that balance \emph{exploration} and \emph{exploitation}. Even though at first glance it may seem that each arm needs to be explored at least once, placing practical limits on large numbers of arms, there is a remedy to this as will be discussed shortly.

\textbf{The C$^2$UCB algorithm.}
Used to solve the contextual combinatorial bandit problem, the C$^2$UCB Algorithm~\cite{c2ucb} models the arms' scores as linearly dependent on their contexts: 
$r_{t}(i) = {\bm{\theta}}^\prime \bm{x}_t(i) + \varepsilon_t(i)$ for unknown zero-mean (subgaussian) random variable $\varepsilon_t$, unknown but fixed parameter $\bm{\theta}\in\mathbb{R}^d$, and known context $\bm{x}_t(i)$. It is crucial to notice the implication that, \textbf{all learned knowledge is contained in estimates of $\bm{\theta}$, which is shared between all arms, obviating the need to explore each arm}. Estimation of $\bm{\theta}$ can be achieved using ridge regression,
with $|s_t|$ new data points $\{(\bm{x}_t(i), r_t(i))\}_{i \in s_t}$ available at round $t$, further \emph{accelerating the convergence rate} of the estimator $\hat{\bm{\theta}}$, over observing only one example as might be na\"ively assumed.

Point estimates on the expected scores can be made with $\bar{r}_t(i) = \hat{\bm{\theta}}_{t}^\prime \bm{x}_t(i)$,
where $\hat{\bm{\theta}}_{t}$ are trained coefficients of a ridge regression on observed rewards against contexts. However, this quantity is oblivious to the variance in the score estimation. Intuitively, to balance out the exploration and exploitation, it is desirable to add an \emph{exploration boost} to the arms whose score we are less sure of (i.e., greater estimate variance). This suggests that the upper confidence bound (UCB) should be used, in place of the expected value, and which can be calculated~\cite{LinUCB} as:
\begin{eqnarray}
\hat{r}_t(i) &=&  \hat{\bm{\theta}}_{t}^\prime \bm{x}_t(i) + \alpha_t \sqrt{\bm{x}_t(i)^\prime\bm{V}_{t-1}^{-1}\bm{x}_t(i)}\enspace,\label{eq:linucb-value}
\end{eqnarray}
where $\alpha_t>0$ is the exploration boost factor, and $\bm{V}_{t-1}$ is the positive-definite $d\times d$ scatter matrix of contexts for the chosen arms up to and including round $t-1$. The first term of $\hat{r}_t(i)$ corresponds to arm $i$'s immediate reward, whereas its second term corresponds to its exploration boost, as its value is larger when the arm is sensitive to the context elements we are less confident of (i.e., the underexplored context dimension). Hence, by using $\hat{r}_t(i)$ in place of $\bar{r}_t(i)$, arms with contexts lying in the underexplored regions of context space are more likely to be chosen, as higher scores yield higher $g$, assuming that $g$ is monotonic increasing in the arm rewards.

Ideally, the super arm $s_t \in \mathcal{S}_t$ is chosen such that $g(s_t, \hat{\bm{r}}_t, \bm{X}_t)$ is maximised. However, it is sometimes computationally expensive to find such super arms. In such cases, it is often good enough to obtain a solution via some approximation algorithm where $g(\hat{\bm{r}}_t, \bm{X}_t, s_t)$ is \linebreak[4] \emph{near} maximum. With this criterion in mind, we now define an \emph{$\alpha$-approximation oracle}.

\begin{definition}
An \emph{$\alpha$-approximation oracle} is an algorithm $\mathcal{A}$ that outputs a super arm $\overline{s}=\mathcal{A}(\bm{r},\bm{X})$ with guarantee $g(\overline{s}, \bm{r},\bm{X})\geq \alpha \cdot \max_s g(s, \bm{r},\bm{X})$, for some $\alpha\in [0,1]$ and given input $\bm{r}$ and $\bm{X}$.  
\end{definition}

Note that knapsack-constrained submodular \linebreak[4] programs are efficiently solved by the greedy algorithm (iteratively select a remaining cost-feasible arm with highest available score) with $\alpha = 1-1/e$.  C$^2$UCB is detailed in Algorithm~\ref{alg:c2ucb}.

\begin{algorithm}[t!]
\caption{The C$^2$UCB Algorithm}
\begin{algorithmic}[1]
\State Input: $\lambda, \alpha_1, \ldots, \alpha_T$
\State Initialize $\bm{V}_0 \gets \lambda \bm{I}_d$, $\bm{b}_0 \gets \bm{0}_d$
\For {$t \gets 1,\ldots,T$}
\State Observe $\mathcal{S}_t$
\State $\hat{\bm{\theta}}_{t} \gets \bm{V}_{t-1}^{-1}\bm{b}_{t-1}$ \Comment{estimate via ridge regression}
\For{$i \in [k]$}
\State Observe context $\bm{x}_t(i)$
\State $\hat{r}_t(i) \gets \hat{\bm{\theta}}_t^\prime \bm{x}_t(i)+ \alpha_t\sqrt{\bm{x}_t(i)^\prime\bm{V}_{t-1}^{-1}\bm{x}_t(i)}$
\EndFor
\State $s_t \gets \mathcal{A}(\hat{\bm{r}}_t, \bm{X}_t)$ \Comment{using $\alpha$-approximation oracle}
\State Play $s_t$ and observe $r_t(i)$ for all $i \in s_t$
\State $\bm{V}_{t} \gets \bm{V}_{t-1} + \sum_{i\in s_t}\bm{x}_t(i)\bm{x}_t(i)^\prime$ \Comment{regression update}
\State $\bm{b}_t \gets \bm{b}_{t-1} + \sum_{i \in s_t}r_t(i)\bm{x}_t(i)$ \Comment{regression update}
\EndFor
\end{algorithmic}
\label{alg:c2ucb}
\end{algorithm}

The performance of a bandit algorithm is usually measured by its \emph{cumulative regret}, defined as the total expected difference between the reward of the chosen super arm $\mathbb{E}[R_t(s_t)]$ and an optimal super arm \linebreak[4] $\max_{s \in \mathcal{S}_t}\mathbb{E}[R_t(s)]$ over $T$ rounds. Such a metric is unfair to C$^2$UCB since its performance depends on the oracle's performance. This suggests assessing C$^2$UCB's performance with a metric using the oracle's performance guarantee as its measuring stick, as follows.
\begin{definition}
Let $\overline{s}$ be a super arm returned by an $\alpha$-approximation oracle as a part of the bandit algorithm, and $\bm{r}^\star_t$ be a vector containing each arms' true expected scores. Then cumulative $\alpha$-regret is the sum of expected instantaneous regret,
$Reg_t^\alpha = \alpha\cdot \max_{s} g(s, \bm{r}^\star_t, \bm{X}_t) - g(\overline{s}_t, \bm{r}^\star_t, \bm{X}_t)$.
\end{definition}

When $g$ is assumed to be monotonic and Lipschitz continuous, \cite{c2ucb} claimed that C$^2$UCB enjoys $\tilde{O}(\sqrt{T})$ $\alpha$-regret.
We have corrected an error in the original proof, as seen in Appendix~\ref{sec:theory}, confirming the $\tilde{O}(\sqrt{T})$ $\alpha$-regret. This expression is sub-linear in $T$, implying that the per-round average cumulative regret approaches zero after sufficiently many rounds. Consequently, online index selection based on C$^2$UCB comes endowed with a \emph{safety guarantee} on worst-case performance: selections become at least as good as an $\alpha$-optimal policy (with perfect access to true  scores); and potentially much better than any fixed policy.
\section{MAB for Online Index Selection}
\label{sec:bandits}
Performant bandit learning for online index tuning demands arms covering important actions and no more, rewards that are observable and for which regret bounds are meaningful, and contexts and oracle that are efficiently computable and predictive of rewards. 
Each workload query is monitored for characteristics such as running time, query predicates, payload, etc. (see Figure~\ref{fig:bandit_overall}). These observations feed into generation of relevant arms and their contexts. The learner selects a desired configuration which is materialised. After query return, the system identifies benefits of the materialised indices, which are then shaped into the reward signal for learning. 

\textbf{Dynamic arms from workload predicates.} 
Instead of enumerating all column combinations, \emph{relevant} arms (indices) may be generated based on queries: combinations and permutations of query predicates (including join predicates), with and without inclusion of payload attributes from the selection clause. 
Such workload-based arm generation drastically reduces the action space,  
and exploits natural skewness of real-life workloads that focus on small subsets of attributes over full tables~\cite{noDBSIGMOD2012}. 
Workload-based arm generation is only viable due to dynamic arm addition (reflecting a dynamic action space) and is allowed by the bandit setting: we may define the set of feasible arms for each round at its start.

\textbf{Context engineering.} 
Effective contexts are predictive of rewards, efficiently computable, and promote generalisation to previously unseen workloads and arms.
We form our context in two parts (see Figure~\ref{fig:bandit_overall}).

\textit{Context Part 1: Indexed column prefix.} We encode one context component per column. However unlike a bag-of-words or one-hot representation appropriate for text, similarity of arms depends on having similar column prefixes; common index columns is insufficient. This reflects a \textbf{\emph{novel bandit learning aspect of the problem.}} 
A context component has value $10^{-j}$ where $j$ is the corresponding column's position in the index, \emph{provided} that the column is included in the index and is a workload predicate column. The value is set to 0 otherwise, including if its presence only covers the payload. Unlike a simple one hot encoding, this context enables the bandit to differentiate between arms with the same set of columns but different ordering, and  reward the columns differently based on the column's position in the index.

\begin{example}
Under the simplest workload (single query) in Figure~\ref{fig:bandit_overall}, our system generates six arms: four using different combinations and permutations of the predicates, two including the payload (covering indices).
Index IX5 includes column C1, but the context for C1 is valued as $0$, as this column is considered only due to the query payload. 
\end{example}

\textit{Context Part 2: Derived statistical information.} 
We represent statistical and derived information about the arms and workload, details available during query execution, and sufficient statistics for unbiased estimates.  
This statistical information includes: a Boolean indicating a covering index, the estimated size of the index  divided by the database size (if not materialised already, 0 otherwise), and the number of times the optimiser has picked this arm in recent rounds. This is shown in Figure~\ref{fig:bandit_overall} under D1, D2 and D3, respectively.

\textbf{Reward shaping.} 
As the goal of physical design tuning tools is to 
minimise end-to-end workload time, we incorporate index creation time and query execution time into the reward for a workload. We omit index recommendation time, as it is independent of arm selection. However, we measure and report recommendation time of the MAB algorithm in our experiments. Recall that MAB depends only on observed execution statistics from implemented configurations and generalisation of the learned knowledge to unseen arms thereafter.

The implementation of the reward for an arm includes the execution time as a \emph{gain} $G_t(i, w_t, s_t)$ for a workload $w_t$ by each arm $i$ under configuration $s_t$. \malinga{Indices can impact the execution time in multiple ways. We split the execution time gain into three components: a) data scan gains ($G_t^{ds}$), b) index maintenance gains ($G_t^{im}$, usually a negative value), and c) other areas of the query plan which can be difficult to attribute to a single index (unclaimed gains) ($G_t^{un}$).}

\malinga{\begin{eqnarray*}&&{G_t(i, w_t, s_t)} \\ &=& G_t^{ds}(i, w_t, s_t) + G_t^{im}(i, w_t, s_t) + G_t^{un}(i, w_t, s_t)\enspace.\end{eqnarray*}}

\malinga{\emph{Data scan gains}: By defining $\mathcal{U}(s, q)$ as the list of indices used by the query optimiser for data access in query $q$  under a given configuration $s$, the data scan gain by index $i$ for query $q$ is defined as: }

\malinga{\begin{eqnarray*}&&G_t^{ds}(i, \{q\}, s_t) \\ &=& \left[C_{tab}(\tau(i), q, \emptyset) - C_{tab}(\tau(i), q, \{i\})\right]\mathbbm{1}_{\mathcal{U}(s,q)}(i)\,, \end{eqnarray*}}

$C_{tab}(\tau(i), q, \emptyset)$ represents the full table scan time for table $\tau(i)$ and query $q$, where $\tau(i)$ is the table which $i$ belongs to.\footnote{Due to the reactive nature of multi-armed bandits, we mostly observe a full table scan time for each table $\tau(i)$ and query $q$. When we do not observe this, we estimate it with the maximum secondary index scan/seek time.} 

\malinga{\emph{Index maintenance gains}: Index maintenance operations can take different forms based on the number of rows updated. The optimiser typically opts for row-wise updates for a small number of rows and index-wise updates otherwise. In the second case, we can easily capture the maintenance gain of an index as each index is updated separately. This is however not straightforward in the case of row-wise updates, where all indices are bulk updated for each row. On these occasions, we compute the total maintenance gain  ($G_t^{im}(\mathcal{V}(s,q), \{q\}, s_t)$) for all secondary indices that require maintenance due to a query $q$ under a given configuration $s$ and equally divide it among the updated indices. $\mathcal{V}(s,q)$ represents the set of indices updated under configuration $s$ by the query $q$.}

\malinga{$$G_t^{im}(\mathcal{V}(s,q), \{q\}, s_t) = \left[C_{im}(q, \emptyset) - C_{im}(q, s)\right]\enspace.$$}

\malinga{where $C_{im}(q, \emptyset)$ and $C_{im}(q, s)$ represent the index maintenance time without secondary indices and index maintenance time under configuration $s$, respectively.}

\malinga{\begin{eqnarray*}&&G_t^{im}(i, \{q\}, s_t) \\ &=& \left[G_t^{im}(\mathcal{V}(s,q), \{q\}, s_t)/\left|\mathcal{V}(s,q)\right|\right]\mathbbm{1}_{\mathcal{V}(s,q)}(i)\enspace. \end{eqnarray*}}

\malinga{\emph{Unclaimed gains}: The use of indices can impact the query plan in very subtle ways which cannot be easily attributed to a single index. For example, introducing a new index can trigger the optimiser to choose a different query plan. Even when the index use provides a faster data scan, new query execution can end up taking more time due to an inefficient nested loop join. Even though this issue arises from the optimiser, MAB needs to synchronise with the optimiser and possibly take corrective actions to trigger a different query plan that can result in better execution time. These gains can be computed by comparing the query running times before and after index creation. We compute the total query gain ($G_t^{to}$) as,}

\malinga{$$G_t^{to}(\{q\}, s_t) = \left[C_{to}(q, \emptyset) - C_{to}(q, s_t)\right]\enspace.$$}

\malinga{where $C_{to}(q, s_t)$ represents the total running time under configuration $s_t$. Once the data scan gain and index maintenance gain is calculated, we calculate the total unclaimed gains for a query by subtracting the total data scan and index maintenance gains from the total query gain. Then we equally divide this cost among participating indices ($\mathcal{U}(s, q) \cup \mathcal{V}(s,q)$).}

The gain for a workload relates to the gain for individual query by:
$$ G_t(i, w_t, s_t) = \sum_{q\in w_t}G_t(i, \{q\}, s_t)\, .$$
By this definition, 
gain $G_t(i,w_t, s_t)$ will be $0$ if $i$ is not used by the optimiser in the current round $t$ and can be negative if the index creation leads to a performance regression \malinga{or if the index incurs a maintenance cost}. Creation time of $i$ is taken as a negative reward, only if $i$ is materialised in round $t$, and is $0$ otherwise:
$$r_{t}(i) = G_t(i,w_t, s_t) - C_{cre}(s_{t-1}, \{i\})\enspace.$$
Minimising the end-to-end workload time, or rather, maximising the end-to-end workload time gained, is the goal of the bandit. As defined earlier, the total workload time $C_{tot}$ is the sum of \emph{execution}, \emph{recommendation} and \emph{creation} times accumulated over rounds. As such, minimising each round's summand is an equivalent problem. Modifying the execution time to the time gain while ignoring the recommendation time yields per-round super arm reward of:
\begin{align*}
    R_t(s_t) &= C_{exc}(w_t, \emptyset) - [C_{exc}(w_t,s_t) + C_{cre}(s_{t-1},s_t)]\\
    &\approx \sum_{i\in s_t} G_t(i,w_t,s_t) - \sum_{i \in s_t}C_{cre}(s_{t-1}, \{i\})\\
    &= \sum_{i\in s_t}r_t(i)\enspace.
\end{align*}
Selection of the execution plan depends on the query optimiser, and as noted, 
the query optimiser may resolve to a sub-optimal query plan. As we show, the bandit is nonetheless resilient as it can quickly recover from any such performance regressions. 
Observed execution times encapsulate real-world effects, \eg the interaction between queries, application properties, run-time parameters, etc. 
Since the end-to-end workload time includes the index creation and query execution times, we are indirectly optimising for both efficiency and the quality of recommendations. 

\begin{algorithm}[t!]
\caption{MAB Simulation for Index Tuning}\label{euclid}
\begin{algorithmic}[1]
\State $\textbf{QS} \gets QueryStore()$ \Comment{keeps query information}
\State $\textbf{C\textsuperscript{2}UCB} \gets InitialiseBandit()$ \Comment{A\ref{alg:c2ucb}, L 1-2}
\While{(TRUE)}
\State $\textbf{queries} \gets getLastRoundWorkload()$ 
\ForAll{$queries$}
\If{(isNewTemplate)} 
\State $QS.add(query)$
\Else
\State $QS.update(query)$
\EndIf
\EndFor
\State $\textbf{QoI} \gets QS.getQoI()$ \Comment{get queries of interest}
\State $\textbf{arms} \gets generateArms(QoI)$
\State $\textbf{X} \gets generateContext(arms, QoI)$
\State $\textbf{s\textsubscript{t}} \gets \textbf{C\textsuperscript{2}UCB}.recommend(arms, X)$ \Comment{A\ref{alg:c2ucb}, L 4-10}
\State $\textbf{C\textsubscript{cre}} \gets materialise(s\textsubscript{t})$
\State $\textbf{C\textsubscript{exc}} \gets executeCurrentWorkload()$ 
\State $\textbf{C\textsuperscript{2}UCB}.updateWeights(C\textsubscript{cre}, C\textsubscript{exc})$ \Comment{A\ref{alg:c2ucb}, L 12-13}
\EndWhile
\end{algorithmic}
\label{algorithm}
\end{algorithm}

\textbf{A greedy oracle for super-arm selection.} Recall that C$^2$UCB leverages a near-optimal oracle to select a super arm, based on individual arm scores~\cite{c2ucb}. As a sum of individual arm rewards, our super-arm reward has a (sub)modular objective function and (as easily proven) exhibits monotonicity and Lipschitz continuity. Approximate solutions to maximise submodular (diminishing returns) objective functions can be obtained with greedy oracles that are efficient and near-optimal~\cite{nemhauser1978analysis}. Our implementation uses such an oracle combined with filtering to encourage diversity. 

Initially, arms with negative scores are pruned. Then arm selection and filtering steps alternate, until the memory budget is reached. In the selection step, an arm is selected greedily based on individual scores. The filtering step filters out arms that are no longer viable under the remaining memory budget, or those that are already covered by the selected arms based on prefix matching. If a covering index is selected for a query, all other arms generated for that query will be filtered out. Note that filtering is a temporary process that only impacts the current round.

\textbf{Bandit learning algorithm.} 
Algorithm~\ref{algorithm} shows the MAB algorithm, which wraps Algorithm~\ref{alg:c2ucb} and handles the domain specific aspects of the implementation. We have divided Algorithm~\ref{alg:c2ucb} into three main parts, initialisation (lines 1-2), arm recommendation (lines 4-10) and weight vector update (lines 12-13). These segments are utilised in  Algorithm~\ref{algorithm} as C\textsuperscript{2}UCB function calls.
 After initialising the bandit, Algorithm~\ref{algorithm} summarises workload information using templates; these track frequency, average selectivity, first seen and last seen times of the queries which help to generate the best set of arms per round (\ie QoI). The context is updated after each round based on the workload and selected set of arms. 
 The bandit then selects the round's set of arms, forming a configuration to be materialised within the database.
 The reward will then be calculated based on  observed execution statistics on a new set of queries, and will be used to update the shared weight. 
 To support shifting workloads, where users' interests change over time, the learner may forget learned knowledge depending on the workload shift intensity (\ie the number of newly introduced query templates). 
 
 \malinga{In our implementation, we perform bandit updates separately for creation time reward and execution time reward (line 13 of Algorithm~\ref{alg:c2ucb}). At the creation cost update, we temporary make all context features 0 except for the context feature that is responsible for the index size. This can be viewed as an innovation of independent interest where we decompose the reward into multiple components and want to direct each reward component feedback to a subset of the features. We term this a \emph{focused update}.
 This idea invites a new flavour of bandits elaborated in the next section.}

\maketeal
\section{Contextual Combinatorial Bandit with Structured Rewards}  %
\begin{figure}[h]
\centering
\centering\includegraphics[width=\columnwidth]{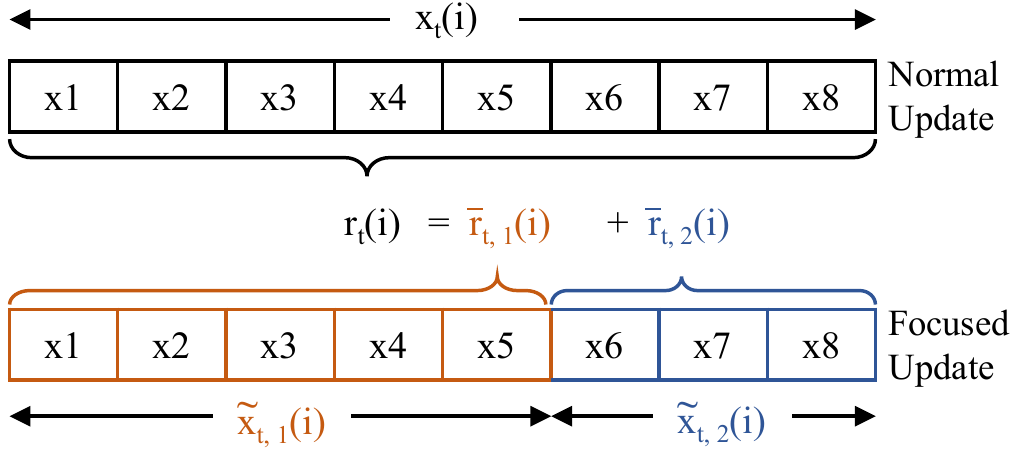}
\caption{Regular contextual updates vs focused update.}
\label{fig:focused-update}
\end{figure}

When rewards can be decomposed into component rewards under two key conditions,  we hypothesise that a \emph{focused updated} can result in faster convergence: (i) when each reward component is directly related to a small subset of context features we create lower dimensional supervised learning problems; and (ii) when each reward component is directly observed we offer more opportunity for bandit feedback. Under focused updates we use each component of the reward to learn part of the weight vector (see Figure~\ref{fig:focused-update}). Indeed for this structured setting we modify the proof the C$^2$UCB to arrive at tighter regret bounds  \rproof{by a factor of $1/\sqrt{n_f}$}, where $n_f$ is the number of reward components (\ie observed number of examples).

\makeorange
\label{sec:proof_mod}
Our approach to structured rewards is by a reduction to C$^2$UCB. We modify the C$^2$UCB's formulation as if two examples are observed for pulled arm $i$ in round $t$ with respective rewards $\bar{r}_{t,1}(i)$ and $\bar{r}_{t,2}(i)$. Throughout both (sub round) observations, the overall arm reward function is fixed as $r_t(i) = \bar{r}_{t,1}(i) + \bar{r}_{t,2}(i)$. This permits learning at a faster rate, while still coordinating an overall arm reward estimate.

\maketeal
\begin{example}
In the setting of this paper, we are motivated by the desire for context part 1 to be completely responsible for execution cost gains---we can learn the negative weight from index creation cost directly into part 2's index size feature. This allows us to switch off the creation cost overhead for already created arms by simply setting the index size context feature to zero. This ability to use domain knowledge to target reward feedback to a subset of context features is a powerful benefit of structured rewards. %
\end{example}

\makeorange
Let $\bar{\bm{x}}_{t,f}(i) \in \mathbb{R}^d$ be the $f$\textsuperscript{th} context of arm $i$ at round $t$, such that $r_t(i) = {\bm{\theta}}_\star'{\bm{x}}_{t}(i)+ \varepsilon_{t} = {\bm{\theta}}_\star'(\bar{\bm{x}}_{t,1}(i) + \bar{\bm{x}}_{t,2}(i)) + \varepsilon_{t,1} +\varepsilon_{t,2}$ for two independent zero-mean (subgaussian) noise random variables $\varepsilon_{t,1}$ and $\varepsilon_{t,2}$ with equal variance\footnote{Equivariance is without loss of generality. Were the  variances to differ, we obtain different expressions for $\bm{V}_t$ and $\bm{b}_t$, weighing data with less variance more: $\bm{V}_t = \lambda \bm{I} + \sum_{\tau = 1}^t \sum_{f=1}^{2} \sum_{i\in S_\tau} (\frac{\sigma_1\sigma_2}{\sigma_f})^2\bar{\bm{x}}_{\tau, f}(i) \bar{\bm{x}}_{\tau, f}(i)'$ and $\bm{b}_t = \sum_{\tau = 1}^{t} \sum_{f=1}^{2} \sum_{i\in S_\tau} (\frac{\sigma_1\sigma_2}{\sigma_f})^2 \bar{\bm{x}}_{\tau, f}(i) \bar{r}_{\tau, f}(i)~$. The hyperparameter $\lambda$ would need different adjustments since $\lambda = \sigma_1^2\sigma_2^2/\gamma^2$. The value of $\gamma^2$ stays the same, serving as the variance for the prior of $\bm{\theta}$.}. \rproof{To see the benefit of the focused update, we further assume that $\varepsilon_{t,1}$ and $\varepsilon_{t,2}$ are each $\frac{R}{\sqrt{2}}$-subgaussian, which makes $\varepsilon_t$ $R$-subgaussian.\footnote{For independent r.v.'s $X$ $R_1$-subgaussian and $Y$ $R_2$-subgaussian, $X+Y$ must be $\sqrt{R_1^2+R_2^2}$-subgaussian.} \eat{In \cite{c2ucb}, it is implied that $R=1$ due to the fact that the factor $R$ is missing from the regret bound, yet we will leave it in terms of $R$ for completeness.}} That the overall context is the sum of the sub-contexts does not change the role of the overall context for its respective arm, ensuring that the UCB in Equation~\eqref{eq:linucb-value} remains the same. We further assume complementary sparse sub-contexts: that overall context ${\bm{x}}_{t}(i)$ is the concatenation of  $\tilde{\bm{x}}_{t,1}(i) \in \mathbb{R}^{d_1}$ and $\tilde{\bm{x}}_{t,2}(i) \in \mathbb{R}^{d_2}$, where $d=d_1+d_2$ and $\bar{\bm{x}}_{t,1}(i)' =
\begin{bmatrix}
    \tilde{\bm{x}}_{t,1}(i)' &\bm{0}_{1\times d_2}
\end{bmatrix}
$ and $\bar{\bm{x}}_{t,2}(i)' =
\begin{bmatrix}
    \bm{0}_{1\times d_1} &\tilde{\bm{x}}_{t,2}(i)'
\end{bmatrix}.
$

To maintain valid learning given two observations per round with equal variances, at the end of round $t$ we generalise our matrix $\bm{V}_t$ and $b_t$ updates to:
\begin{align*}
    \bm{V}_t &= \lambda \bm{I} + \sum_{\tau = 1}^t \sum_{f=1}^{2} \sum_{i\in S_\tau} \bar{\bm{x}}_{\tau, f}(i) \bar{\bm{x}}_{\tau, f}(i)'\\
    \bm{b}_t &= \sum_{\tau = 1}^{t} \sum_{f=1}^{2} \sum_{i\in S_\tau} \bar{\bm{x}}_{\tau, f}(i) \bar{r}_{\tau, f}(i)~.
\end{align*}
Notice that this is different from C$^2$UCB's definition of $\bm{V}_t$ and $\bm{b}_t$, and hence a new regret analysis is warranted.

We exploit the fact that Theorem 4.2 in \cite{c2ucb} is true regardless of the super arm $S_t$. Therefore, solely for the purpose of modifying the aforementioned theorem, we re-index the context and rewards such that $\bar{\bm{x}}_{t,f}(i) = \bar{\bm{x}}_{t}(i+kf)$ and $\bar{r}_{t,f}(i) = \bar{r_{t}}(i+kf)$, and we construct the effective super arm $S'_t = \{i': i'=i+kf, i\in S_t, f\in\{0,1\}\}$. As such, our definition of $\bm{V}_t$ and $\bm{b}_t$ can now be rewritten as:
\begin{align*}
    \bm{V}_t &= \lambda \bm{I} + \sum_{\tau = 1}^t \sum_{i\in S'_\tau} \bar{\bm{x}}_{\tau}(i) \bar{\bm{x}}_{\tau}(i)'\\
    \bm{b}_t &= \sum_{\tau = 1}^{t} \sum_{i\in S'_\tau} \bar{\bm{x}}_{\tau}(i) \bar{r}_{\tau}(i)~,
\end{align*}
which is syntactically the same as the definition given in \cite{c2ucb}. We need to be more careful in concluding the theorem, however, since it contains an intermediate step involving $\det(\bm{V}_t)$ as defined in \cite{abbasi2011improved}. Assuming that $\|\bm{x}_t(i)\|\leq 1$, we bound $\det(\bm{V}_t)$ as follows:
\begin{align*}
    \det(\bm{V}_t) &\leq \left( \frac{\Tr(\bm{V}_t)}{d} \right)^d\\
    &= \left( \frac{\Tr(\lambda \bm{I}_d + \sum_{\tau = 1}^t \sum_{i\in S'_\tau} \bar{\bm{x}}_{\tau}(i) \bar{\bm{x}}_{\tau}(i)')}{d} \right)^d\\
    &= \left( \frac{\Tr(\lambda \bm{I}_d) + \sum_{\tau = 1}^t \sum_{i\in S'_\tau} \Tr(\bar{\bm{x}}_{\tau}(i) \bar{\bm{x}}_{\tau}(i)')}{d} \right)^d\\
    &= \left( \frac{\lambda d + \sum_{\tau = 1}^t \sum_{i\in S'_\tau} \|\bar{\bm{x}}_{\tau}(i)\|_2^2}{d} \right)^d\\
    &= \left( \frac{\lambda d + \sum_{\tau = 1}^t \sum_{i\in S_\tau} \left(\|\bar{\bm{x}}_{\tau,1}(i)\|_2^2 + \|\bar{\bm{x}}_{\tau,2}(i)\|_2^2\right)}{d} \right)^d\\
    &= \left( \frac{\lambda d + \sum_{\tau = 1}^t \sum_{i\in S_\tau} \|{\bm{x}}_{\tau}(i)\|_2^2}{d} \right)^d\\
    &\leq \left( \frac{\lambda d + \sum_{\tau = 1}^t \sum_{i\in S_\tau} 1}{d} \right)^d\\
    &\leq \left( \frac{\lambda d + tk}{d} \right)^d~,
\end{align*}
where we have used AM-GM Inequality for the first inequality and the fact that the overall context is the first and second individual relevant contexts concatenated to arrive at the last equality.
Finally, using the fact that $\bm{V}_{t-1} \preceq \bm{V}_{t}$\rproof{, that the noise is $\frac{R}{\sqrt{2}}$-subgaussian} and Theorem 2 from \cite{abbasi2011improved}, Theorem 4.2 from \cite{c2ucb} becomes:
\begin{align*}
    \|\hat{\bm{\theta}}_t - \bm{\theta}_\star\|_{\bm{V}_{t-1}} &\leq \|\hat{\bm{\theta}}_t - \bm{\theta}_\star\|_{\bm{V}_{t}}\\
    &\leq \rproof{\frac{R}{\sqrt{2}}}\sqrt{2\log \left( \frac{\det(\bm{V}_t)^{1/2}}{\delta\det(\lambda \bm{I}_d)^{1/2}}\right)} + \lambda ^{1/2} S\\
    & \leq \rproof{\frac{R}{\sqrt{2}}}\sqrt{d\log\left(\frac{1+tk/\lambda}{\delta}\right)} + \lambda^{1/2}S~,
\end{align*}
with probability at least $1-\delta$, which is the same as that in \cite{c2ucb}, with the exception of the definition of $\bm{V}_{t}$.

Conveniently, Lemma 4.1 from \cite{c2ucb} is written in terms of $\bm{V}_{t-1}$ and $\alpha_t$, thus the proof follows exactly \rproof{besides the choice of $\alpha_t$}, rewritten below for convenience:
\begin{lemma}
If $\alpha_t =\rproof{\frac{R}{\sqrt{2}}} \sqrt{d\log\left(\frac{1+tk/\lambda}{\delta}\right)} + \lambda^{1/2}S$, then we have
$$0 \leq \hat{r}_{t}(i) - r_{t}^\star(i) \leq 2\alpha_t\|\bm{x}_{t}(i)\|_{\bm{V}_{t-1}^{-1}}$$
holds simultaneously for all $t\geq 0$ and $i\in[k]$ with probability at least $1-\delta$.
\end{lemma}

We have provided the correction of the proof of Lemma 4.2 from \cite{c2ucb} in Appendix~\ref{sec:theory}. This proof can be used by changing the definition of the matrix $\bm{X}_T$ into:
$$\bm{X}'_{T} = \begin{bmatrix}
    \bar{\bm{x}}'_{T,1}(s_{(1,T)}) \\
    \vdots\\
    \bar{\bm{x}}'_{T,1}(s_{(|S_{T}|,T)})\\
    \bar{\bm{x}}'_{T,2}(s_{(1,T)})\\ \vdots \\ \bar{\bm{x}}'_{T,2}(s_{(|S_{T}|,T)})~.
\end{bmatrix}\enspace.$$
Then we rewrite
\begin{align*}
    \det(\bm{V}_T)
    &= \det\left(\bm{V} + \sum_{t=1}^T\sum_{f=1}^{2}\sum_{i\in s_t}\bar{\bm{x}}_{t,f}(i)\bar{\bm{x}}_{t,f}(i)'\right)\\
    &= \det\left(\bm{V} + \sum_{t=1}^{T-1}\sum_{f=1}^{2}\sum_{i\in s_t}\bar{\bm{x}}_{t,f}(i)\bar{\bm{x}}_{t,f}(i)' \right.+\\
    &\qquad\qquad\left.\sum_{f=1}^{2}\sum_{i\in s_T} \bar{\bm{x}}_{T,f}(i)\bar{\bm{x}}_{T,f}(i)'\right)\\
    &= \det\left(\bm{V}_{T-1} + \bm{X}_{T}\bm{X}_{T}'\right)\enspace,
\end{align*}
and the rest of the proof follows very similarly, with slight difference in the dimension of $\bm{X}_T$, changing from $|s_T|$ into $2|s_T|$. Finally, the third last equality requires us to find the trace of the matrix of interest, which is $\Tr(\bm{X}_T'\bm{V}_{T-1}^{-1}\bm{X}_T)= \sum_{f=1}^{2}\sum_{i\in s_T} \bm{x}_{T,f}(i)\bm{V}_{T-1}^{-1}\bm{x}_{T,f}(i)$~, which in turn gives us our new determinant inequality $$\det(\bm{V}_T) \geq \det(\mathbf{V}_{T-1})\left( 1 + \sum_{f=1}^{2}\sum_{i\in s_T} \|\bm{x}_{T,f}(i)\|^2_{\bm{V}_{T-1}^{-1}}\right)~.$$ Therefore, we have our modification of Lemma 4.2 of \cite{c2ucb} as follows:
\begin{lemma}
    Let $\bm{V} \in \mathbb{R}^{d\times d}$ be a positive definite matrix, $s_t \subseteq \{1,\cdots,k\}$ where $|s_t|\leq \ell$ for $t=1,2,\dots$, and $\bm{V}_T=\bm{V}+\sum_{t=1}^T\sum_{f=1}^{2}\sum_{i\in s_t}\bar{\bm{x}}_{t,f}(i)\bar{\bm{x}}_{t,f}(i)'$. Then, if $\forall t,i \, \,\,\lambda\geq\ell$ and $||\bm{x}_{t}(i)||_2\leq 1$ for concatenated context $\bm{x}_{t}(i) = \bar{\bm{x}}_{t,1}(i) + \bar{\bm{x}}_{t,2}(i)$ where $\bar{\bm{x}}_{t,1}(i)' =
\begin{bmatrix}
    \tilde{\bm{x}}_{t,1}(i)' &\bm{0}_{1\times d_2}
\end{bmatrix}
$ and $\bar{\bm{x}}_{t,2}(i)' =
\begin{bmatrix}
    \bm{0}_{1\times d_1} &\tilde{\bm{x}}_{t,2}(i)'
\end{bmatrix}
$, we have
    \begin{align*}
        \sum_{t=1}^T\sum_{i\in s_t}\|\bm{x}_{t}(i)\|^2_{\bm{V}_{t-1}^{-1}}&=\sum_{f=1}^{2}\sum_{t=1}^T\sum_{i\in s_t}\|\bm{x}_{t,f}(i)\|^2_{\bm{V}_{t-1}^{-1}}\\ &\leq 2 \log \det \bm{V}_T - 2 \log \det \bm{V}\\
        &\leq 2d \log ((\Tr(\bm{V})+T\ell)/d) - \\
        &\qquad2\log \det \bm{V}\enspace.
    \end{align*}
\end{lemma}

Since there is no modification on the objective function, and since all the theorems and lemma required to arrive at the final regret bound are the same, the regret bound for the modified C\textsuperscript{2}UCB stays the same, as stated in \cite{c2ucb}:
\begin{align*}
    \sum_{t=1}^TReg_t^\alpha &\leq  C \rproof{\frac{R}{\sqrt{2}}}\sqrt{8Td \log\left(1 + \frac{Tk}{d\lambda}\right)}~\cdot \\
    &\quad\quad\quad\left(\sqrt{d\log\left(\frac{1+Tk/\lambda}{\delta}\right)} + \sqrt{\lambda}S\right)~.
\end{align*}

\rproof{Notice that in cases where there are $n_f$ examples per arm in each round instead of only two, the regret will generalise into:
\begin{align*}
    \sum_{t=1}^TReg_t^\alpha &\leq  C \rproof{\frac{R}{\sqrt{n_f}}}\sqrt{8Td \log\left(1 + \frac{Tk}{d\lambda}\right)}~\cdot \\
    &\quad\quad\quad\left(\sqrt{d\log\left(\frac{1+Tk/\lambda}{\delta}\right)} + \sqrt{\lambda}S\right)~,
\end{align*}
which has a factor of $\frac{1}{\sqrt{n_f}}$, $n_f\in \mathbb{N}$ compared to the original C\textsuperscript{2}UCB where $n_f=1$.}
\color{black}
\section{Experimental Methodology}
\label{sec:evaluation}
We evaluate our MAB framework across a range of widely used analytical and HTAP industrial \linebreak[4] benchmarks, comparing it to a state-of-the-art physical design tool shipped with a commercial database product referred to as the Physical Design Tool (PDTool). This is a mature product, proven to outperform other physical design tools available on the market \cite{DBTest2012PhysicalDesigners,magicmirror}.
We also test our framework against a highly tuned RL agent, which is a popular technique adopted in recent database tuning research~\cite{no_dba,COREIL,partitioningCarsten,pavlo}.

\begin{figure*}[t]
\centering
\begin{minipage}{0.19\textwidth}
\centering\includegraphics[width=\textwidth]{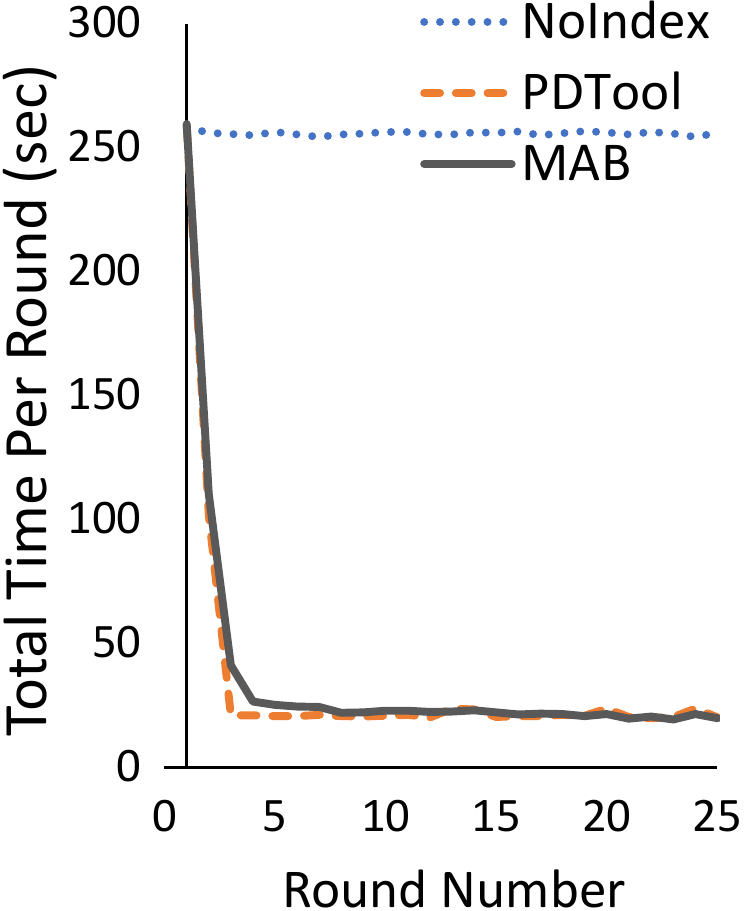}\\ \textbf{(a)}
\end{minipage}\hfill
\begin{minipage}{0.19\textwidth}
\centering\includegraphics[width=\textwidth]{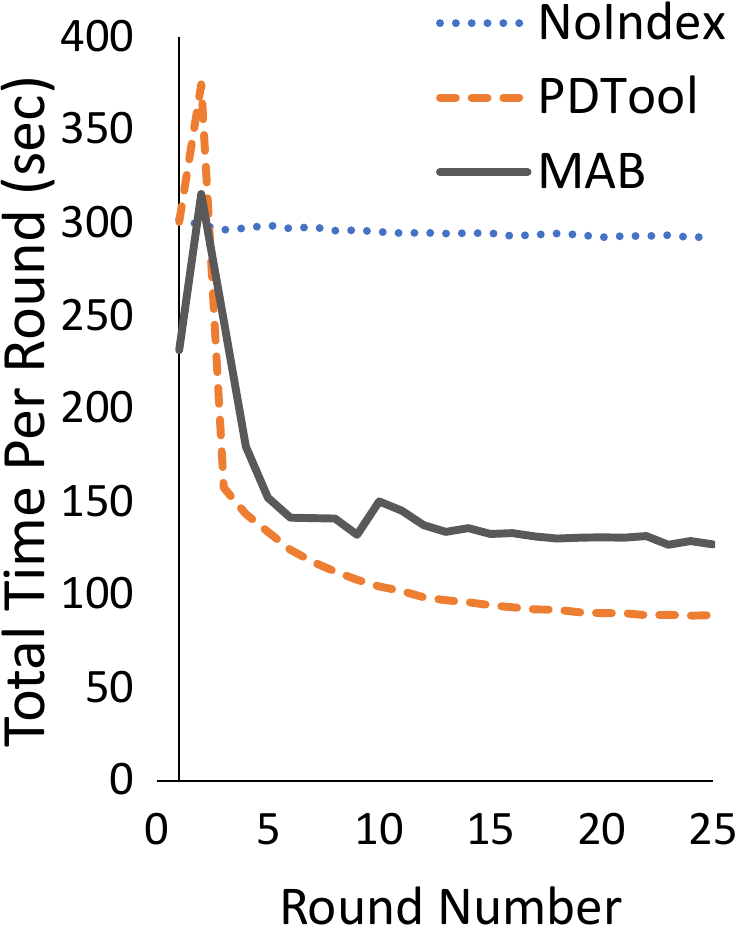}\\ \textbf{(b)}
\end{minipage}\hfill
\begin{minipage}{0.19\textwidth}
\centering\includegraphics[width=\textwidth]{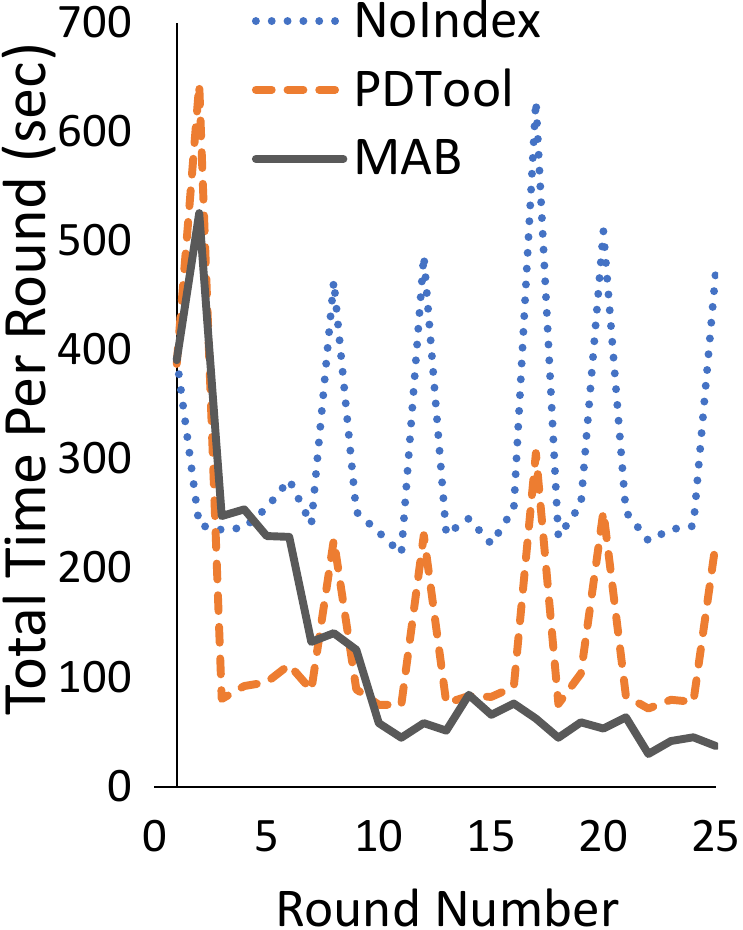}\\ \textbf{(c)}
\end{minipage}\hfill
\begin{minipage}{0.19\textwidth}
\centering\includegraphics[width=\textwidth]{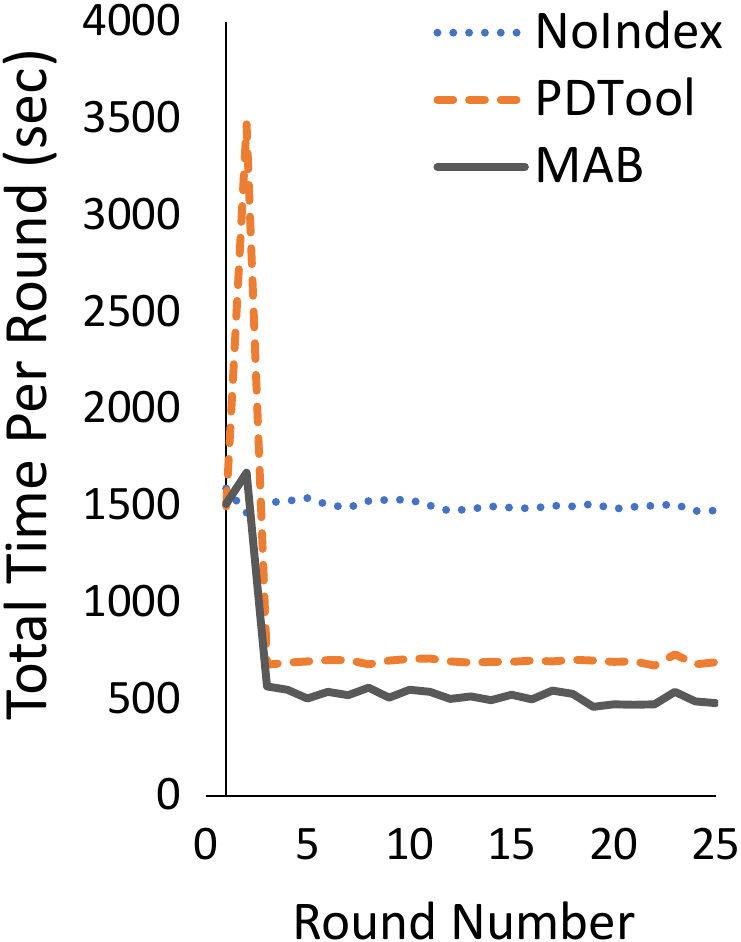}\\ \textbf{(d)}
\end{minipage}\hfill
\begin{minipage}{0.19\textwidth}
\centering\includegraphics[width=\textwidth]{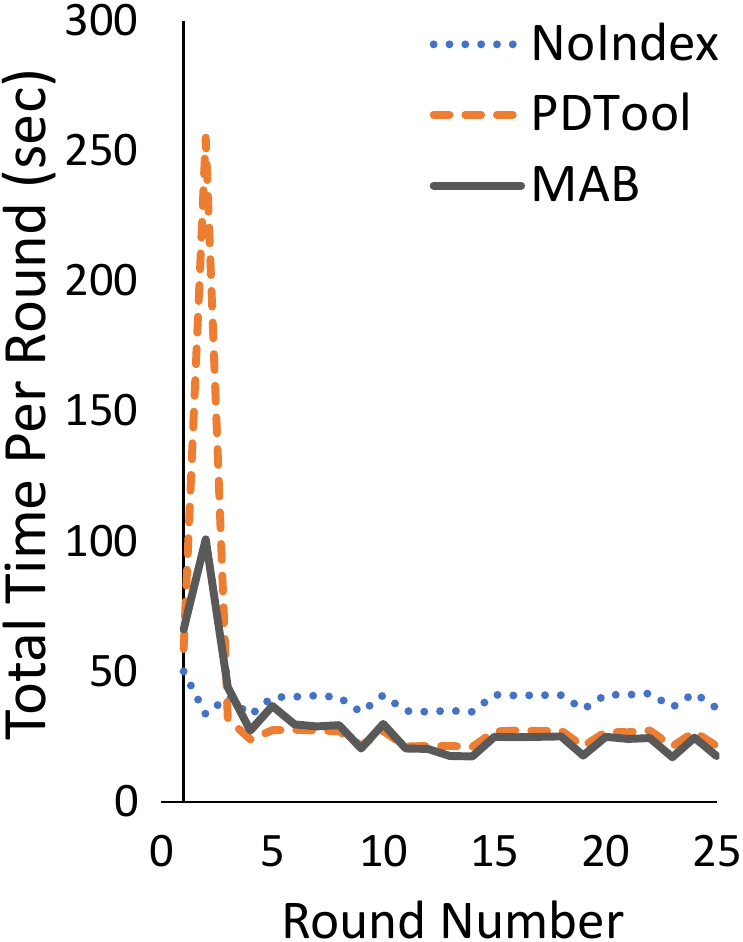}\\ \textbf{(e)}
\end{minipage}\\
\caption{MAB vs. PDTool convergence for \emph{static} workloads: (a) SSB, (b) TPC-H, (c) TPC-H Skew, (d) TPC-DS and (e) IMDb.}
\label{fig:static-convergence}
\end{figure*}

\begin{figure}[t]
\centering
\includegraphics[width=\columnwidth]{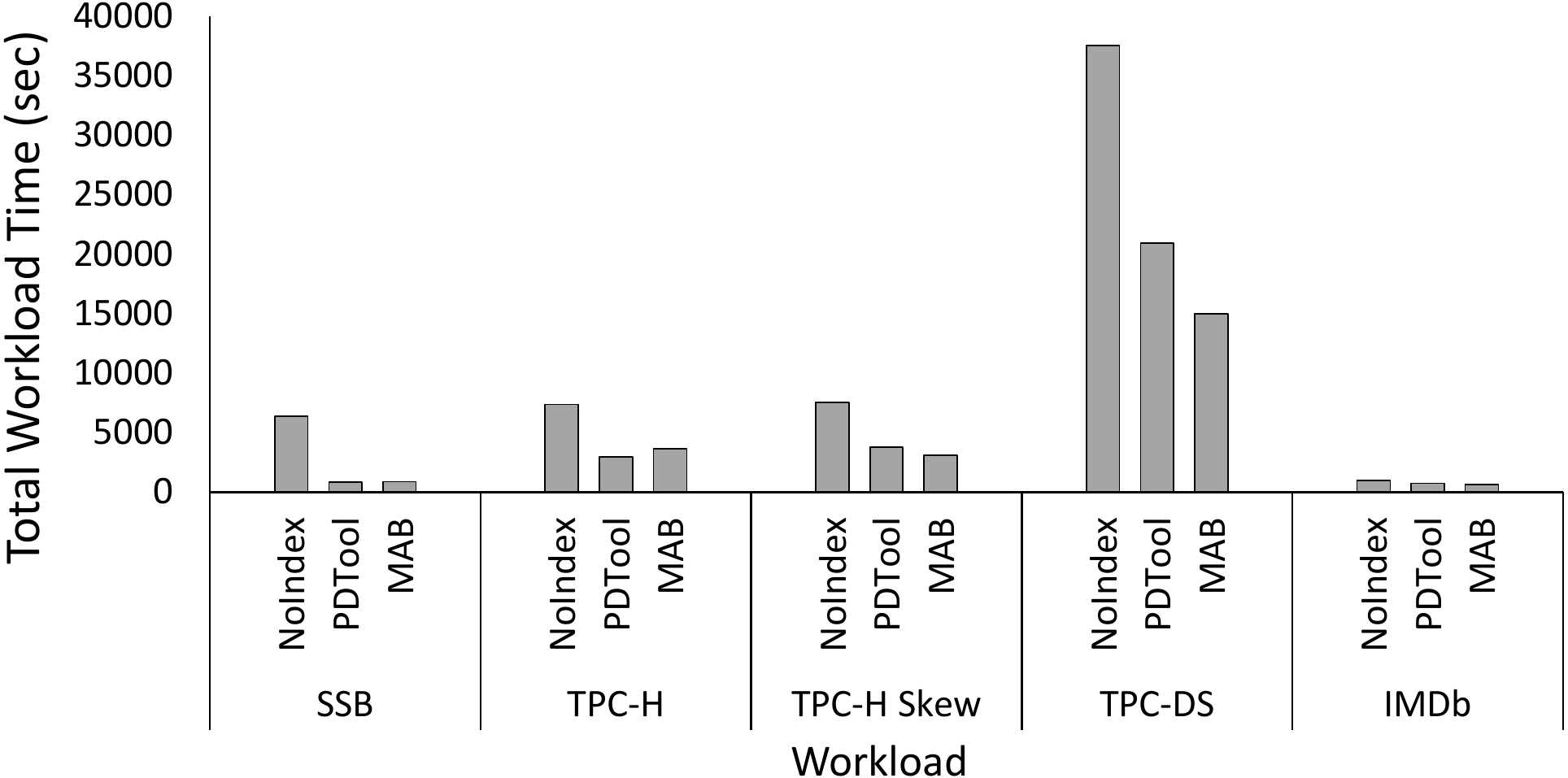}\\
\caption{MAB vs. PDTool total end-to-end workload time for \emph{static} workloads.}
\label{fig:static-total}
\end{figure}

\textbf{Benchmarks.}
We use five publicly available \malinga{analytical processing} benchmarks: TPC-H (with uniform distribution)~\cite{tpch} and TPC-H Skew~\cite{tpchskew} with zipfian factor 4, allowing the reader to understand the impact of data skewness when all the other aspects are kept identical; TPC-DS~\cite{tpcds}, which probes solution fitness under a large number of candidate configurations; SSB~\cite{SSB} with easily achievable high index benefits; and finally, Join Order Benchmark (JOB) with IMDb dataset (a real-world dataset)~\cite{HowGoodQO} (henceforth referred to as IMDb) a challenging workload for index recommendations, with index overuse leading to performance regressions. 

\malinga{For HTAP performance testing we use  CH-BenCH\-mark \cite{chbenchmark,difallah2013oltp}, TPC-H and TPC-H Skew benchmarks. CH-BenCH\-mark provides a complex mixed workload, combining TPC-C and TPC-H benchmarks. The \linebreak[4]  CH-BenCHmark schema is composed of unmodified \linebreak[4] TPC-C schema and three tables (Supplier, Region, Nation) from TPC-H. Its workload is composed of TPC-C~\cite{tpcc} transactional workload and modified 22 TPC-H~\cite{tpch} queries adapted to the CH-BenCHmark schema.}

\malinga{While CH-BenCHmark provides a uniform dataset, we are not aware of any HTAP benchmarks with skewed data generation. While there are some OLTP benchmarks with skewed datasets~\cite{difallah2013oltp}, they do not provide the required OLAP complexity for the index selection problem to be interesting. Due to the limitations in existing benchmarks we decided to extend the TPC-H skew benchmark to include INSERT, DELETE and UPDATE queries to mimic a skewed HTAP \linebreak[4] benchmark~\cite{tpchhtap}. Data generation tool for TPC-H skew already provides the functionality to generate data for inserts and deletes. In our extension we additionally perform updates on existing records using the data generated for the new inserts. To highlight the impact of skewness on the overall performance, we report comparable HTAP results on TPC-H database as well.}  

Unless stated otherwise, all experiments use scale factor (SF) 10, resulting in approximately 10GB of data per workload, except in the case of the IMDb dataset which has a fixed size of 6GB.\footnote{\malinga{CH-BenCHmark does not scale with the SF parameter like most of the other benchmarks we use. It uses a number of warehouses (similar to TPC-C benchmark) as a scaling parameter. For our experiments we use 137 warehouses which generate approximately 10GB dataset.}} We consider three broad types of workloads, allowing us to compare different aspects of the recommendation process:
\begin{enumerate}

    \item {\emph{Static}}: The workload sequence is known in advance, and repeating over time (modelling workloads used for reporting purposes). In the absence of dynamic environment complexities, this simpler setting allows us to single-out the effectiveness (the ability to find a good configuration) and the efficiency (the search overhead) of the MAB search strategy.
    
    \item \emph{Dynamic shifting}: The region of interest shifts over time from one group of queries to another (modelling data exploration). This experiment evaluates the adaptation speed to workload shifts and the cost of exploration when adapting.

    \item \emph{Dynamic random}: A query sequence is chosen entirely at random (modelling more dynamic settings, such as cloud services). Dynamic random experiments test the delicate balance between swift and careful adaptation under returning workloads, which can lead to unwanted index oscillations.
\end{enumerate}

Across experiments, each group of templatized quer\-ies is invoked over rounds, producing different query instances. For static and dynamic settings, PDTool is invoked every time after the first round of new queries, with those queries given as the training workload, since this workload will become representative of future \linebreak[4] rounds. This setting is somewhat unrealistic and favour\-able for PDTool, since in real-life the PDTool will seldom truly have knowledge of the representative workload (\ie what is yet to arrive in the future), advantaging the PDTool in our experiments. However, it presents a viable comparison against the workload-oblivious \linebreak[4] MAB. Bandits do not use any workload information ahead of time, but instead observe a workload sequence and react accordingly. 

\textbf{Physical design tuning parameters.}
Both PD\-Tool and MAB are given a memory budget approximately equal to the size of the data (1x; 10GB for SF 10 datasets and 6GB for IMDb dataset) for the creation of secondary indices. We have experimented with different memory budgets ranging from 0.25x to 2x (since benefits of additional memory seem to diminish beyond a 2x limit) under TPC-H and TPC-H skew benchmarks, and observed the same patterns throughout that range.\footnote{Both tools converge to the same execution cost by the final round, when enough memory was given to fit the entire useful configuration.} We have naturally picked the middle of the active region (1x) as our default memory budget. All these workloads come with original primary and foreign keys that influence the choice of indices. We grant the aforementioned memory budget on top of this. 

In search of the best possible design, we do not constrain the running time of PDTool, with one exception: In TPC-DS dynamic random, PDTool was uncompetitive due to long running times,\footnote{A single PDTool invocation took around 8 hours (default limit). The total recommendation time was around 40 hours, which is not competitive compared to the end-to-end workload time of 4 hours under MAB.} hence the PDTool running time of each invocation was restricted to 1 hour. All proposed indices are materialised and queries invoked over the same commercial DBMS in both cases (MAB and PDTool).

\textbf{Metrics.}
In addition to reporting total end-to-end workload time for all rounds, we also report the total workload time per round used to demonstrate the convergence of different tools. Additionally, we present the total workload time broken down by recommendation time (when invoking the PDTool or the MAB framework), index creation time, and workload execution time. For completeness, we show original query times, without any secondary indices (denoted as NoIndex).
In addition to convergence graphs of individual benchmarks, we present a summary graph with total end-to-end workload time for all rounds under MAB and PDTool tuning of SSB, TPC-H (uniform), TPC-H skew, TPC-DS and IMDb benchmarks.

\textbf{Hardware.}
All experiments are performed on a server equipped with 2x 24 Core Xeon Platinum 8260 at 2.4GHz, 1.1TB RAM, and 50TB disk (10K RPM) running Windows Server 2016. 
We report cold runs, clearing database buffer caches prior to every query execution. 

\section{Experimental Results}
In this section, we report on wide ranging empirical comparisons of MAB and PDTool, while supporting both  analytical and HTAP workloads. Finally, we present results against a well tuned reinforcement learning agent. 

\subsection{\malinga{MAB vs  PDTool Under Analytical Workloads}}
\label{sec:analytical-evaluation}

\begin{figure*}[t]
\centering
\begin{minipage}{0.19\textwidth}
\centering\includegraphics[width=\textwidth]{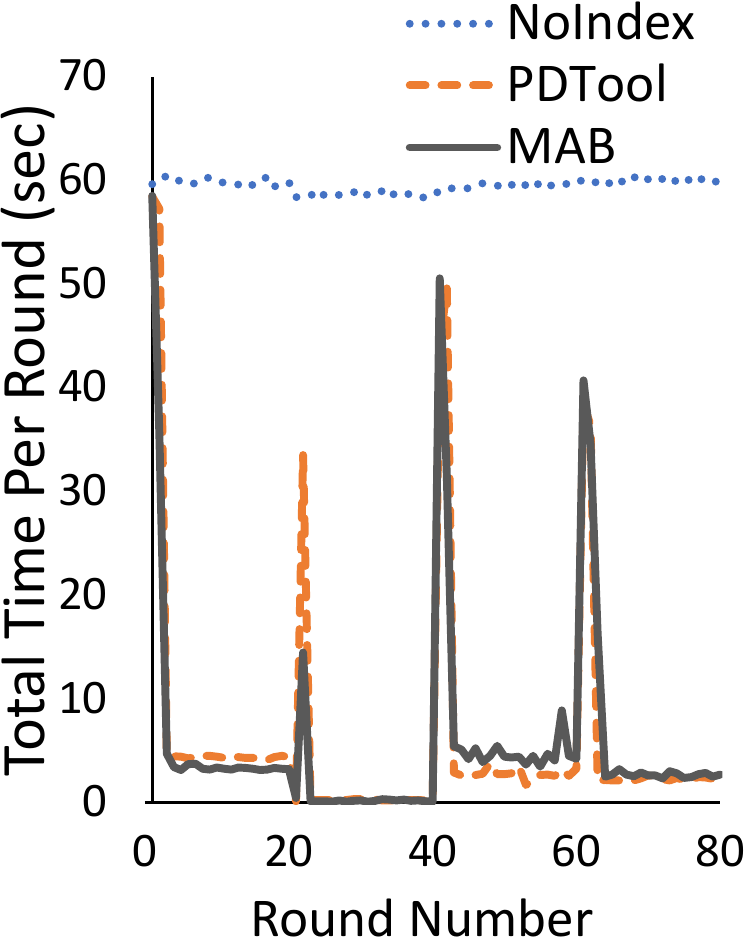}\\ \textbf{(a)}
\end{minipage}\hfill
\begin{minipage}{0.19\textwidth}
\centering\includegraphics[width=\textwidth]{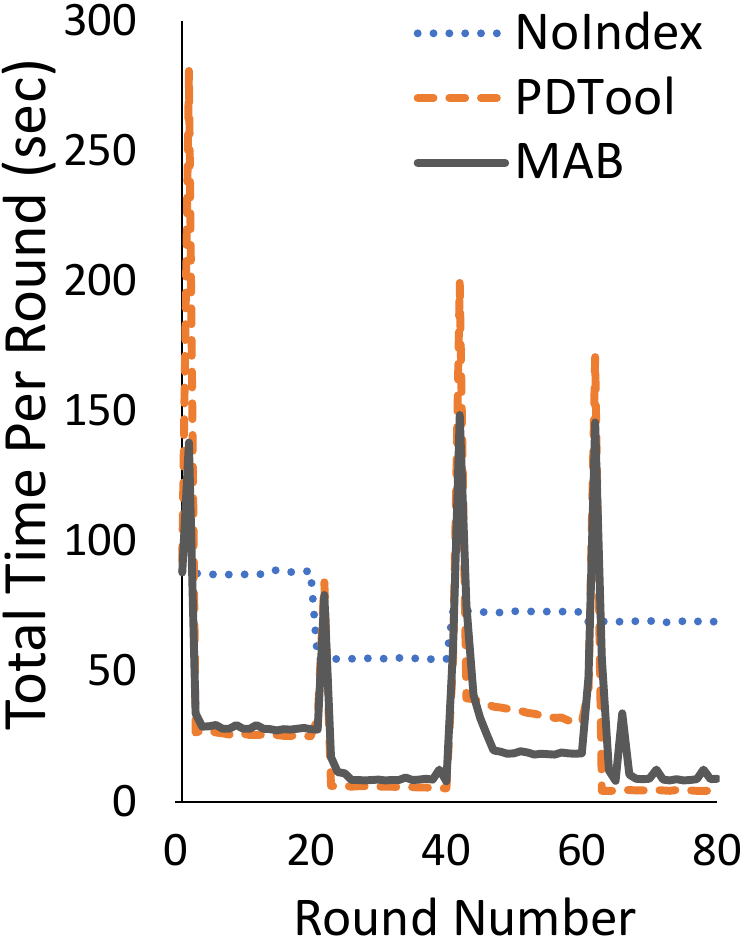}\\ \textbf{(b)}
\end{minipage}\hfill
\begin{minipage}{0.19\textwidth}
\centering\includegraphics[width=\textwidth]{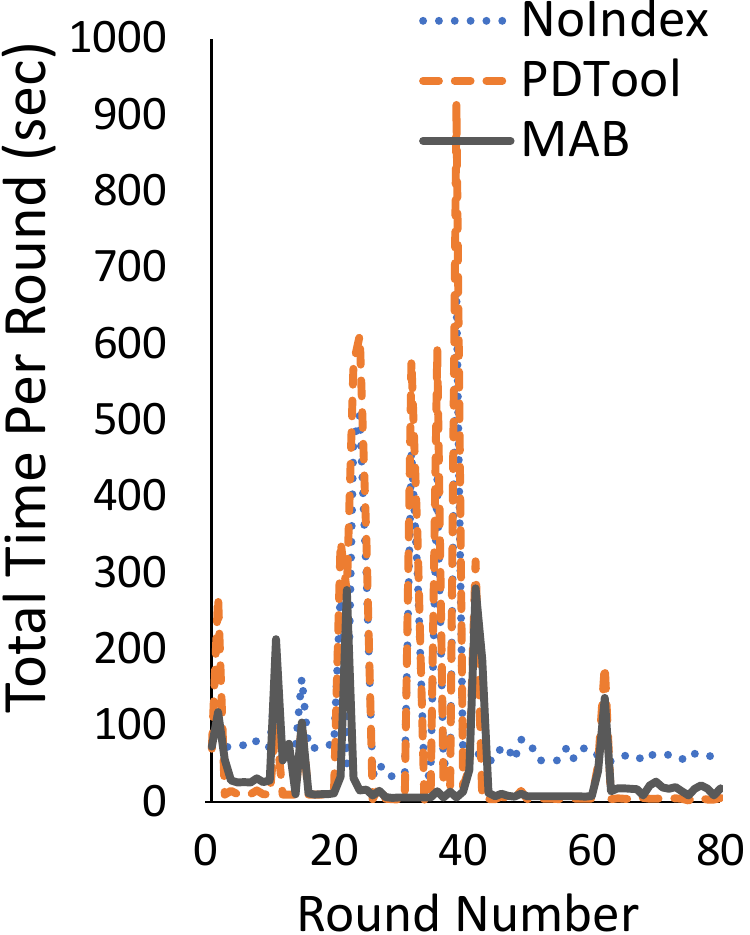}\\ \textbf{(c)}
\end{minipage}\hfill
\begin{minipage}{0.19\textwidth}
\centering\includegraphics[width=\textwidth]{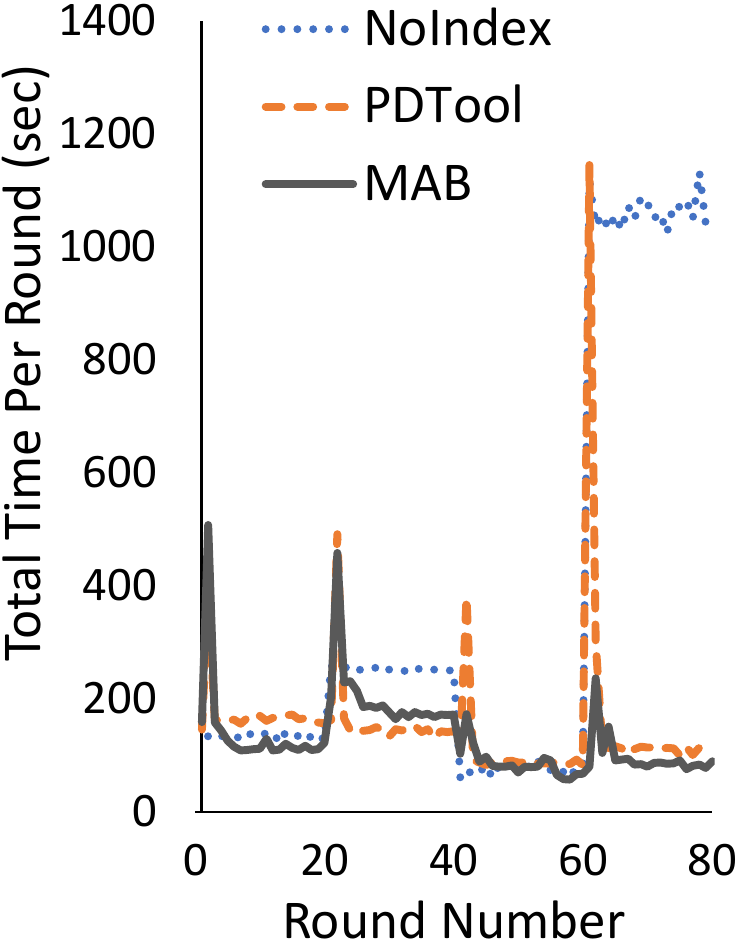}\\ \textbf{(d)}
\end{minipage}\hfill
\begin{minipage}{0.19\textwidth}
\centering\includegraphics[width=\textwidth]{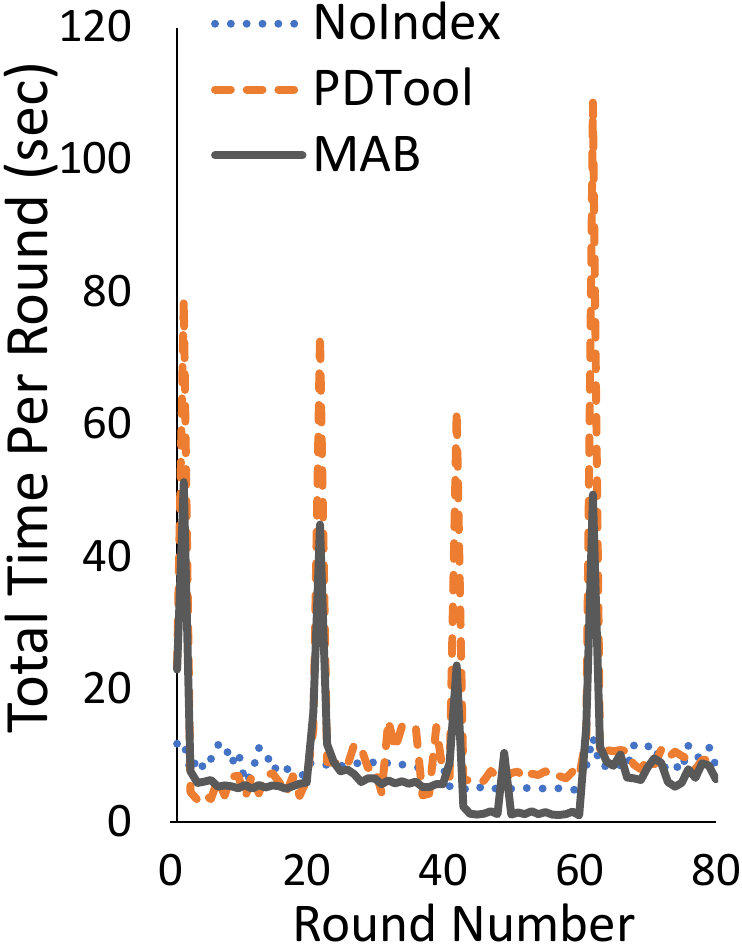}\\ \textbf{(e)}
\end{minipage}\\
\caption{MAB vs. PDTool Convergence for \emph{dynamic shifting} workloads: (a) SSB, (b) TPC-H, (c) TPC-H Skew, (d) TPC-DS and (e) IMDb.}
\label{fig:dynamic-convergence}
\end{figure*}

\begin{figure}[t]
\centering
\includegraphics[width=\columnwidth]{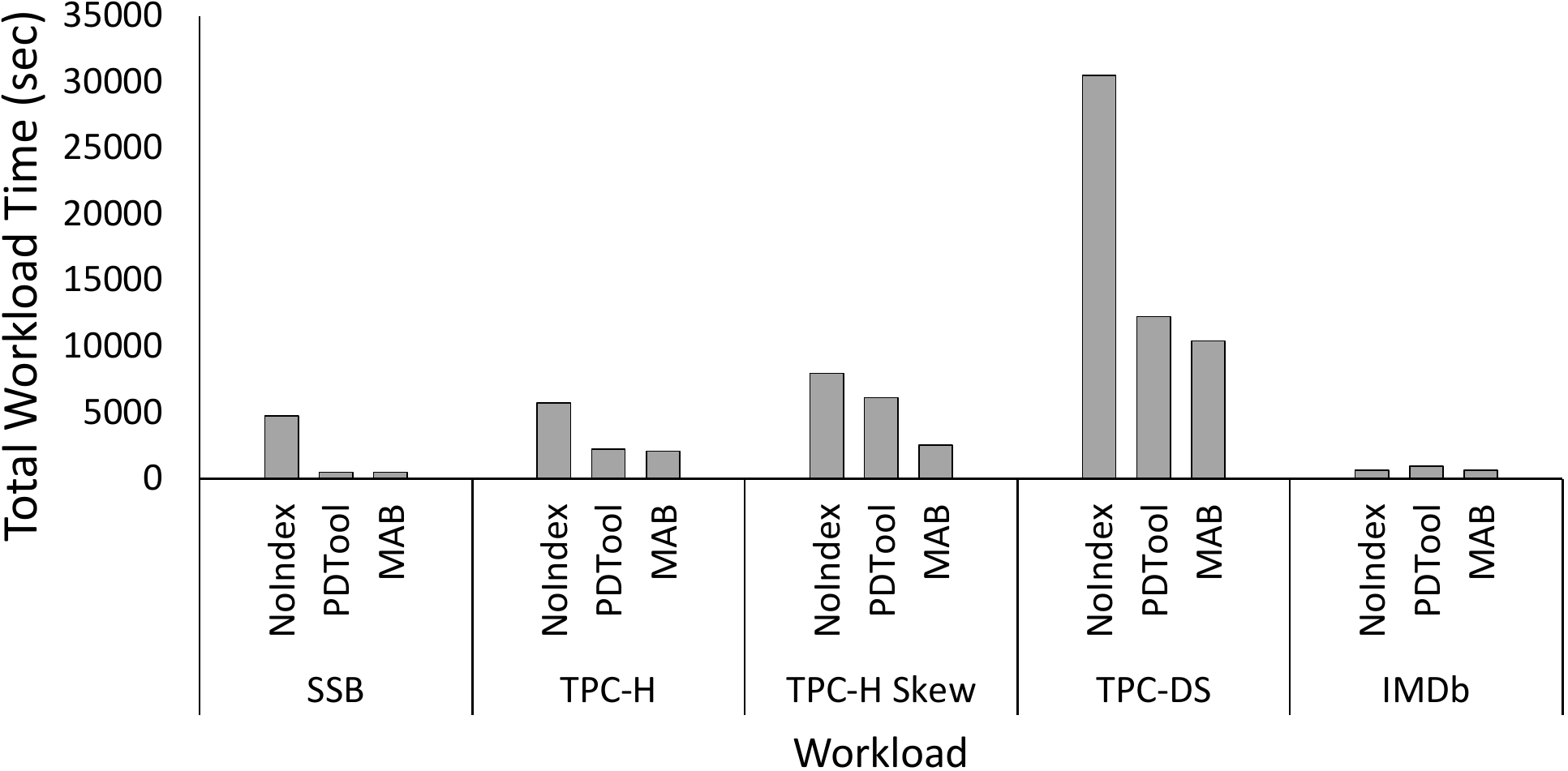}
\caption{MAB vs. PDTool total end-to-end workload time for \emph{dynamic shifting} workloads.}
\label{fig:dynamic-total}
\end{figure}

\textbf{Static workloads.} Static workloads over uniform data\-sets are the best case for offline physical design tools, as a pre-determined workload sequence may perfectly represent future queries. However, when underlying data is skewed, recommendations based on a pre-determined workload alone can have unfavourable outcomes. While used for reporting, static workloads do not reflect modern dynamic workloads (\eg data exploration)~\cite{StratosExploration}. In static workloads,
all query templates in the benchmark (22, 13, 99 and 33 templates for TPC-H, SSB, TPC-DS and IMDb, respectively) are invoked once every round, each with a different query instance of the template, for a total of 25 rounds, providing sufficient time for   convergence.

Figure~\ref{fig:static-total} displays overall workload time (including recommendation and index creation time) for all 25 rounds under MAB and PDTool. For skewed datasets (TPC-H Skew, TPC-DS and IMDb) MAB outperforms PDTool. MAB shows over 17\%, 28\% and 11\% performance gain against PDTool, under TPC-H Skew, TPC-DS, IMDb benchmarks, respectively. Under uniform \linebreak[4] datasets (TPC-H and SSB), both MAB and PDTool provide significant performance gains over NoIndex (over 50\% and 85\%, respectively), while PDTool outperforms the MAB (by 19\% and 5\%). This is not surprising since for \malinga{fully analytical,} uniform, static experiments usually align with PDTool assumptions where the future can be perfectly represented by a pre-determined workload. 

Convergence plots in Figure~\ref{fig:static-convergence}(a--e), show MAB's gradual improvement over 25 rounds. Both MAB and PDTool have large spikes after the first round for all the workloads.
For both tools, this is due to recommendation and creation of indices. However, MAB might drop proposed indices and create new ones later on, generating relatively smaller spikes in subsequent rounds. Nonetheless, MAB efficiently balances the exploration of new indices, reducing exploration with time.

\emph{What is the best search strategy?} 
Comparison of execution times in the final round of the static experiment provides a clear idea about the benefit of using execution cost guided search. As evident from Figure~\ref{fig:static-convergence}(a--e), in 4 out of 5 cases, MAB converges to a better configuration than PDTool. MAB provides over 5\%, 84\%, 31\% and 19\% better execution time by the last round (25\textsuperscript{th}) compared to PDTool under SSB, TPC-H Skew, TPC-DS and IMDb, respectively.

For TPC-H skew, PDTool misses a vital index on $Orders.O\_custkey$. This index boosts the performance of some queries (Q22 in particular) which MAB correctly detects and materialises. Missing this leads to large execution times in a few rounds including the last round (8, 12, 17, 20 and 25) for PDTool. These experiments illustrate the risk of relying on the query optimiser and imperfect statistics as a single source of truth. 

The only case when MAB is outperformed by the PDTool is under TPC-H (PDTool delivers over 21\% better execution time by the last round); different indices are proposed, as our current MAB framework does not support an index merging phase employed by some physical design tools~\cite{chaudhuri1999merging}. Instead, MAB uses individual queries to propose index candidates. We plan to address index merging in future work. 

\begin{figure*}[t]
\centering
\begin{minipage}{0.19\textwidth}
\centering\includegraphics[width=\textwidth]{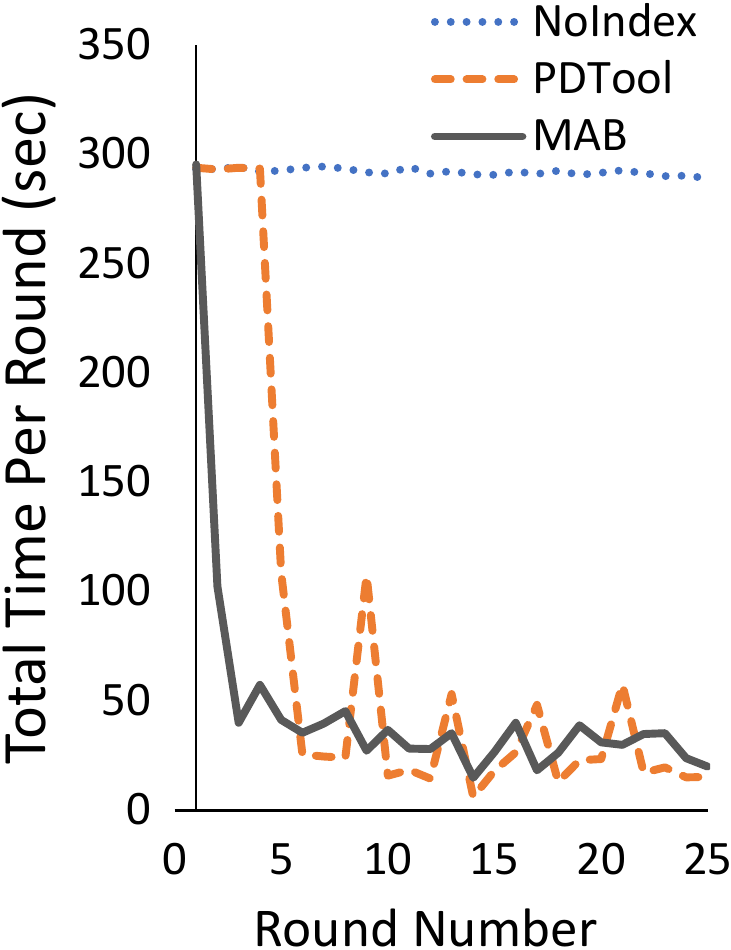}\\ \textbf{(a)}
\end{minipage}\hfill
\begin{minipage}{0.19\textwidth}
\centering\includegraphics[width=\textwidth]{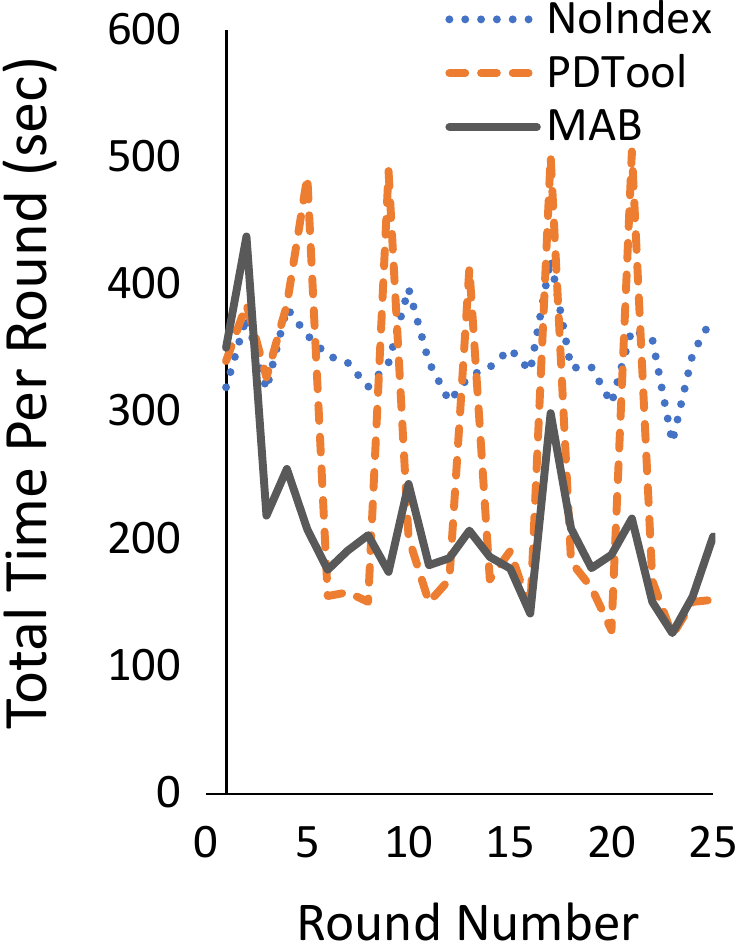}\\ \textbf{(b)}
\end{minipage}\hfill
\begin{minipage}{0.19\textwidth}
\centering\includegraphics[width=\textwidth]{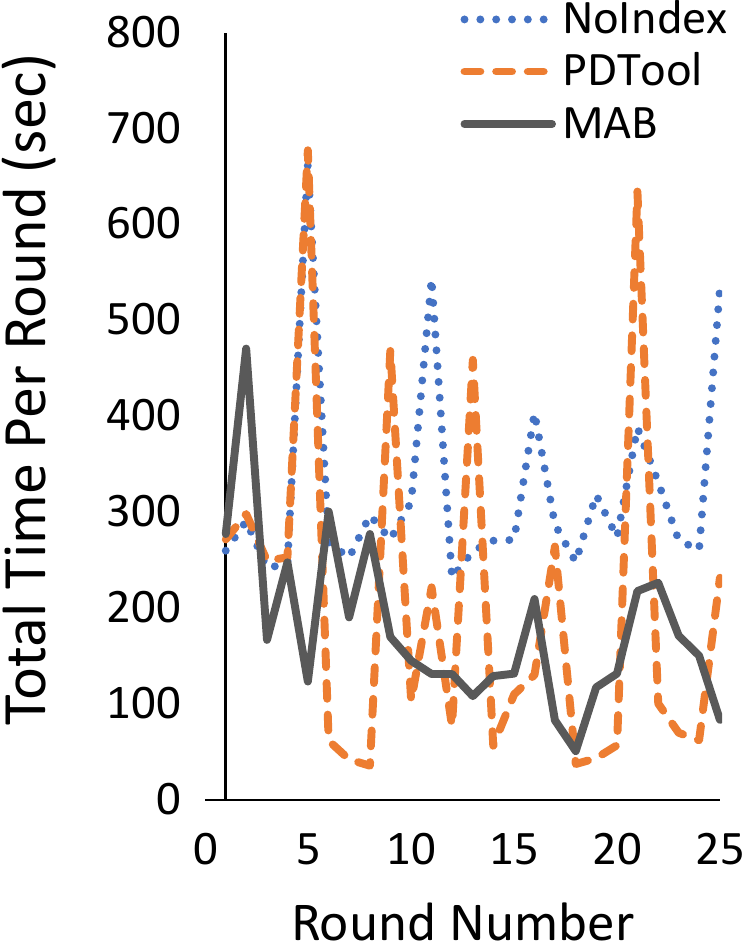}\\ \textbf{(c)}
\end{minipage}\hfill
\begin{minipage}{0.19\textwidth}
\centering\includegraphics[width=\textwidth]{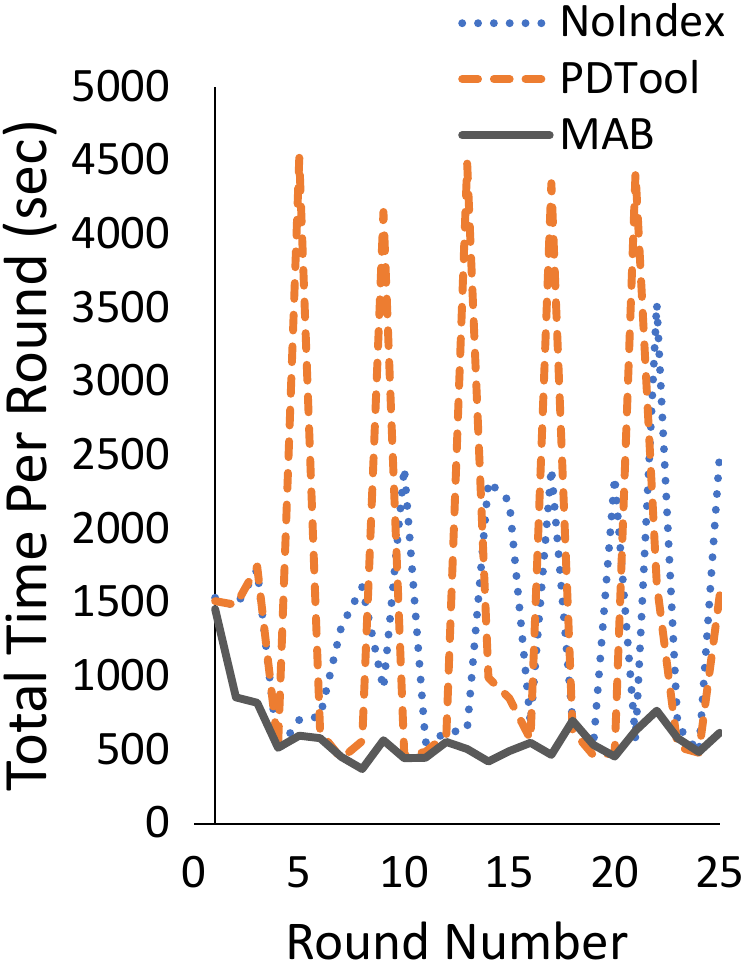}\\ \textbf{(d)}
\end{minipage}\hfill
\begin{minipage}{0.19\textwidth}
\centering\includegraphics[width=\textwidth]{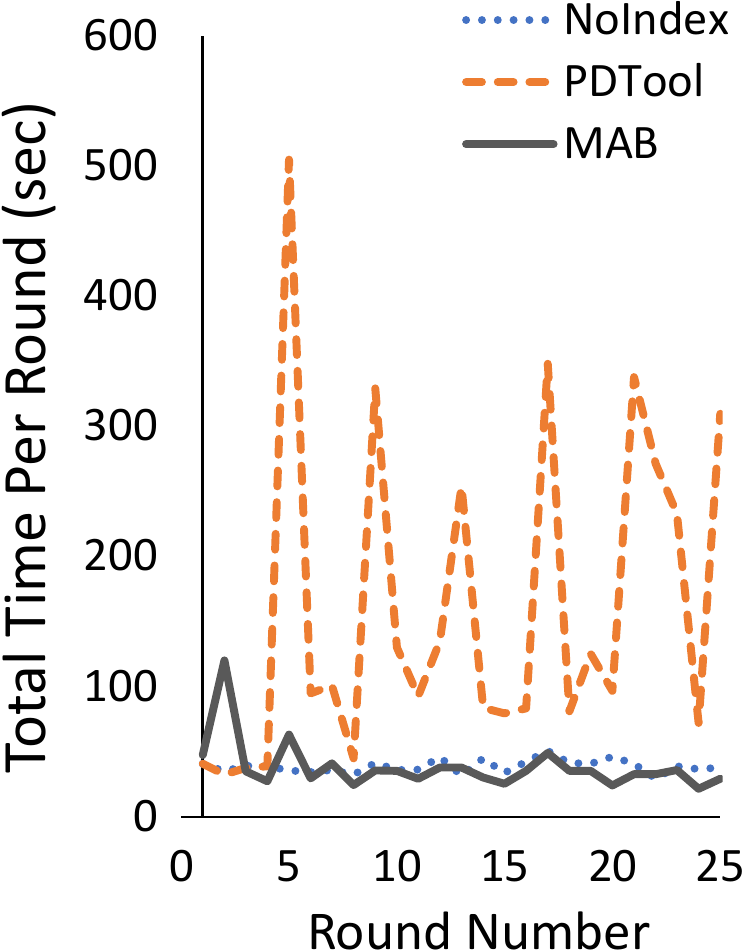}\\ \textbf{(e)}
\end{minipage}\\
\caption{MAB vs. PDTool Convergence for \emph{dynamic random} workloads: (a) SSB, (b) TPC-H, (c) TPC-H Skew, (d) TPC-DS and (e) IMDb}
\label{fig:random-convergence}
\end{figure*}

\begin{figure}[t]
\centering
\includegraphics[width=\columnwidth]{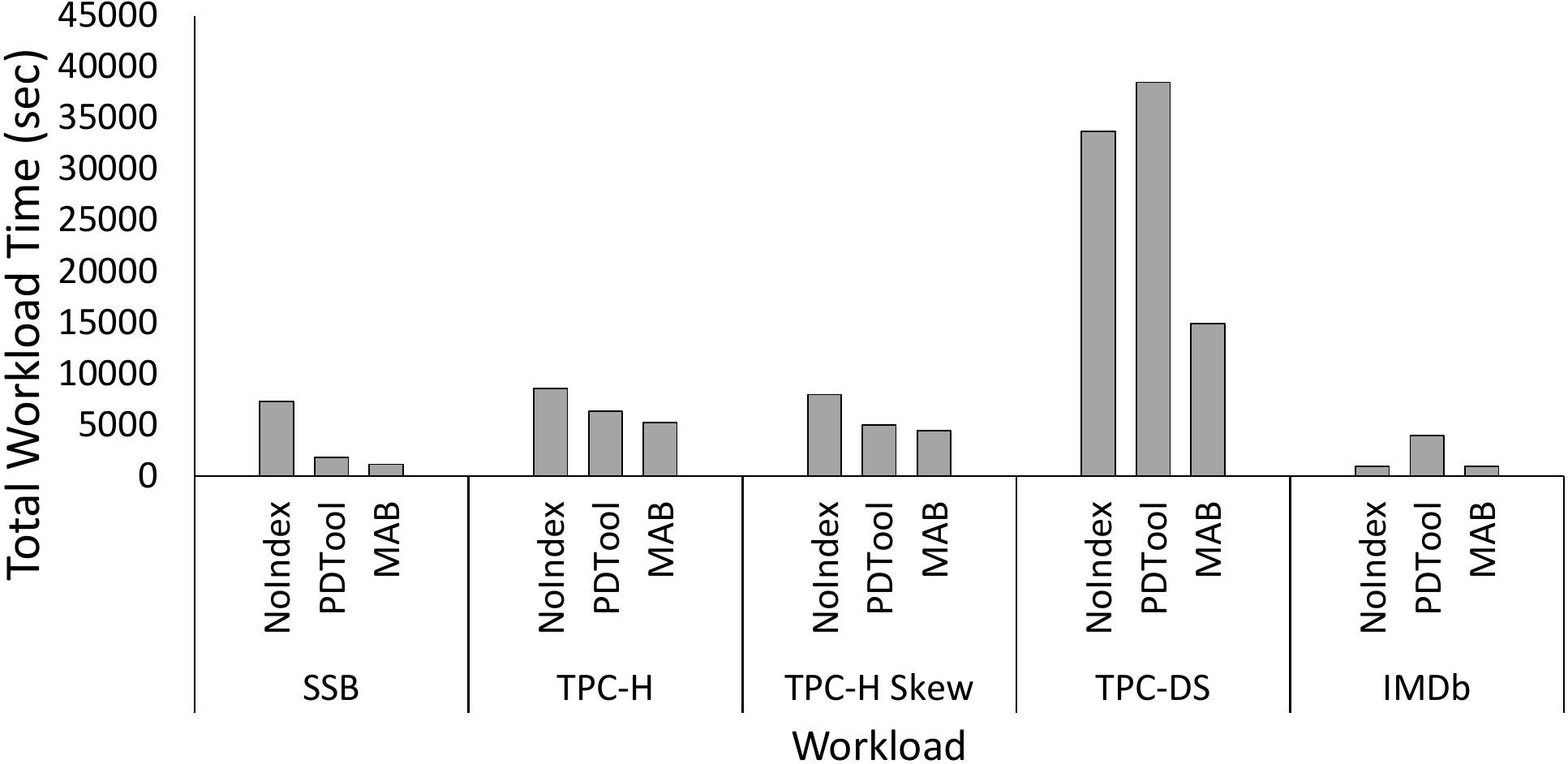}\\ 
\caption{MAB vs. PDTool total end-to-end workload time for \emph{dynamic random} workloads.}
\label{fig:random-total}
\end{figure}

\textbf{Dynamic shifting workloads.} Under the dynamic shifting workloads, all query templates in the benchmark are randomly divided into 4 equal-sized groups.
A group of query templates is then executed for 20 rounds, after which the workload switches to a new group of unseen queries (no overlap with the previous queries). 
When the workload switches, PDTool is invoked and trained on the new sequence of queries (whose templates will be used in the next 19 rounds).\footnote{This relaxation assumes a DBA with knowledge that the previous workload will not be repeated, placing PDTool at an advantage. In reality, proposing training workloads might be much more challenging for dynamic workloads.} Thus, PDTool is invoked four times in total (in rounds 2, 22, 42, 62). On the other hand, the MAB framework does not assume any workload knowledge.

Figure~\ref{fig:dynamic-total} displays MAB's end-to-end workload time as substantially lower compared to the alternatives, under all benchmarks. MAB provides over 3\%, 6\%, 58\%, 14\% and 34\% speed-up compared to PDTool, under SSB, TPC-H, TPC-H Skew, TPC-DS and IMDb, respectively. 

Interestingly, NoIndex performs better than PDTool against the IMDb workload. PDTool has a higher total workload time as well as higher execution time compared to NoIndex. NoIndex provides 3.5\% (24 seconds) speedup in execution time over PDTool. This is mainly due to misestimates of the optimiser~\cite{das2019automatically}. As an example (out of many), query 18 takes less than 1 second under NoIndex, whereas with the created indices by PDTool some instances of this query take around 7--8 seconds due to a suboptimal plan chosen favouring the index usage.
This affects both MAB and PDTool, but MAB identifies the indices with a negative impact based on the reward and drops them. For the IMDb workload which does not get much support from indices, MAB provides 3\% total performance gain and 26\% execution time gain compared to NoIndex.

One can easily observe the workload shifts in Figure~\ref{fig:dynamic-convergence}(a--e) due to the spikes in rounds 2, 22, 42, and 62. For PDTool, this is due to the invocation of PDTool and index creation after the workload shifts. Similar spikes can be seen in the MAB line with automatic detection of workload shifts. Further random spikes can be observed, for PDTool, from rounds 20-40 in TPC-H skew and rounds 30-40 under IMDb, due to the issues discussed in the previous paragraphs (Q22 in TPC-H Skew, Q18 in IMDb).\\[-0.8em]

\textbf{Dynamic random workloads.} We simulate modern data analytics workloads that are truly ad-hoc in nature. For instance, cloud providers, hosting millions of databases, seldom can detect representative queries, since they frequently change~\cite{das2019automatically}. In such cases, it is common to invoke the PDTool periodically (\eg nightly or weekly), with the training workload taken to be \linebreak[4]queries observed since last invocation. In this setting, we invoke the PDTool every 4 rounds, using queries from the last 4 rounds as the representative workload. 
In the dynamic random setting, the number of total training queries in the complete sequence is similar to the number of queries we had in the static setting. However, we have no control over the selection of queries for the workload and they are chosen completely randomly. The sequence is then divided into 25 equal-sized rounds. In all cases, the round-to-round repeat workload was between 45-54\%.

As shown in Figure~\ref{fig:random-total}, again we see a considerably lower total workload time of MAB compared to PDTool. MAB provides over 37\%, 17\%, 11\%, 61\% and 75\% speed-up compared to PDTool, under SSB, TPC-H, TPC-H Skew, TPC-DS and IMDb, respectively.  
It is notable that in Figure~\ref{fig:random-total}, the total workload time of PDTool climbs higher than NoIndex on two occasions, in TPC-DS and IMDb. In IMDb, this is due to the same issue discussed previously under dynamic shifting workloads (due to the optimiser's misestimates, favouring the usage of sub-optimal indices, \eg IMDb Q18). 
While \linebreak[4]PDTool has a much better execution time than NoIndex under TPC-DS (execution time of 5.3h under PDTool vs 9.3h under NoIndex), due to high recommendation time (5.1 hours, see Table~\ref{tab:full}), PDTool ends up with a higher total workload time. Under these 2 benchmarks (TPC-DS and IMDb), MAB provides over 55\% and 1.5\% performance gain over NoIndex, respectively. In Figure~\ref{fig:random-convergence}(a--e), we can see five major spikes for PDTool due to the tuning invocations (in rounds 5, 9, 13, 17, 21).

\begin{table*}[ht]
\begin{center}
\centering
\caption{Total time breakdown (min): the best choice is in bold text.}
\begin{tabular}{|l|l|l|l|l|l|l|l|l|l|} 
\hline
\multicolumn{2}{|l|}{\multirow{2}{*}{Workload}} & \multicolumn{2}{l|}{Recommendation} & \multicolumn{2}{l|}{Creation} & \multicolumn{2}{l|}{Execution} & \multicolumn{2}{l|}{Total} \\ 
\cline{3-10}
\multicolumn{2}{|l|}{} & PDTool (\#) & MAB & PDTool & MAB & PDTool & MAB & PDTool & MAB \\ 
\hline
\parbox[t]{2mm}{\multirow{5}{*}{\rotatebox[origin=c]{90}{Static}}} & SSB & 0.34 (0.34) & \textbf{0.02} & \textbf{0.95} & 1.86 & \textbf{12.9} & 13.15 & \textbf{14.19} & 15.03 \\ 
\cline{2-10}
 & TPC-H & 0.6 (0.6) & \textbf{0.08} & \textbf{2.45} & 5.66 & \textbf{46.35} & 55.64 & \textbf{49.4} & 61.38 \\ 
\cline{2-10}
 & TPC-H Sk. & 0.58 (0.58) & \textbf{0.11} & \textbf{8.37} & 19.82 & 54.17 & \textbf{32.06} & 63.12 & \textbf{51.99} \\ 
\cline{2-10}
 & TPC-DS & 44.86 (44.86) & \textbf{1.53} & \textbf{1.45} & 5.94 & 302.63 & \textbf{242.15} & 348.94 & \textbf{249.62} \\ 
\cline{2-10}
 & IMDB & 0.34 (0.34) & \textbf{0.31} & \textbf{1.1} & 1.3 & 11.01 & \textbf{9.42} & 12.41 & \textbf{11.03} \\ 
\hline
\parbox[t]{2mm}{\multirow{5}{*}{\rotatebox[origin=c]{90}{Dynamic}}} & SSB & 1.28 (0.32) & \textbf{0.05} & \textbf{1.5} & 2.21 & \textbf{5.42} & 5.69 & 8.2 & \textbf{7.95} \\ 
\cline{2-10}
 & TPC-H & 1.55 (0.32) & \textbf{0.12} & \textbf{9.36} & 9.74 & 26.35 & \textbf{25.14} & 37.25 & \textbf{35} \\ 
\cline{2-10}
 & TPC-H Sk. & 1.65 (0.41) & \textbf{0.16} & \textbf{14.98} & 20.96 & 85.49 & \textbf{21.44} & 102.11 & \textbf{42.56} \\ 
\cline{2-10}
 & TPC-DS & 11.13 (2.78) & \textbf{1.66} & \textbf{6.08} & 16.48 & 187.08 & \textbf{155.65} & 204.29 & \textbf{173.79} \\ 
\cline{2-10}
 & IMDB & 3.09 (0.77) & \textbf{0.29} & \textbf{1.59} & 2.24 & 11.21 & \textbf{7.93} & 15.89 & \textbf{10.46} \\ 
\hline
\parbox[t]{2mm}{\multirow{5}{*}{\rotatebox[origin=c]{90}{Random}}} & SSB & 2.83 (0.57) & \textbf{0.02} & \textbf{1.77} & 2.37 & 26.59 & \textbf{16.83} & 30.85 & \textbf{19.22} \\ 
\cline{2-10}
 & TPC-H & 7.55 (1.51) & \textbf{0.08} & 14.68 & \textbf{7.06} & 84.14 & \textbf{80.43} & 106.37 & \textbf{87.57} \\ 
\cline{2-10}
 & TPC-H Sk. & 3.3 (0.66) & \textbf{0.08} & \textbf{31.74} & 34.68 & 48.71 & \textbf{39.44} & 83.75 & \textbf{74.2} \\ 
\cline{2-10}
 & TPC-DS & 310.22 (62.04) & \textbf{1.4} & \textbf{8.23} & 19.81 & 323.57 & \textbf{227.02} & 642.01 & \textbf{248.24} \\ 
\cline{2-10}
 & IMDB & 14.74 (2.94) & \textbf{0.28} & 2.72 & \textbf{1.14} & 48.55 & \textbf{14.47} & 66.01 & \textbf{15.89} \\
\hline
\multicolumn{10}{@{}p{120mm}}{\footnotesize \# The average time of a single PDTool invocation}
\end{tabular}
\label{tab:full}
\end{center}
\end{table*}

\begin{table}[ht]
\centering
\caption{Total end-to-end workload time for static workloads under different database sizes (min).}
\begin{tabular}{|l|l|l|l|}
\hline
Workload & SF & PDTool & MAB \\ \hline
\multirow{3}{*}{TPC-H} & 1 & \textbf{2.02} & 2.03 \\ \cline{2-4} 
 & 10 & \textbf{49.4} & 61.38 \\ \cline{2-4} 
 & 100 & 891.01 & \textbf{793.40} \\ \hline
\multirow{3}{*}{\begin{tabular}[c]{@{}l@{}}TPC-H \\ Skew\end{tabular}} & 1 & 4.17 & \textbf{3.83} \\ \cline{2-4} 
 & 10 & 63.12 & \textbf{51.99} \\ \cline{2-4} 
 & 100 & 2640.64 & \textbf{1219.33} \\ \hline
\end{tabular}
\label{tab:db-size}
\end{table}

\begin{table}[ht]
\centering
\caption{Recommendation times (min) vs. workload complexity.}
\begin{tabular}{|l|l|l|l|}
\hline
Workload & SSB & TPC-H & TPC-DS \\ \hline
PDTool & 0.84 & 1.36 & 44.86 \\ \hline
MAB & 0.05 & 0.14 & 1.53 \\ \hline
\end{tabular}
\label{tab:workload-complexity}
\end{table}

\subsubsection{The Impact of Database Size} 
\label{sec:db-size-impact}
To examine the impact of database size, we run TPC-H uniform and TPC-H Skew static experiments on SF 1, 10 and 100 databases. As previously discussed, under SF 10, MAB performed better in the case of TPC-H Skew and PDTool performed better on TPC-H (see Table~\ref{tab:db-size}). The impact of sub-optimal index choices is even more evident for larger databases, leading to a huge gap between total workload times of MAB and PDTool for TPC-H Skew (44 hours in the former vs 20 hours in the latter case). In TPC-H, PDTool results in a higher total workload time (14.8 hours vs. 13.2 hours for MAB). 
This is mainly due to sub-optimal optimiser decisions, where the optimiser favours the usage of indices (coupled with nested loops joins) when alternative plans would be a better option. For instance, under the recommended indices from PDTool, some instances of Q5 run longer than 8 minutes (using index nested loops join), where others finish in 1.5 minutes (using a plan based on hash joins). We notice that, with larger database sizes, execution time dominates contributing more than 91\% to the total workload time. We observe faster and more accurate convergence of MAB under larger databases, due to a clear difference between rewards for different arms, highlighting MAB's excellent potential benefits for larger databases. 

\subsubsection{Hypothetical Index Creation vs Actual Index Creation} \label{sec:exploration-exploitation}

Managing the exploration-exploitation balance under a large number of candidate indices, with an enormous number of combinatorial choices, is non trivial. PDTool explores using the ``what-if" analysis, which comes under the tool's recommendation time, whereas MAB explores using index creations. 

\textbf{Cost of hypothetical index creation:} When \linebreak[4]
 analysing PDTool's average invocation times in dynamic shifting (small workloads) and dynamic random (large workloads) settings, it becomes evident that PDTool invocation cost grows noticeably with the training workload size, under all benchmarks (see Table~\ref{tab:full}). As an example, PDTool tuning of the TPC-DS benchmark grows from 3 minutes in the dynamic shifting setting (25-query workload) to 1 hour in the dynamic random setting (400-query workload). Furthermore, multiple invocations required in dynamic random and shifting settings aggravate the problem further for PDTool (see Table~\ref{tab:full}). On the other hand, PDTool recommendation time rapidly increases with the complexity of the workloads. In an experiment with 100 query workloads from SSB, TPC-H and TPC-DS (with the complexity of SSB $<$ TPC-H $<$ TPC-DS), it is evident that the complexity of a workload has a considerable impact on the PDTool recommendation time (see Table~\ref{tab:workload-complexity}).

MAB recommendation times stay significantly lower and stable despite the workload shifts and changes in complexity or size (see Tables~\ref{tab:full} and \ref{tab:workload-complexity}). In all experiments, MAB takes less than 1\% of the total workload time for recommendation, except for IMDb where it takes around 2\% (due to low total workload time and a high number of query templates). More than 80\% of this recommendation time is spent on the initial setup (1\textsuperscript{st} round) and the continuous overhead is negligible.

\textbf{Cost of actual index creation:} While actual execution statistics based search allows the MAB to converge to better configurations, as a down side, MAB spends more time on index creation (see Table~\ref{tab:full}). For instance, under TPC-H and TPC-H Skew static experiments, MAB spends 5.6 and 19.8 minutes on index creation where PDTool only spends 2.4 and 8.3 minutes, respectively. Under skewed data, rewards show more variability which delays the convergence for MAB. This leads to higher exploration and greater creation costs. While MAB is still competitive due to efficient exploration, we consider ways to improve its convergence in future work.

\textbf{Final verdict:} Comparing the total of recommendation and index creation times (henceforth referred to as \emph{exploration cost}) between MAB and PDTool presents a clear picture about these two exploration methods. From Table~\ref{tab:full} we can observe that, in most cases (9 out of 15) MAB archives a better exploration cost compared to PDTool when running analytical workloads. However when the workload is small (\eg dynamic shifting) PDTool tends to perform better. TPC-DS, with the highest number of candidate indices among these benchmarks (over 3200 indices), provides a great test case for exploration efficiency. Under TPC-DS, MAB exploration cost is significantly lower in shifting and random settings, and marginally higher in the static setting. Despite the efficient exploration, MAB does not sacrifice recommendation quality in any way (better execution costs in 12 out of 15 cases, with significantly better execution costs under all cases of TPC-DS). 

This efficient exploration is promoted by the linear reward-context relationship along with C$^2$UCB's weight sharing (Section \ref{sec:background}), resulting in a small number of parameters to learn. An arm's identity becomes irrelevant and context (Section \ref{sec:bandits}) becomes the sole determining factor of each arm's expected score, which allows MAB to predict the UCB of a newly arriving arm with known context \emph{without} trying it even once.
\subsubsection{The Impact of Round Size} In the original TPC-H Skew static experiment (1x), each bandit round includes all the benchmark templates (22 queries). To analyse the impact of the round size (bandit invocation frequency), we conduct experiments with single-query (1 query), 0.5x (11 queries) and 2x (44 queries) round sizes on TPC-H Skew benchmark. All three round sizes converge to the same performant configurations by the last round. We observe a faster convergence with small round sizes, resulting in lower execution costs in the first few rounds. While the execution cost gain from 1x to 0.5x is noticeable, dividing the round further (single query) does not provide a considerable benefit compared to the added creation and recommendation overhead. With larger round sizes, we observe lower creation costs due to less frequent bandit updates (see Table~\ref{tab:round-size}). MAB performs better under all round-sizes compared to PDTool. A DBA can decide on the round size (bandit invocation frequency) based on the application and DBA's primary goal (faster convergence vs low creation cost). We leave auto-tuning of this parameter as an interesting future work avenue.
\begin{table}
\centering
\caption{TPC-H Skew benchmark under different round sizes (min)}
\begin{tabular}{|l|l|l|l|l|}
\hline
Round size & Rec. & Creation & Execution & Total\\ \hline
Single Query & 1.11 & 27.77 & 30.16 & 59.04 \\ \hline
0.5x & 0.13 & 22.39 & 30.39 & 52.92 \\ \hline
1x & 0.11 & 19.82 & 32.06 & 51.99 \\ \hline
2x & 0.08 & 12.66 & 43.53 & 56.27 \\ \hline
\end{tabular}
\label{tab:round-size}
\end{table}

\begin{figure}[t]
\centering
\includegraphics[width=\columnwidth]{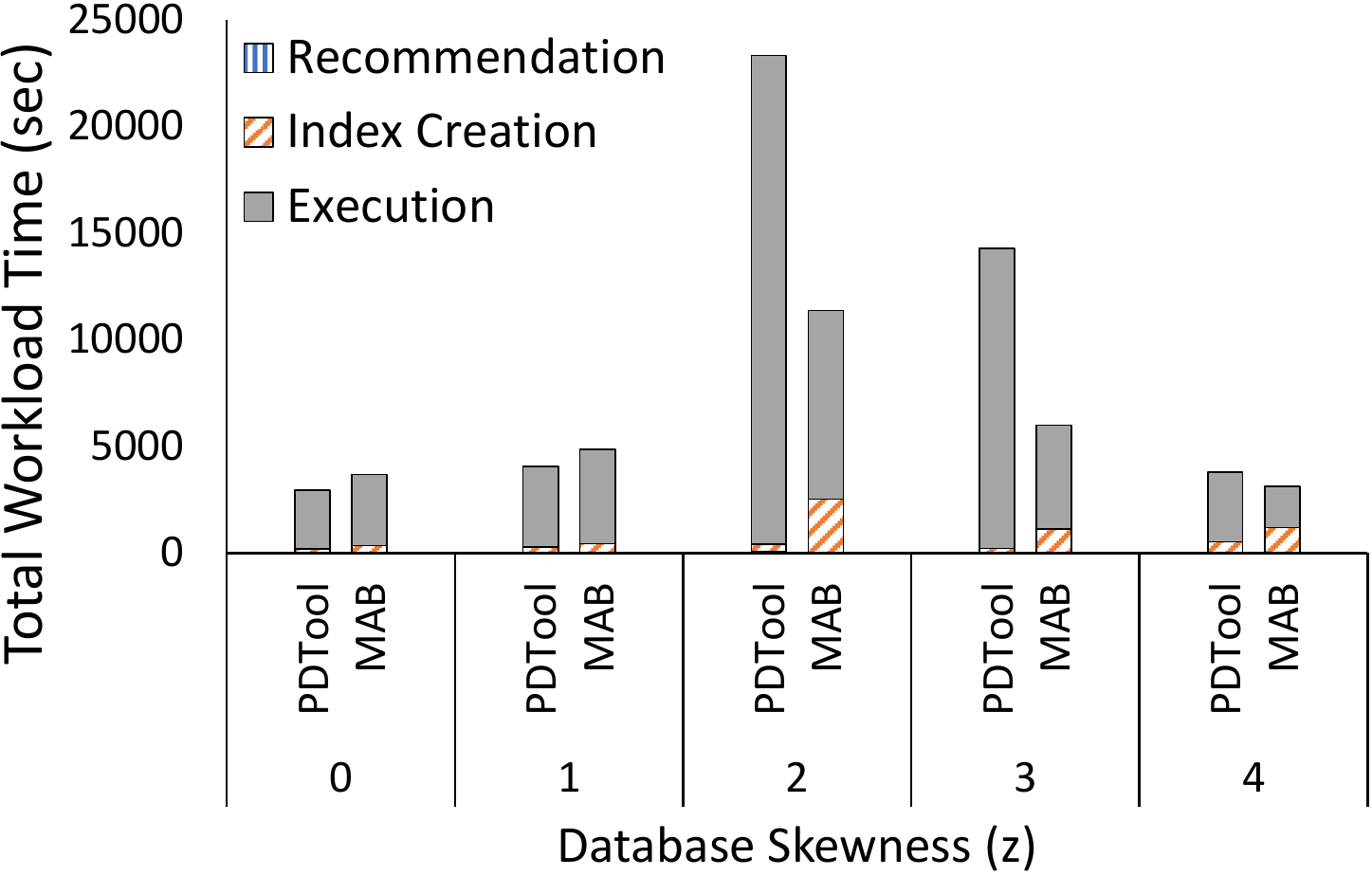}\\
\caption{MAB vs. PDTool total end-to-end workload time under TPC-H skew \emph{static} workloads with different Zipfian factors (z)}
\label{fig:skewness}
\end{figure}

\malinga{\subsubsection{The Impact of Data Skew in Analytical Workloads}
While PDTool outperforms MAB in TPC-H (uniform) by 19\% in total workload time, MAB outperforms PD\-Tool by 17\% on TPC-H skew with Zipfian factor 4. To further investigate the impact of the degree of data skew, we experiment with different Zipfian factors ranging from 1 to 3. As shown in Figure~\ref{fig:skewness} under Zipfian factors 2 and 3, MAB showed over 51\% and 58\% performance gain against PDTool, respectively. Whereas under Zipfian factor 1, PDTool outperformed the MAB by 16\%. PDTool missing the index $Orders.O\_custkey$ appears to be more costly with Zipfian factors 2 and 3, mainly affecting Q22.}

\malinga{To provide further explanation, in the Zipfian distribution, frequency of an item is inversely proportional to its rank in the frequency ordered list. For example, with z=4, the most frequent customer in the order table is repeated 86k times, whereas the second most frequent customer is only repeated 17k times (80\% drop). Similarly, for z=2, the frequency decreases from 227k to 101k (46\% drop). The percentage drop in frequency between adjacent ranks increases with the Zipfian factor. 
We observed that the running time of Q22 depends on the \emph{frequency of the most frequent customer} in the order table (which is 50k, \emph{227k}, 156k, 86k for z values 1 to 4, respectively), more than the skewness, which explains the results shown in Figure~\ref{fig:skewness}.}

\subsubsection{Optimal Configurations} With many possible configurations for even the smallest databases, finding an optimal configuration under given memory budget, while considering performance regressions,  is a non trivial task. However, when the memory budget limitation is lifted, we can estimate optimal configurations with a set of covering indices. This configuration occupies around 40 GB of space for both TPC-H and TPC-H Skew. Under this budget, PDTool and MAB both converged to the optimal configuration by the final round.

\maketeal
\subsection{MAB vs PDTool Under HTAP Workloads}
\label{sec:htap}

In this section, we demonstrate  MAB's ability to efficiently tune indices under HTAP workloads, which are well known for their complexity.

\textbf{Static workloads.}
In Section \ref{sec:analytical-evaluation}, we observed better performance from PDTool under fully analytical static workloads on uniform datasets (TPC-H and SSB), which serves as the best case for offline tuning tools such as PDTool. In this section, we show that, even under repeating workloads on uniform datasets, the addition of transactional queries can create problems for PDTool. To  illustrate the impact of transactional queries, we use a series of CH-BenCHmark static experiments varying the transactional to analytical ratio. In each of these experiments, we keep the analytical component constant with 22 adapted TPC-H queries (\emph{a set of analytical queries}) while changing the size of the transactional component.

\begin{figure}[t]
\centering
\centering\includegraphics[width=\columnwidth]{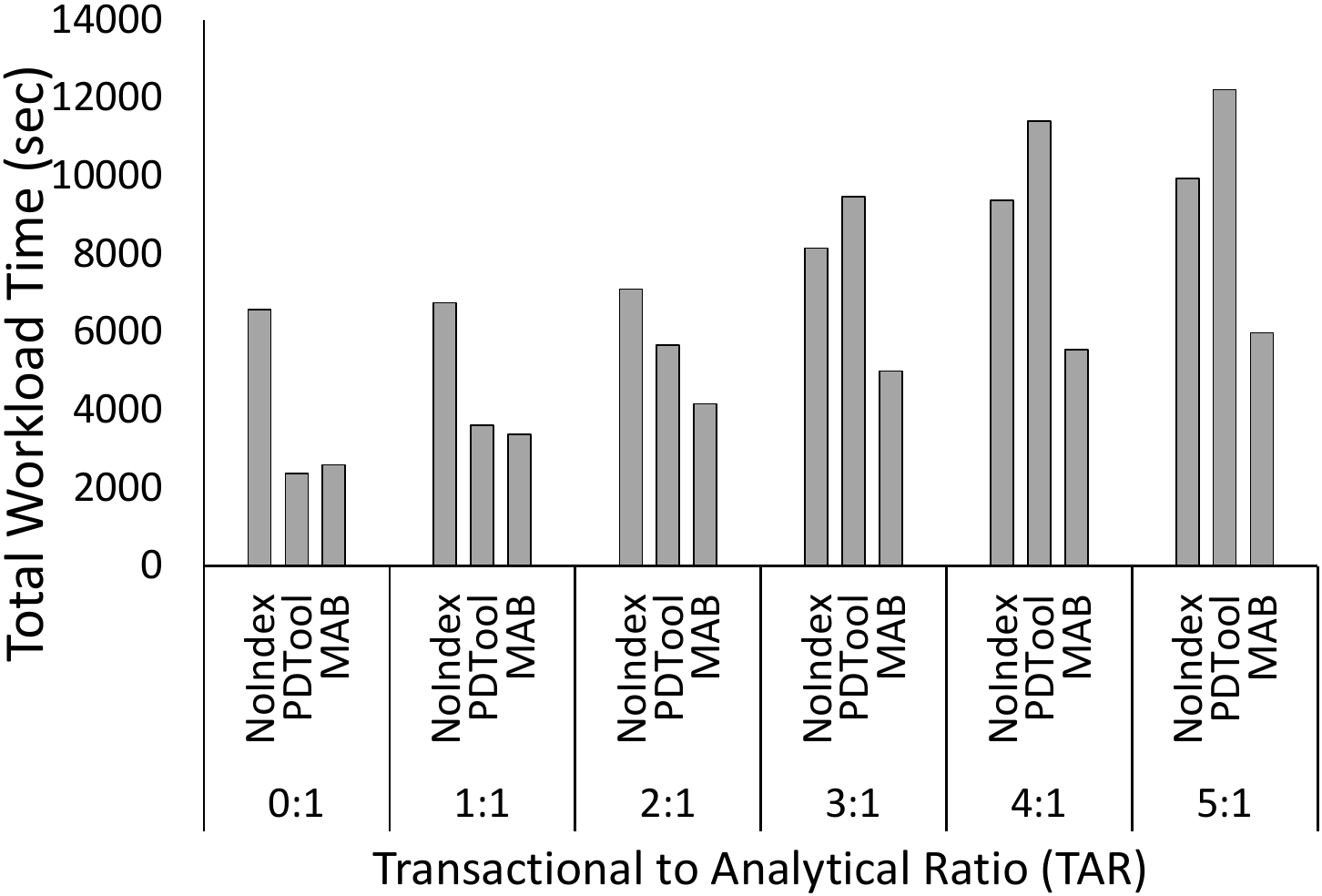}
\caption{MAB vs. PDTool total workload time under CH-BenCHmark for \emph{static} workloads with a range of different TARs}
\label{fig:ch_workload_mix}
\end{figure}

The transactional component of the workload is composed of 5 transactions (new-order, payment, order-status, delivery and stock-level) with a pre-specified transaction mixture (44\%, 44\%, 4\%, 4\% and 4\%, respectively). The smallest transactional workload adhering to the specified TPC-C transaction mixture  comprises 11 new order transactions, 11 payment transactions, an order-status transaction, a delivery transaction and a stock-level transaction (approximately 650 queries). This smallest transactional workload is henceforth referred to as \emph{a set of transactional queries}. We define the transactional to analytical ratio (henceforth referred to as \emph{TAR}) as the ratio between transactional and analytical query sets. As an example, 5:1 TAR is composed of one analytical query set (22 TPC-H queries) and 5 transactional query sets (\ie 55 new order transactions, 55 payment transactions, 5 order-status transactions, 5 delivery transactions and 5 stock-level transactions,  resulting in approximately 3300 transactional queries per round).

\begin{figure}[t]
\centering
\begin{minipage}{0.48\columnwidth}
\centering\includegraphics[width=\columnwidth]{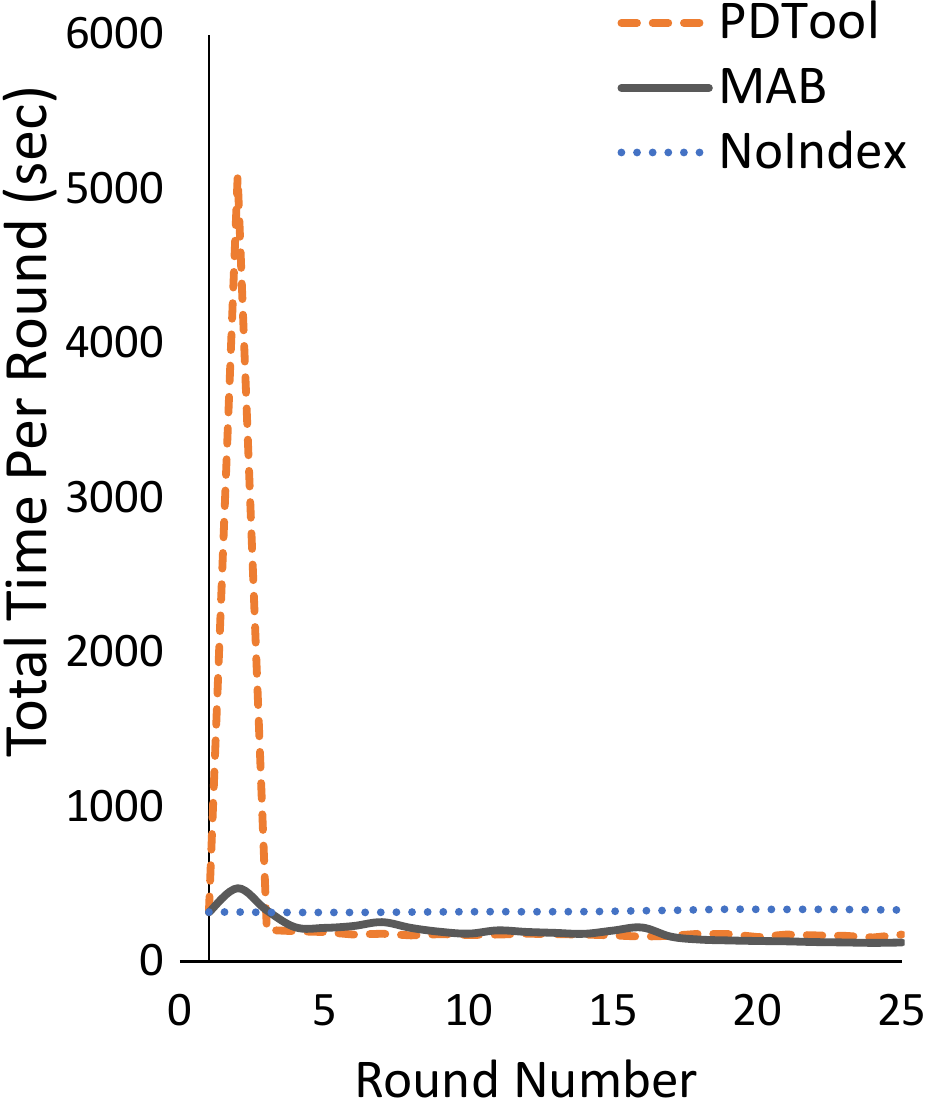}\\ \textbf{(a)}
\end{minipage}\hfill\hspace{0.5em}
\begin{minipage}{0.48\columnwidth}
\centering\includegraphics[width=\columnwidth]{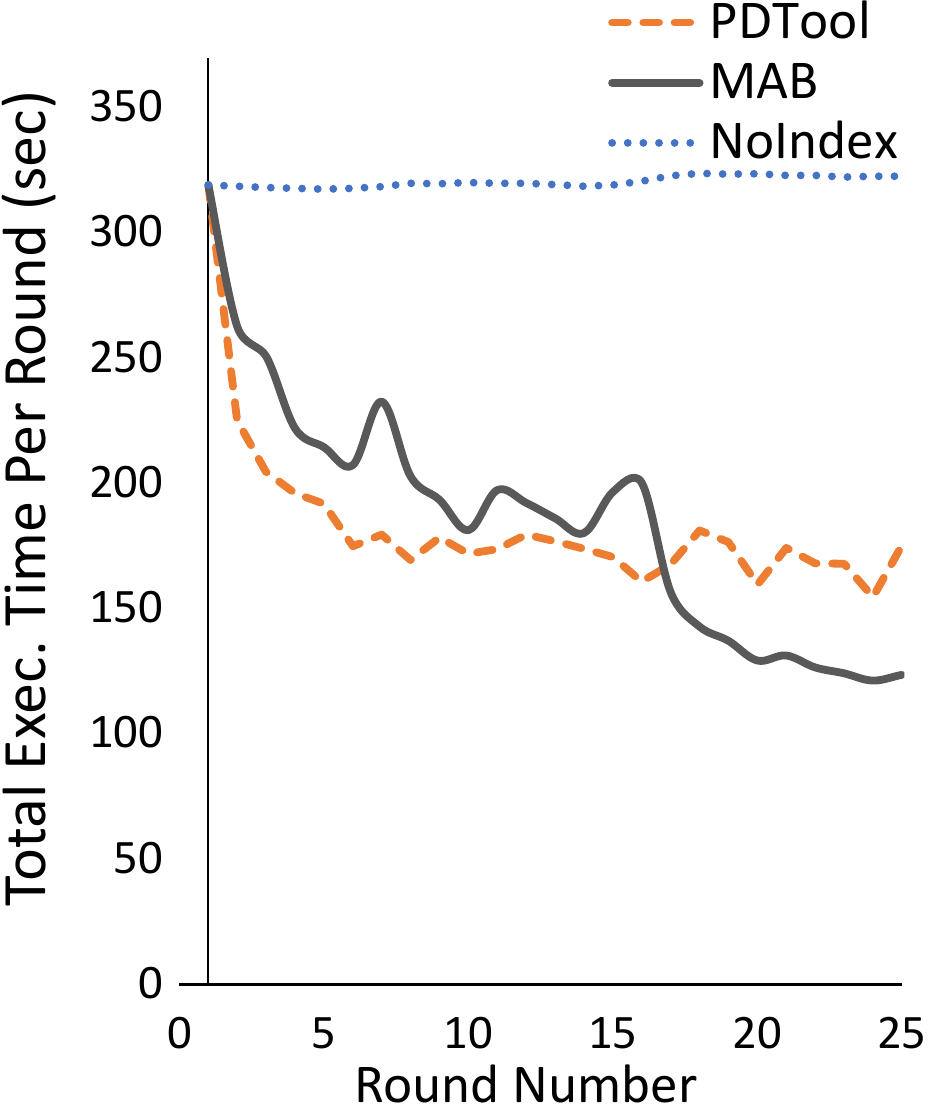}\\ \textbf{(b)}
\end{minipage}\\
\caption{MAB vs. PDTool convergence under CH-BenCHmark for \emph{static} workloads with 3:1 TAR: (a) End-to-end workload time, (b) Total execution time.}
\label{fig:ch_total}
\end{figure}

\begin{figure}[t]
\centering
\begin{minipage}{0.48\columnwidth}
\centering\includegraphics[width=\columnwidth]{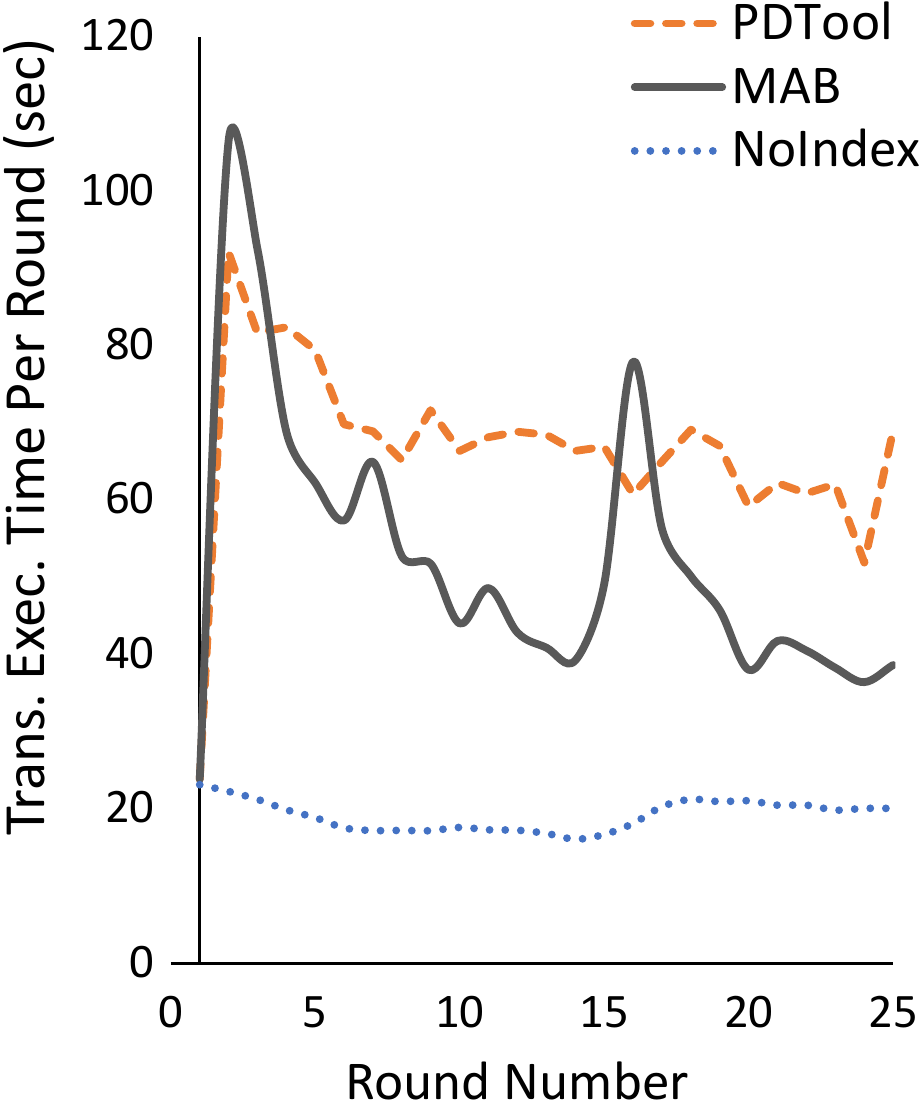}\\ \textbf{(a)}
\end{minipage}\hfill\hspace{0.5em}
\begin{minipage}{0.48\columnwidth}
\centering\includegraphics[width=\columnwidth]{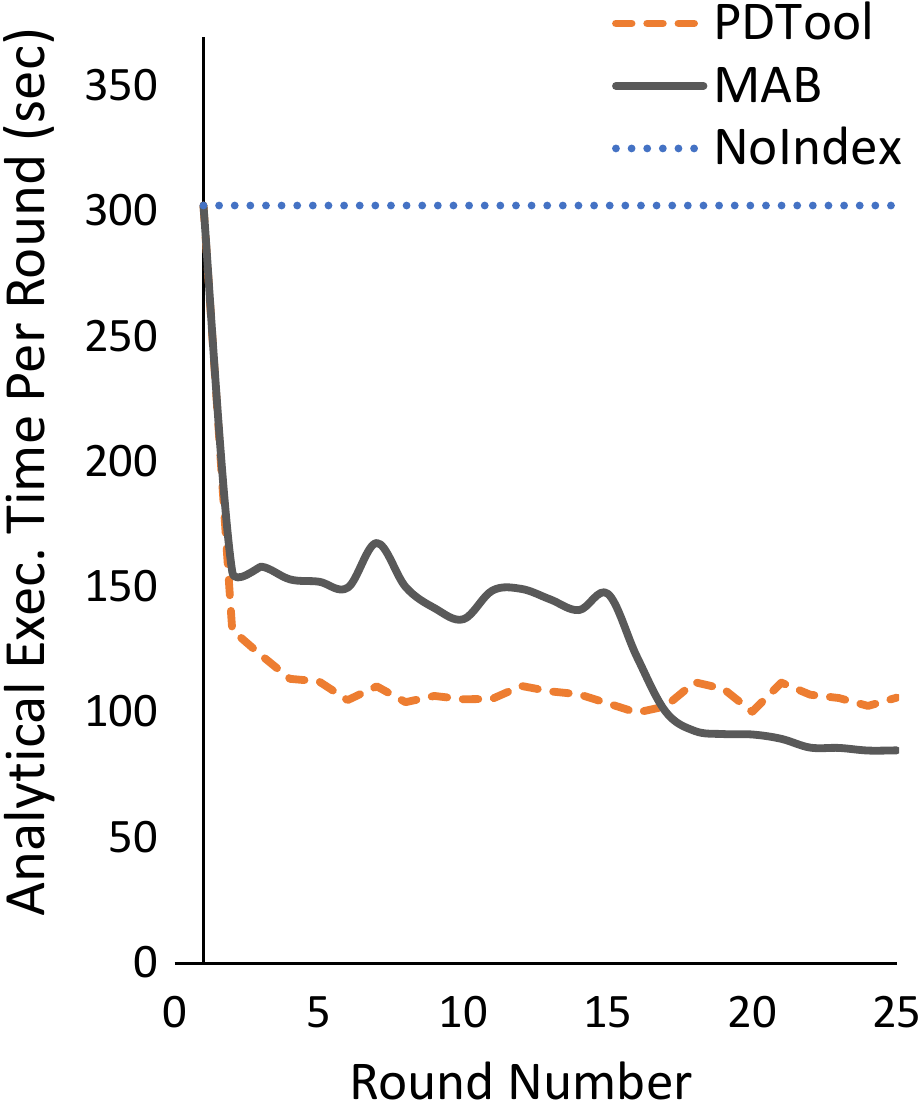}\\ \textbf{(b)}
\end{minipage}\\
\caption{MAB vs. PDTool convergence under CH-BenCHmark for \emph{static} workloads with 3:1 TAR: (a) Transactional Execution cost, (b) Analytical Execution cost.}
\label{fig:ch_analytical_transactional}
\end{figure}

\begin{table*}
\centering
\caption{HTAP: Total time breakdown (min): the best choice is in bold text.}
\label{table:htap_summary}
\begin{tabular}{|l|r|r|r|r|r|r|r|r|r|r|r|r|} 
\hline
Workload & \multicolumn{2}{l|}{Recommendation} & \multicolumn{2}{l|}{Index
  Creation~} & \multicolumn{2}{l|}{Execution} & \multicolumn{2}{l|}{Analytical} & \multicolumn{2}{l|}{transactional} & \multicolumn{2}{l|}{Total} \\ 
\hline
 & \multicolumn{1}{l|}{MAB} & \multicolumn{1}{l|}{PDT} & \multicolumn{1}{l|}{MAB} & \multicolumn{1}{l|}{PDT} & \multicolumn{1}{l|}{MAB} & \multicolumn{1}{l|}{PDT} & \multicolumn{1}{l|}{MAB} & \multicolumn{1}{l|}{PDT} & \multicolumn{1}{l|}{MAB} & \multicolumn{1}{l|}{PDT} & \multicolumn{1}{l|}{MAB} & \multicolumn{1}{l|}{PDT} \\ 
\hline
CH & \textbf{0.1} & 79.15 & 6.67 & \textbf{1.86} & \textbf{76.6} & 76.72 & 55.02 & \textbf{48.74} & \textbf{21.58} & 27.98 & \textbf{83.37} & 157.73 \\ 
\hline
TPC-H & \textbf{0.14} & 14.09 & 9.41 & \textbf{6.43} & 87.41 & \textbf{81.82} & 57.42 & \textbf{53.4} & 29.99 & \textbf{28.42} & \textbf{96.96} & 102.35 \\ 
\hline
TPC-H Sk. & \textbf{0.15} & 13.92 & 14.79 & \textbf{12.74} & \textbf{77.29} & 102.29 & \textbf{37.19} & 65.45 & 40.11 & \textbf{36.84} & \textbf{92.22} & 128.94 \\
\hline
\end{tabular}
\end{table*}

As evident from Figure~\ref{fig:ch_workload_mix}, in transaction-heavy \linebreak[4] workloads, MAB performs much better than PDTool providing up to 51\% speed-up (5:1) in total workload time, whereas PDTool performs better in fully analytical workloads providing up to 8\% speed-up (0:1). This result is similar to the results observed under the static setting with TPC-H and SSB workloads where the data is uniformly distributed, and the workload is fully analytical. At 1:1 TAR, which is the first ratio that introduces the transactional component, MAB starts to take the lead providing a 4\% performance gain in total workload time. MAB manages to reach the same last round execution time as PDTool; however, due to better execution times in early rounds, PDTool provides 7.7\% total execution time speed up over MAB. From 2:1 TAR onwards, MAB dominates the PDTool providing 26\%, 47\%, 51\% and 51\% total workload time speed-up over PDTool, under 2:1, 3:1, 4:1, 5:1 TARs, respectively. PDTool struggles to perform better than NoIndex from 3:1 TAR onwards due to the heavy recommendation costs incurred by PDTool, yet PDTool is still superior to NoIndex in execution cost.

To further understand the results, we dive into the 3:1 TAR experiment, which has enough transactional queries to demonstrate the importance of both analytical gain and transactional overhead. As shown in the convergence graphs in Figure~\ref{fig:ch_total}, despite the high recommendation cost in the first round, PDTool provides much better per round total workload times compared to NoIndex. However, MAB converges to an even better total workload time by the last round. As clearly visible from Figure~\ref{fig:ch_total} (b), MAB converges to a better configuration (providing 29.4\% better execution time by the last round) compared to PDTool.

Balancing the configuration fitness between transactional workloads and analytical workloads is the prime concern of index tuning in HTAP environments. How both tools achieve this balance can be better understood by breaking the execution cost into analytical and transactional components. As Figure~\ref{fig:ch_analytical_transactional} demonstrates, MAB configuration provides better execution time for both analytical and transactional workload components (19.9\% and 43.9\% better execution times by 25\textsuperscript{th} round for analytical and transactional workloads, respectively). In initial rounds, MAB performance is inferior to PDTool in transactional execution time but it quickly learns the negative impact of indices on the transactional workload. By the 4\textsuperscript{th} round, MAB surpasses PDTool in transactional execution time by dropping indices with negative rewards. Note that, while removing the unnecessary indices, MAB makes sure not to impact the analytical execution times by keeping the high reward indices intact. MAB performs several configuration changes in rounds 15--17, which results in a sudden oscillation in transactional execution time, but these configuration changes allow the bandit to find a superior configuration in both analytical and transactional execution costs. There is some variability in transactional execution costs even with the same number of transactions in each round, as the number of queries per round can be different (\eg different new orders can have a different number of items in a transaction, leading to a different number of queries).

\subsubsection{The Impact of Data Skew in HTAP Workloads}
We now experiment with TPC-H and TPC-H Skew HTAP workloads to demonstrate the impact of the addition of transactional queries to well-known benchmarks that we already examined in Section~\ref{sec:analytical-evaluation}. %
We experiment with a similar number of transactional queries as in the 3:1 TAR CH-BenCHmark experiment. The OLTP part of the workload is composed of 6 templates (two insert templates, two delete templates and two update templates). We use the original data generation tools to generate the insert and delete data tuples for ORDER and LINEITEM tables. Additionally, we use the same tool's insert data to populate our update queries over the same tables. The transactional workload used here is simpler than CH-BenCHmark, but sufficient to demonstrate the impact HTAP workloads have on the overall performance.

\begin{figure}[t]
\centering
\centering\includegraphics[width=\columnwidth]{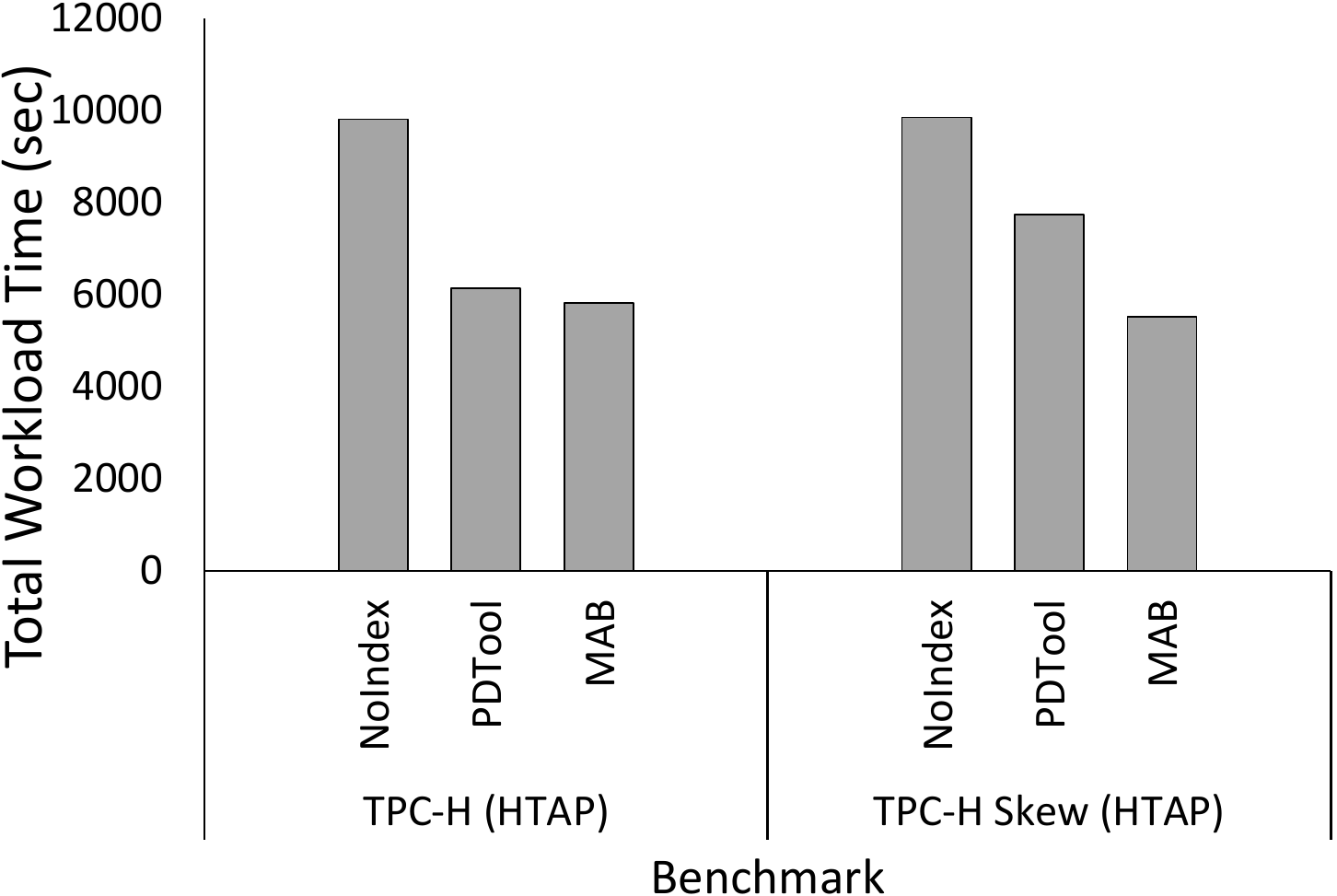}
\caption{MAB vs. PDTool total end-to-end workload time for static TCP-H and TPC-H skew HTAP workloads.}
\label{fig:static_htap_skewness}
\end{figure}

As shown in Figure~\ref{fig:static_htap_skewness}, MAB performs better in both TPC-H and TPC-H skew HTAP workloads. Interestingly, MAB manages to achieve 5.2\% better total workload time even under the TPC-H benchmark, whereas PDTool previously outperformed MAB by 19\% under the fully analytical static setting. MAB converges to a similar performant configuration, comparing the final round execution cost, whereas, under the fully analytical setting, there was a considerable gap between MAB and PDTool execution times even at the last round. However, due to MAB's longer execution times in the first few rounds, PDTool achieves a 6\% better total execution time. Due to the higher recommendation time in PDTool, PDTool ends up having a higher total workload time. 
In TPC-H Skew, MAB dominates the PDTool on all ends, having 28.4\% better total workload time and 24\% better execution time.  
All HTAP results are summarised in Table~\ref{table:htap_summary}.

\subsubsection{Dynamic HTAP Workloads}

\begin{figure}[t]
\centering
\centering\includegraphics[width=\columnwidth]{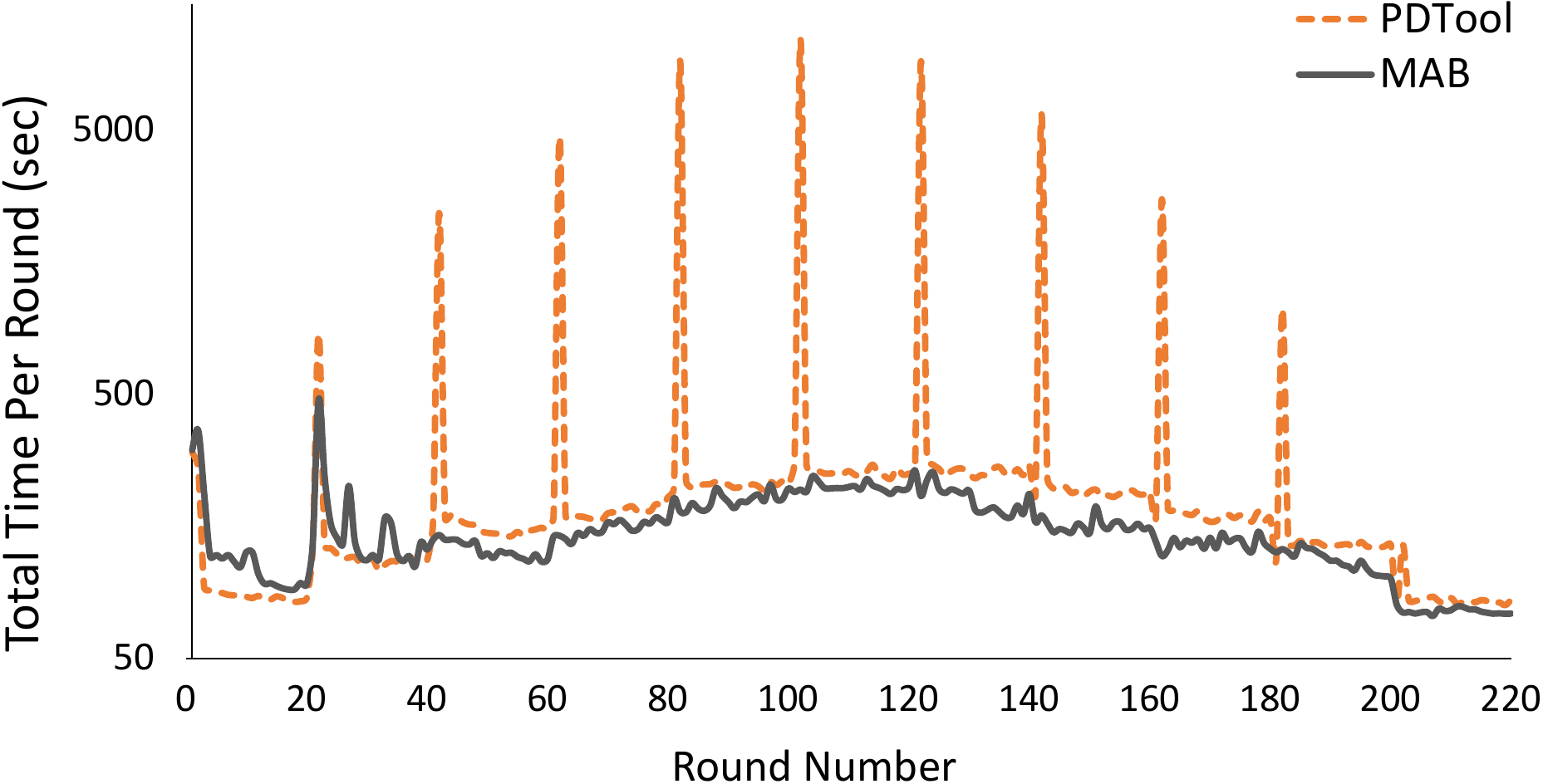}
\caption{MAB vs. PDTool total-workload time convergence under CH-BenCHmark for \emph{dynamic} workloads with different transaction levels \textbf{(log y-axis)}}
\label{fig:ch_dynamic_con_total}
\end{figure}

\begin{figure}[t]
\centering
\centering\includegraphics[width=\columnwidth]{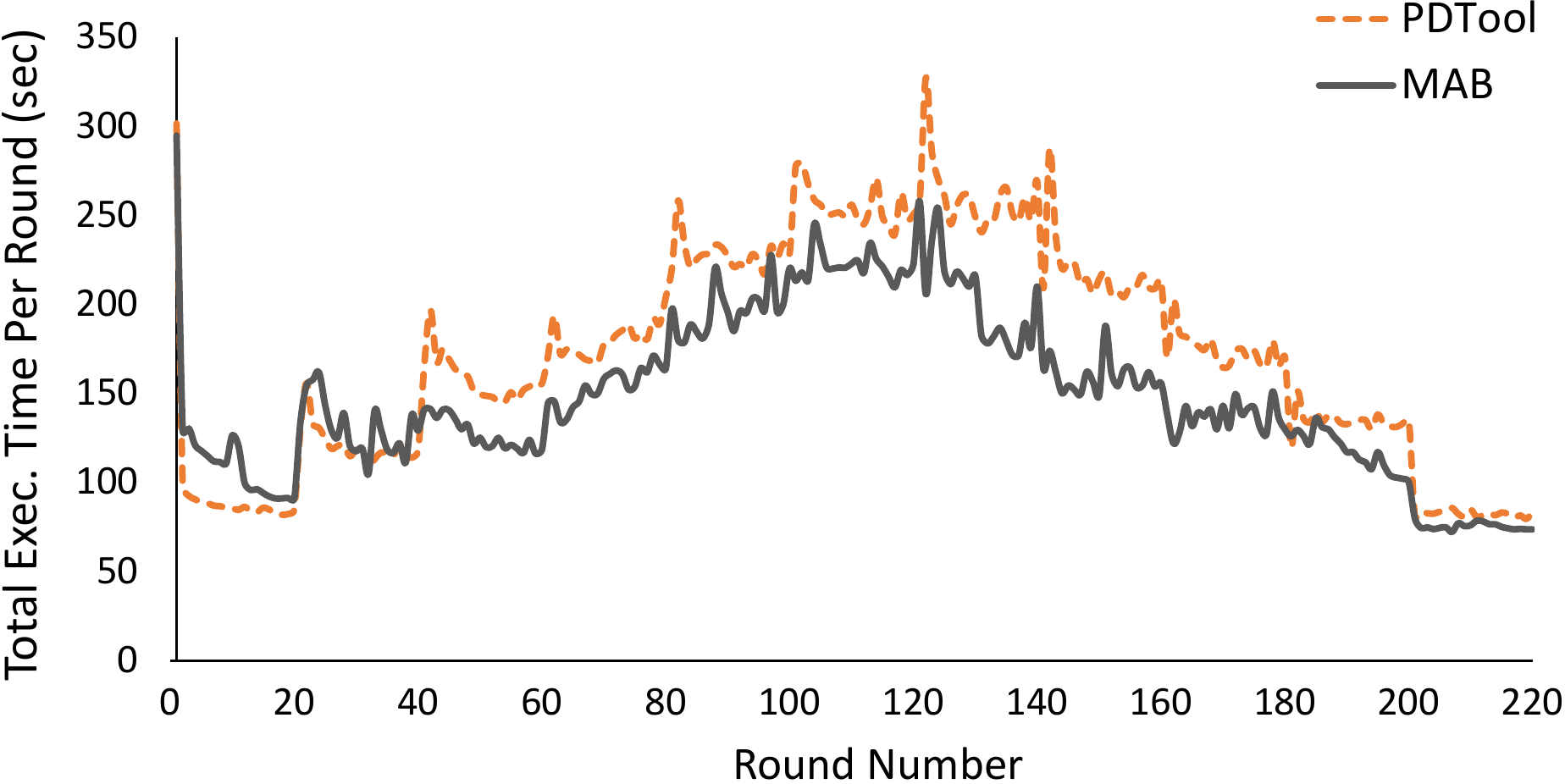}
\caption{MAB vs. PDTool \emph{total execution time} convergence under CH-BenCHmark for dynamic workloads with different transaction levels}
\label{fig:ch_dynamic_con_execution}
\end{figure}

\begin{figure}[t]
\centering
\centering\includegraphics[width=\columnwidth]{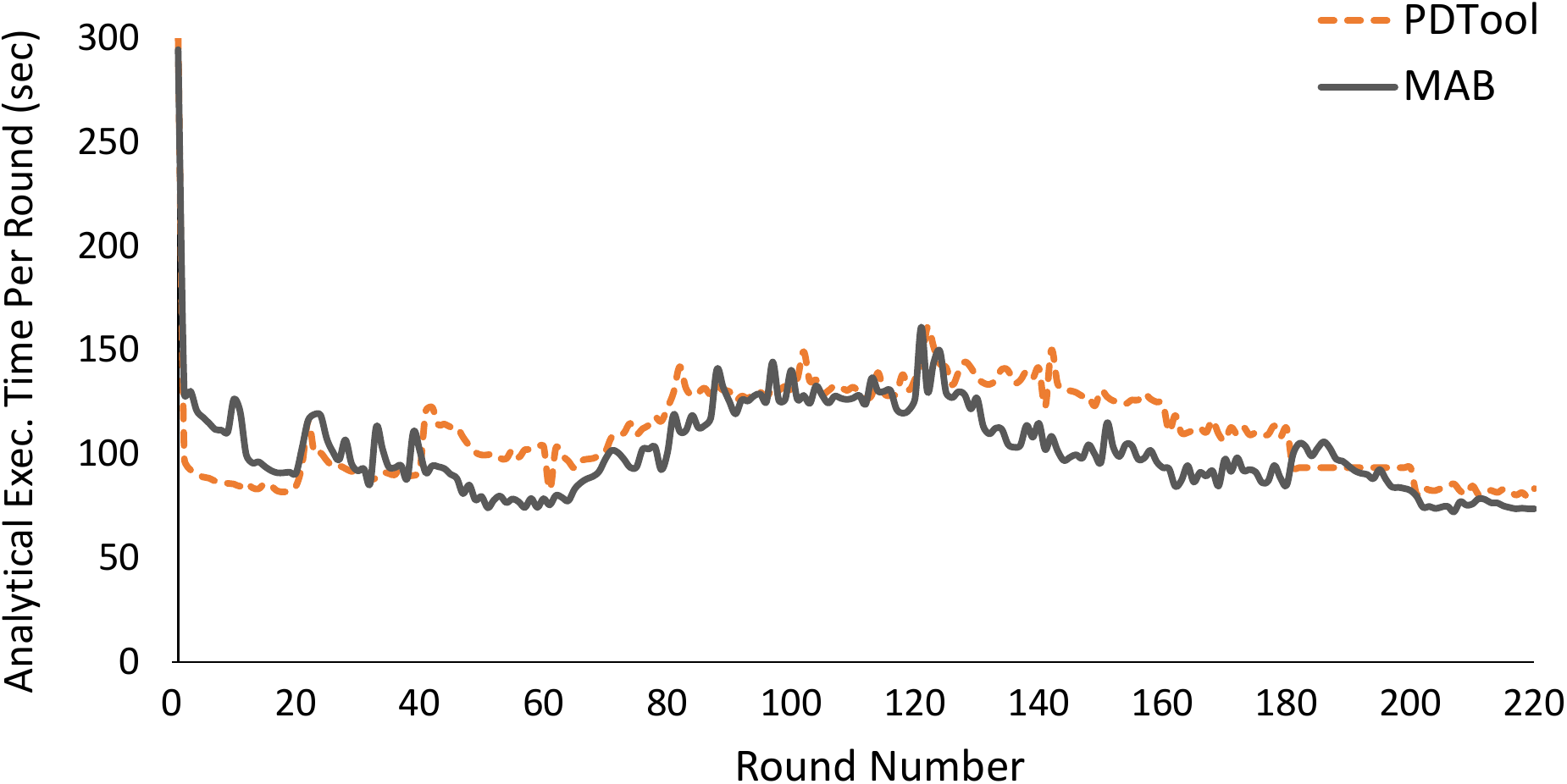}
\caption{MAB vs. PDTool \emph{analytical execution time} convergence under CH-BenCHmark for dynamic workloads with different transaction levels}
\label{fig:ch_dynamic_con_analytical_exec}
\end{figure}

\begin{figure}[t]
\centering
\centering\includegraphics[width=\columnwidth]{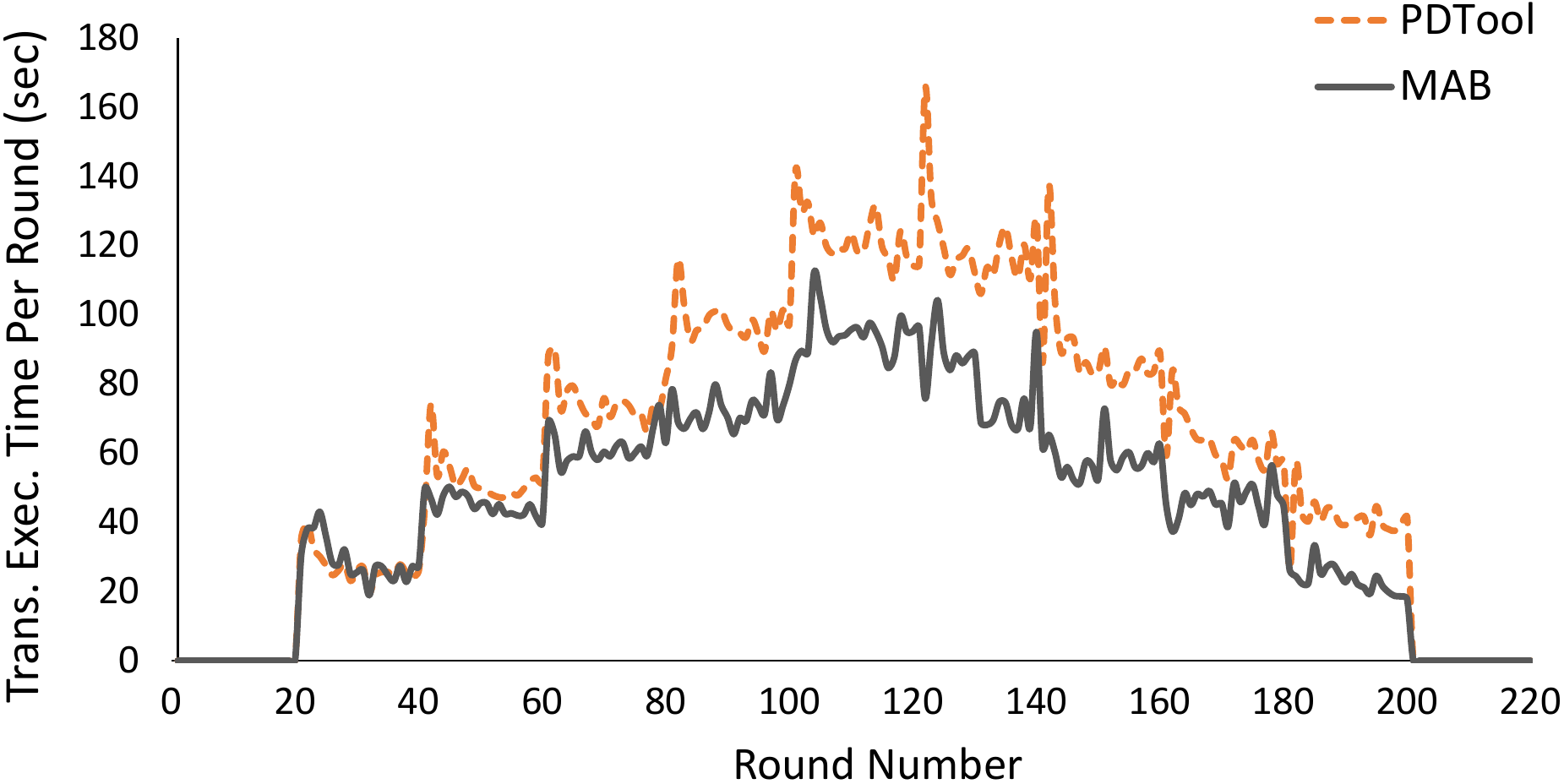}
\caption{MAB vs. PDTool \emph{transactional execution} time convergence under CH-BenCHmark for dynamic workloads with different transaction levels}
\label{fig:ch_dynamic_transactional_exec}
\end{figure}

In this experiment, we gradually increase and decrease the transactional workload size over the rounds. We start with 0:1 TAR, which is purely analytical, and then we add transactional workload sets one by one till we reach 5:1 TAR. Afterwards, we gradually reduce the transactions workload sets one by one to reach 0:1 TAR again. We run 20 rounds in each TAR.

Soon after each workload change, PDTool is invoked with the new workload from the previous round, which is a good representation of the next 19 rounds. It is essential to provide the workload from at least a complete single round as PDTool takes the transactional to analytical ratio into account when making recommendations. However, as observable from Figure~\ref{fig:ch_dynamic_con_total}, each invocation of PDTool takes a substantial amount of time for larger workloads in higher transaction levels. 

We can observe that MAB performs most of the configuration changes at the start of the experiment and then after the first workload change (0:1 $\rightarrow$ 1:1). It is understandable that MAB needs to explore extensively at the start of the experiment. However, a similar level of exploration for the first workload change might not be as intuitive, given that we do not see such exploration from MAB for the rest of the experiment. At round 21, MAB is exposed to transactional queries for the first time in this experiment. MAB performs more exploration and learns the negative impact of indices in these rounds. We can see MAB doing much better after round 40. MAB provide a 57.8\% speed up in the total workload time compared to PDTool. A significant portion of this speed-up is attributed to MAB's lower recommendation cost compared to PDTool. 

To compare the different configurations proposed by two tools, we plot execution time over the rounds in Figure~\ref{fig:ch_dynamic_con_execution}. After the first two transaction levels (i.e., after round 40), MAB always manages to lock into a better configuration providing faster execution time. MAB provides an 11\% speed-up in total execution cost.

In the entire experiment, we go through the same TAR two times (except for 5:1 TAR), which results in similar workloads. However, from Figure~\ref{fig:ch_dynamic_con_execution}, it is noticeable that PDTool reaches higher execution costs in the descending part of the experiment (after round 120) compared to the ascending part of the experiment (rounds 1 to 120).  Configurations proposed by the PDTool depend on the existing secondary indices present in the system and the underlying data. Continuous additions and deletions change the underlying data, somewhat invalidating the statistics used by the optimiser. Furthermore, at each PDTool invocation, the system has a different starting set of secondary indices, impacting PDTool recommendations. Therefore the recommendations proposed for similar workloads in ascending and descending parts of the graph are different. For example, rounds 80--100 and 120--140 run a 4:1 TAR workload, nonetheless PDTool proposes two very different configurations for these two sections of the experiment. While it proposes only 18 indices at round 81, 27 indices are proposed in round 121, with only 11 indices being shared between 2 configurations. On the other hand, MAB  converges quickly in the later rounds of the experiment, taking advantage of the already obtained knowledge. 

To further analyse MAB's gain in the dynamic experiment, we need to break down the execution time into analytical and transactional components. As one can observe from Figures~\ref{fig:ch_dynamic_con_analytical_exec} and \ref{fig:ch_dynamic_transactional_exec} MAB is obtaining the gain mainly from the transactional workload. MAB provides 4.5\% better analytical execution cost and 22.6\% better transactional execution cost compared to the PDTool. As observable from Figure~\ref{fig:ch_dynamic_con_analytical_exec}, MAB is obtaining a noticeable analytical gain from 2:1 and 3:1 TARs. In the analytical heavy workloads (0:1, 1:1 TARs), PDTool records a better or similar analytical execution time. MAB opts for the transactional friendly configuration for transactional heavy workloads (4:1, 5:1 TARs), reducing thereby the analytical execution time gain. On the transactional end, MAB leads in almost all workloads. As expected, transactional execution cost gain increases for the transactional heavy workloads. Again we see a clear increase in transactional workload execution times for PDTool in the descending part of the graph. This is due to the different configurations proposed by the PDTool, as discussed above.

\subsubsection{Space Savings Under HTAP Workloads}
In the case of index tuning of HTAP workloads, more indices can result in higher running time for transactional queries. Consequently, a minimal index set can be optimal for a transaction-heavy workload. While such a configuration might be suboptimal for the analytical component of the workload, a minimal configuration can result in a better total query execution time due to the significant savings obtained from avoiding index maintenance activities stemming from transactional queries. In our experiments, we observed that PDTool usually exploits the entire given memory budget, resulting in a higher transactional execution cost.\footnote{In our experiments, we have observed that PDTool sometimes goes over the given budget due to errors in index size estimations. On the other hand, MAB initially estimates the index size based on the statistics and corrects the estimate after indices are materialised for the first time.} 

On the flip side, MAB learns the negative impact of indices on transactional queries and dynamically adjusts the configuration. This behaviour was not visible when the workload was fully analytical and is only observable in transaction-heavy workloads. We notice that MAB context helps the bandit to find smaller configurations with higher execution cost gains. MAB context includes the index size as a context feature, and this is typically learned as a negative weight due to high negative rewards from index creation operations. This property then forces the bandit to choose the smallest arms that provide the best gains in execution cost.

\begin{figure}[t]
\centering
\begin{minipage}{0.48\columnwidth}
\centering\includegraphics[width=\columnwidth]{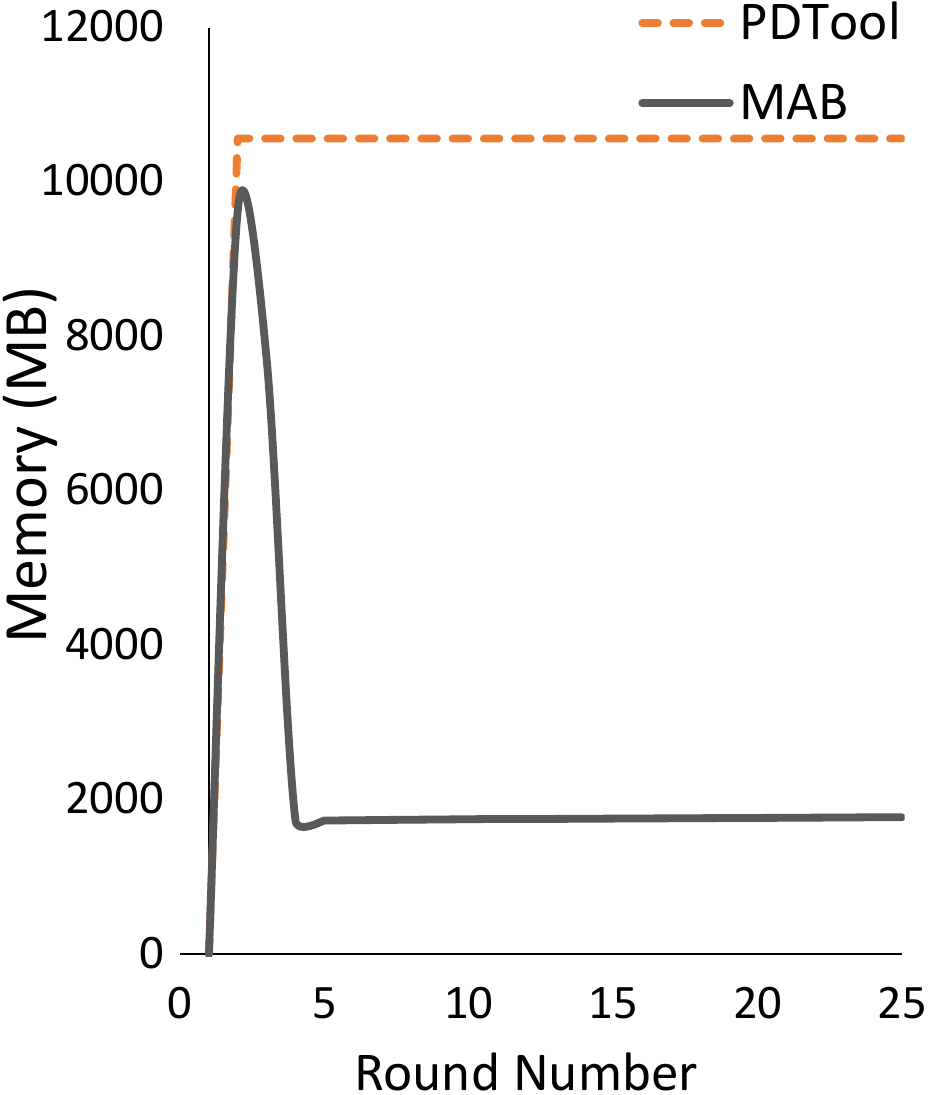}\\ \textbf{(a)}
\end{minipage}\hfill\hspace{0.5em}
\begin{minipage}{0.48\columnwidth}
\centering\includegraphics[width=\columnwidth]{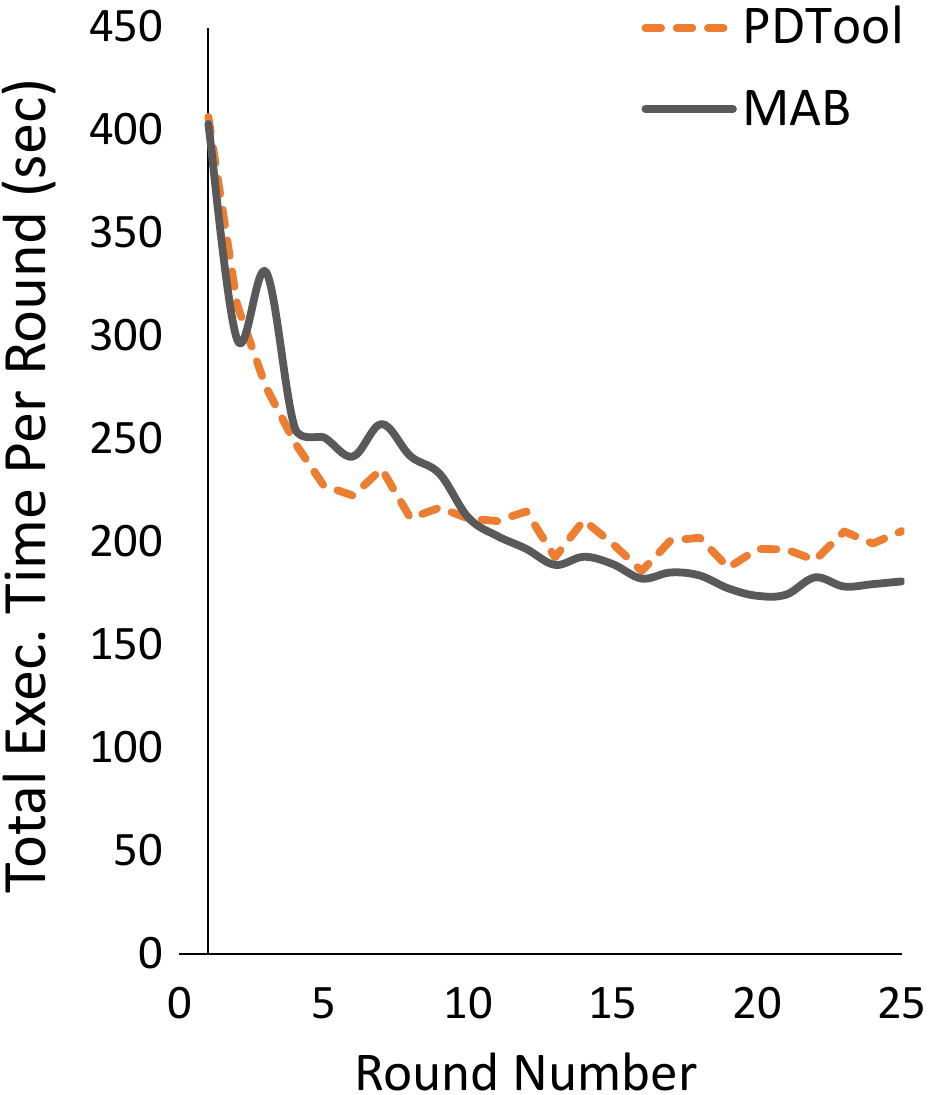}\\ \textbf{(b)}
\end{minipage}\\
\caption{MAB vs. PDTool convergence under CH-BenCHmark for \emph{static} workloads with 5:1 TAR: (a) Memory use, (b) Total execution time.}
\label{fig:ch_memory1}
\end{figure}

At 5:1 TAR, MAB provides a configuration that yields an 83\% memory saving while achieving an 8.8\% execution time gain by the last round (see Figure~\ref{fig:ch_memory1}). This execution cost gain is smaller than the gain we observed under transaction level 3, as the usefulness of indices reduces when the workload becomes transaction heavy.

\begin{figure}[t]
\centering
\begin{minipage}{0.48\columnwidth}
\centering\includegraphics[width=\columnwidth]{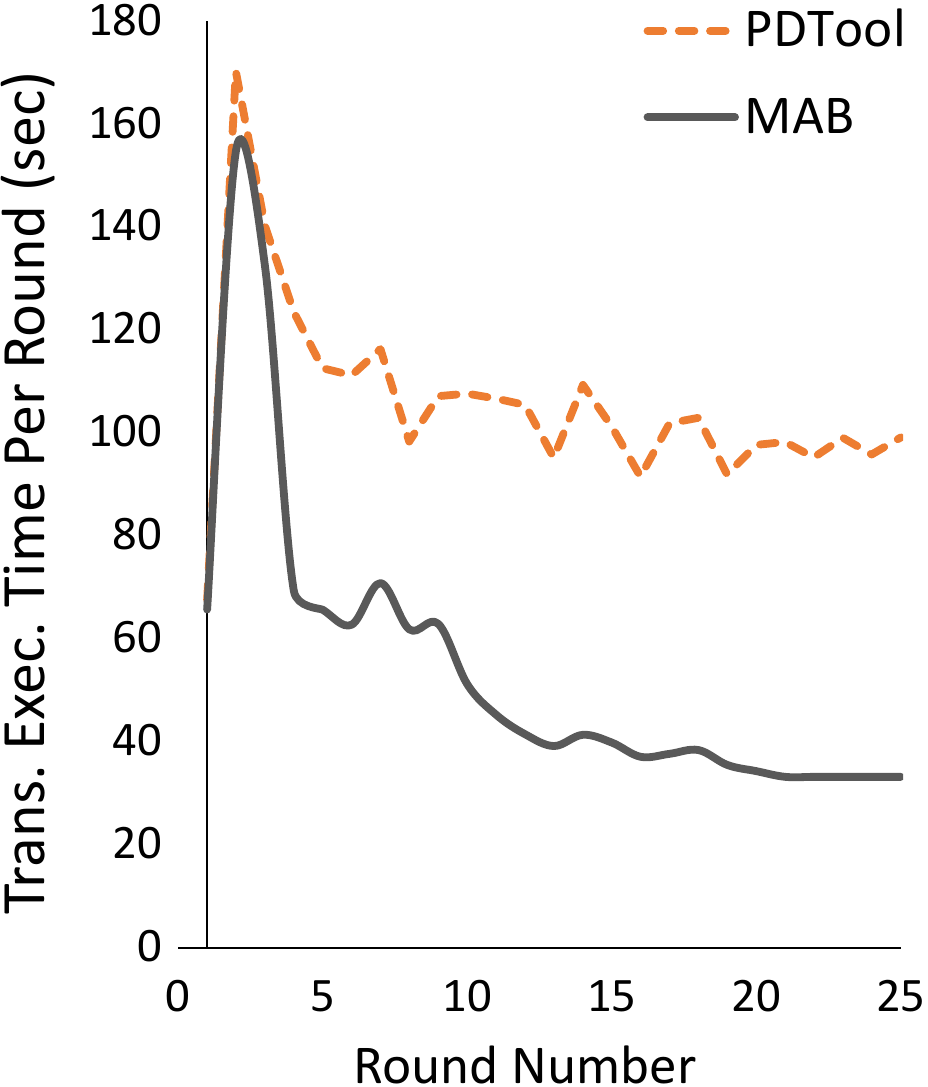}\\ \textbf{(a)}
\end{minipage}\hfill\hspace{0.5em}
\begin{minipage}{0.48\columnwidth}
\centering\includegraphics[width=\columnwidth]{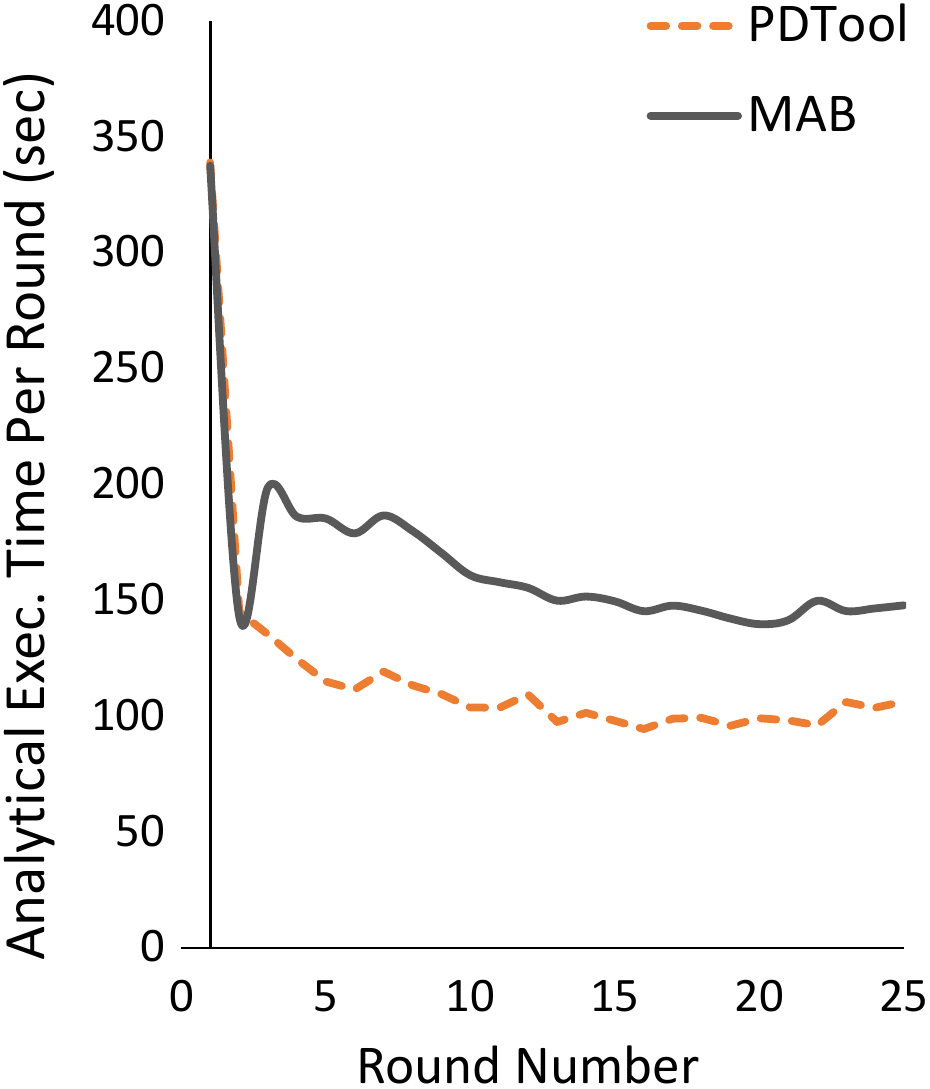}\\ \textbf{(b)}
\end{minipage}\\
\caption{MAB vs. PDTool convergence under CH-BenCHmark for \emph{static} workloads with 5:1 TAR: (a) Transactional Execution cost, (b) Analytical Execution cost.}
\label{fig:ch_memory2}
\end{figure}

While it might be counter-intuitive for an index tuning tool to use less memory to provide better performing configurations, it can be easily understood by observing the analytical and transactional execution times of both tools by the last round (See Figure~\ref{fig:ch_memory2}). Comparing the last round configurations of both tools, PDTool provides an analytical friendly configuration that provides a 27\% speed-up in analytical execution time (around 40 second gain per round). On the other hand, MAB locks into a smaller configuration that is more suitable for transactional queries providing a 60\% speed-up in transactional execution cost (around 60 second gain per round). Ultimately MAB provides an 8.8\% speed-up in total execution time even after a significant memory saving.

\color{black}

\begin{figure*}[t]
\centering
\begin{minipage}{0.26\textwidth}
\centering\includegraphics[width=\textwidth]{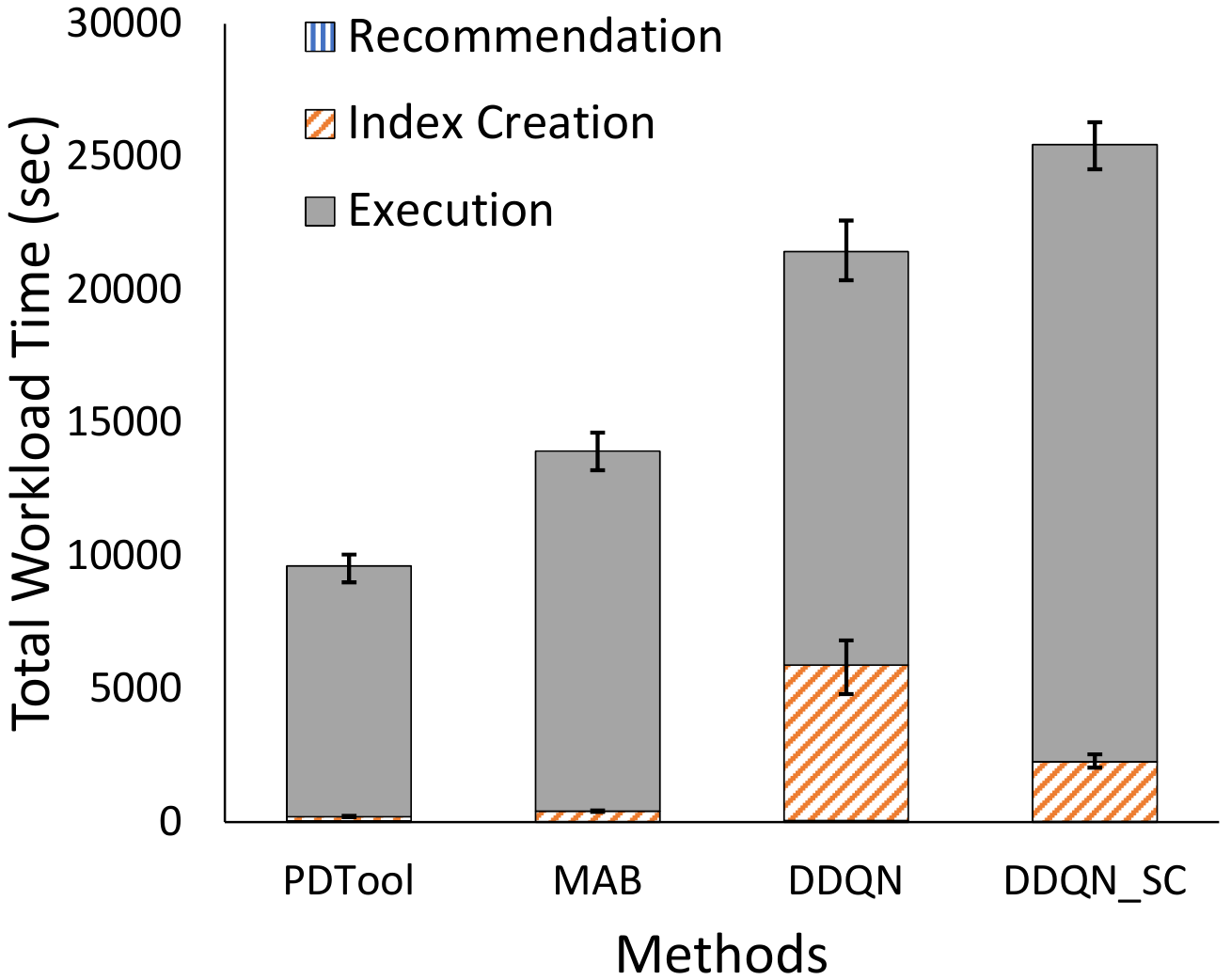}\\ \textbf{(a)}
\end{minipage}\hfill
\begin{minipage}{0.26\textwidth}
\centering\includegraphics[width=\textwidth]{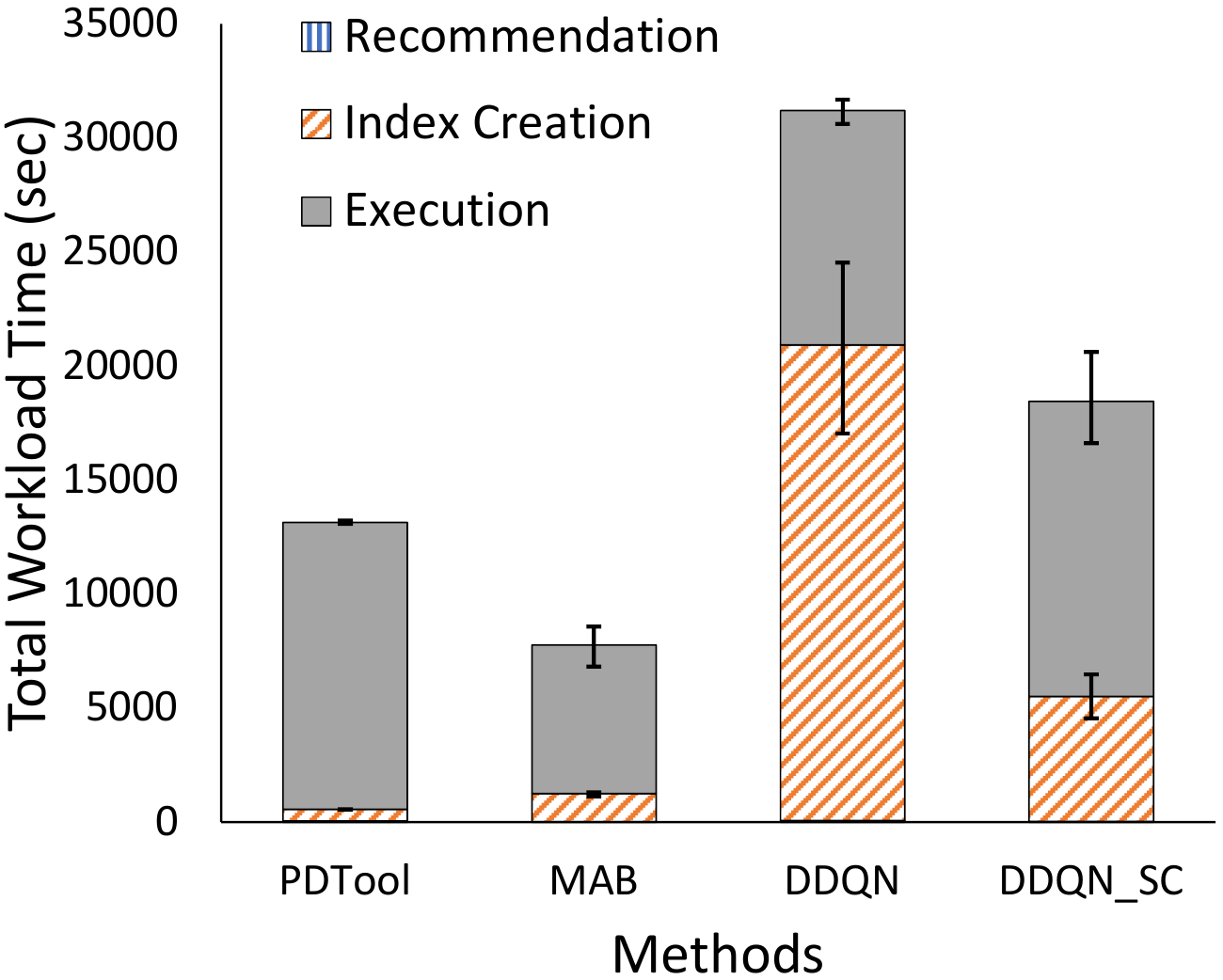}\\ \textbf{(b)}
\end{minipage}\hfill
\begin{minipage}{0.202\textwidth}
\centering\includegraphics[width=\textwidth]{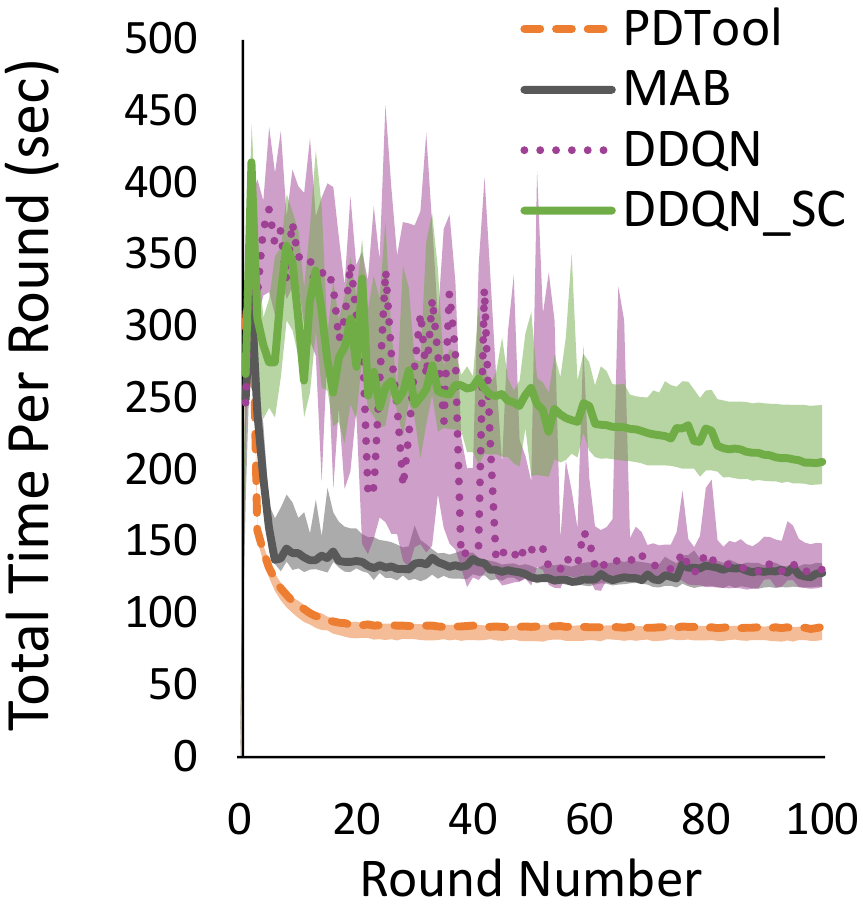}\\ \textbf{(c)}
\end{minipage}\hfill
\begin{minipage}{0.202\textwidth}
\centering\includegraphics[width=\textwidth]{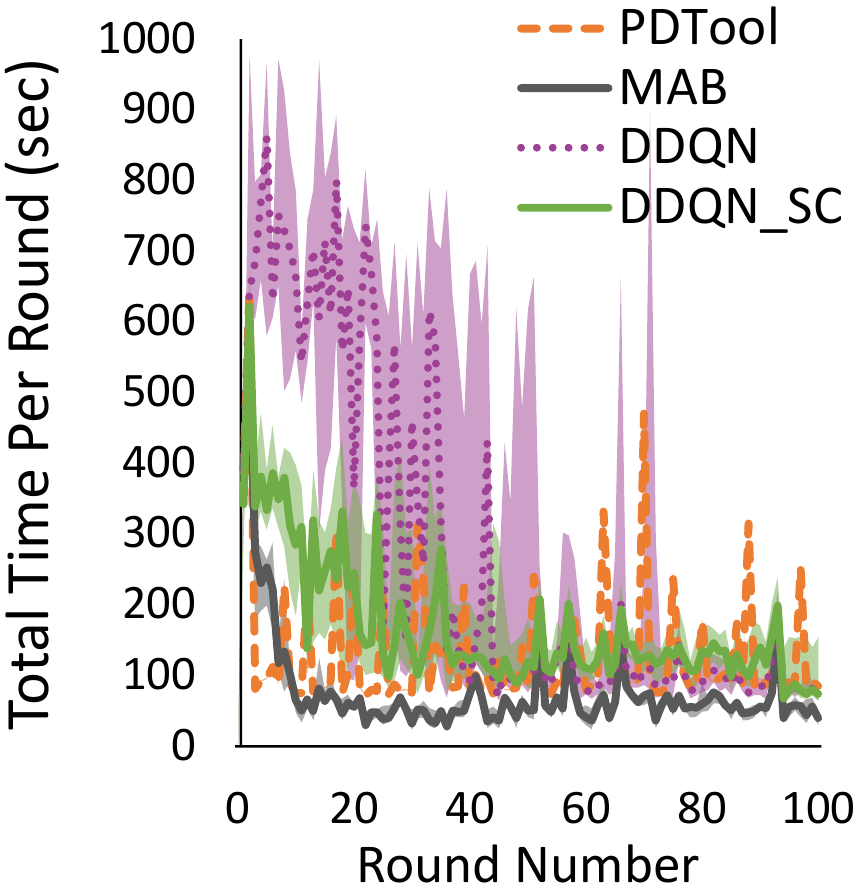}\\ \textbf{(d)}
\end{minipage}\hfill
\\
\caption{DDQN vs. MAB for static workloads: (a) End-to-end workload time for TPC-H, (b) End-to-end workload time for TPC-H Skew, (c) TPC-H convergence, (d) TPC-H Skew convergence.}
\label{fig:rl_comparison}
\end{figure*}

\subsection{Why Not (General) Reinforcement Learning?}\label{sec:not-rl}

Past efforts have considered more general reinforcement learning (RL) for physical design tuning~\cite{no_dba,COREIL}. 
Compared to most MAB approaches, deep RL invites over parameterisation, which can slow convergence (see Figure~\ref{fig:rl_comparison}),
whereas MAB typically provides better convergence, simpler implementation, and \emph{safety guarantees} via strategic exploration and knowledge transfer (see Section~\ref{sec:background}). Due to its randomisation, RL can also suffer from performance volatility as compared to C$^2$UCB, a deterministic algorithm.

\textbf{Experimental setup.}
The above intuition is supported by experiments with more general RL, where we evaluate the popular DDQN RL agent~\cite{van2016DDQN}.
We run the static 10GB TPC-H and TPC-H Skew analytical benchmarks over 100 rounds and present results in Figure~\ref{fig:rl_comparison}.
For a fair comparison, we combine all of MAB's arms' contexts as DDQN state. We also present the same set of candidate indices to the DDQN. For the DDQN's neural network hyperparameters, we followed the experiment of~\cite{no_dba} by setting 4 hidden layers, with 8 neurons each. The discount factor $\gamma$ is set to $0.99$ and the exploration parameter is set to $1$ at the first sample, decaying to 0 with exponential rate reaching 0.01 in the 2400$^\text{th}$ sample. One sample corresponds to one index chosen by the agent. In the beginning of the round, if the agent decides to explore, then the choice of the set of indices will be randomly made for that entire round. 
These experiments are repeated ten times, reporting either average value (Figure \ref{fig:rl_comparison} (a) and (b)) or median ((c) and (d)) along with inter-quartile range. For completeness, we include the case of only using single column indices (DDQN-SC in Figure~\ref{fig:rl_comparison}), as originally proposed  in~\cite{no_dba}.

\textbf{Evaluation.} Due to DDQN-SC's reduced search space in some scenarios, it might not be possible to find an optimal configuration for a workload. This is evident from Figure~\ref{fig:rl_comparison} (a) and (b) where DDQN shows 33\% and 21\% speedup, compared to DDQN-SC, in execution time under TPC-H and TPC-H Skew, respectively. Interestingly, under TPCH-Skew, where the demand for exploration is higher, DDQN-SC has a lower total workload time than DDQN  due to the noticeably small index creation times of single column indices (1.5 hours vs 5.8 hours, respectively). Under both TPC-H and TPC-H Skew, MAB performs significantly better, providing 35\% and 58\% speed-up against the better RL alternative, respectively.

\textbf{No state transitions.}
A strength of more general reinforcement learning is its ability to take into account (random) state transitions when actions are taken. However, the importance of state transition in online index selection is unclear. While modelled state could include the collection of indices that exist in the system, actions (\ie choosing an index) deterministically govern the ensuing state. State could also model characteristics of the next round's queries. However these queries do not depend on the prior action, thus it is appropriate to take successor query state as independent of action, as promoted by bandits. Hence, adopting more general RL provides no clear benefit over MAB, while imposing delay to convergence as demonstrated in Figure~\ref{fig:rl_comparison} (c) and (d), and passing over 
MAB-style performance guarantees (see Section~\ref{sec:background}).

\textbf{Hyperparameter search space.}
Deep RL is notorious for challenging  hyperparameter tuning. For example, in this experiment, we have to decide: the number of layers of the neural network, the neurons per layer, activation functions and loss, the exploration parameter $\varepsilon$, and discount factor $\gamma$. C$^2$UCB has just $\lambda$---which becomes less relevant as rounds are observed---and $\alpha$ which controls exploration.

\textbf{Volatility of deep RL.}
Most deep RL algorithms randomise to explore vast state-action spaces. 
This is not the case with C$^2$UCB. Extending UCB,  deterministic C$^2$UCB is capable of identifying underexplored arms through their context vectors. The only (rare and as such not strictly necessary) case when  C$^2$UCB is random is where the MAB must tie-break arms. A more significant cause of stability of our MAB is its small parametrisation compared to deep learner-based RL.
Combined, the stable MAB yields a more consistent result, as can be seen in Figure~\ref{fig:rl_comparison}(a) and (b). Much wider variance on the DDQN plot demonstrates how the performance of DDQN can vary significantly, compared to the narrow error bars on the MAB, which demonstrates the algorithm's stability.

\section{Related Work}
\label{sec:related}
\malinga{\textbf{HTAP.}
HTAP workloads are composed of online transaction processing (OLTP) workloads and online analytical processing (OLAP) workloads. While most existing analytical systems depend on data pipelines writing to a separate data warehouse for OLAP queries, such an approach limits the user from running analytics on fresh operational data. Research has targeted hybrid environments that can cater to OLTP and OLAP queries. The last few years have witnessed the emergence of HTAP focused database architectures, platforms and databases~\cite{Appuswamy2017TheCF,ArulrajHTAP,AthanassoulisColumnLayout,BatchDB,TiDB,SnappyData,HyPer}, commercial tools~\cite{OracleHTAP,SQLServerHTAP,F1Lightning}, benchmarks~\cite{chbenchmark,difallah2013oltp,HTAPBench}. This rapid growth of research and commercial interest in HTAP environments highlights a pain point of efficiently processing analytical and transactional queries over the same dataset.}

\textbf{Automated physical design tuning.}
Most commercial DBMS vendors nowadays
 offer physical design tools in their products~\cite{DTA2005,DB2_IntegratedApproach,OracleAdvisor}.
These tools rely heavily
on the query optimiser 
to compare benefits of different design
structures without materialisation~\cite{ChaudhuriWhatIf}.
Such an approach is ineffective when base data statistics are unavailable, skewed, or change dynamically~\cite{SurajitDecade}.
In these dynamic environments, the problem of physical design is aggravated: 
a) deciding \emph{when} to call a tuning process is not straightforward;
and b) deciding \emph{what} is a representative training workload is a challenge.

\textbf{Online physical design tuning.} Several research groups have recognised these problems and have offered lightweight solutions to physical design tuning \cite{ToTuneOrNot,OnlineApproachAutoAdmin,QUIET,COLT2}. While such solutions are more flexible and need not know the workload in advance, they are typically limited in terms of applicability to new unknown workloads (generalisation beyond past), and do not come with theoretical guarantees that extend to actual runtime conditions. Moreover, by giving the optimiser a central role, the tools remain susceptible to its mistakes~\cite{AIMeetsAI}. \cite{das2019automatically} extends~\cite{DTA2005} with the use of additional components, in a narrowed scope of index selection to mimic an online tool. This takes corrective actions against the optimiser mistakes through a validation process.

\textbf{Adaptive and learning indices.}
Another dimension of online physical design tuning is database
cracking and adaptive indexing that smooth the creation cost of indices 
by piggybacking on query execution~\cite{StratosCracking,SelfTuningIndexes}. 
Recent efforts have gone a step further and  
proposed replacing data structures with learned models that are smaller in size and faster to query~\cite{learningIndicesKraska,fitingTreeKraska,dataCalculatorStratos}. 
Such approaches are complementary to our efforts: once the data structures (or models) are materialised inside a DBMS, the MAB framework can be used to automate the decision making as to which data structure should be used to speed-up query analysis.

\textbf{Learning approaches to optimisation and tuning.}
Recent years have witnessed new machine learning approaches to automate decision-making processes within databases.
For instance, reinforcement learning approaches have been 
used for query optimisation and join ordering~\cite{cuttlefish,trummer2018skinnerdb,deepLearningVictor,handsFreeOlga}. In~\cite{AIMeetsAI},  regression has been used to successfully mitigate the optimiser's cost misestimates as a path toward more robust index selection. \cite{AIMeetsAI} shows promising results when avoiding query regressions. 
However, this classifier incurs up to 10\% recommendation time, impacting recommendation cost in all cases, especially where recommendation cost already dominates the cost for PDTool (\eg TPC-DS, IMDb).

When it comes to tuning, the closest approaches employ variants of RL for index selection or partitioning~\cite{no_dba,partitioningCarsten,COREIL} or configuration tuning~\cite{pavlo}.  \cite{COREIL} describes RL-based index selection, which depends solely on the recommendation tool for query-level recommendations and is affected by decision combinatorial explosion, both issues addressed in our work. Unlike its more general counterpart (RL), MABs have advantages of faster convergences as demonstrated in Section~\ref{sec:not-rl}, simple implementation, and theoretical guarantees.  
There has been recent interest in using bandits for database tasks such as monitoring, query optimisation and join ordering~\cite{grushka2020using,bao,banditjoin}.
\section{Discussion and Future Avenues}
\label{sec:discussion}
This paper scratches the surface of the numerous opportunities for applying bandit learners to performance tuning of databases. We now discuss a rich research vision for the area.

\textbf{Multi-tenant environments}.
A crucial advantage of the MAB setting is theoretical guarantees on the fitness of proposed indices to observed run-time conditions. This is critical for production systems in the cloud and multi-tenant environments~\cite{AIMeetsAI,sqlvm,das2019automatically}, where analytical modelling is impossible due to unpredictable changes in run-time conditions. 
The MAB approach on the contrary eschews the optimiser and modelling completely, choosing indices based on observed query performance and is thus equally applicable to these challenging environments.

\textbf{Beyond index choices}.
Despite focusing solely on the task of index selection in this paper, the MAB framework is equally applicable to other physical design choices, such as materialised views selection, statistics collection, or even selection of design structures that are a mix of traditional and approximate data structures, such as learned models~\cite{learningIndicesKraska} or other fine-grained design primitives~\cite{dataCalculatorStratos}. Furthermore, MABs can be used in other areas in databases which require strategic exploration under theoretical guarantees like dynamic memory allocation, query optimisation, and database monitoring.

\textbf{Cold-start problem.}
Under the current setup, \linebreak[4] MAB starts without any secondary indices or knowledge about their benefits, forming a cold-start problem and leading to higher creation costs. While MAB is already superior against PDTool even with the creation cost burden (see Section~\ref{sec:exploration-exploitation}), even faster convergence and better creation costs can be provided by pre-training models in hypothetical rounds (using what-if) or workload forecasting~\cite{forecastingPavlo} to improve context quality.

\textbf{Opportunities for bandit learning.}
The increas\-ed search space of possible design choices calls for advancements to bandits algorithms and theory, where unbound\-ed/infinite numbers of arms will be increasingly important. Similarly, various flavours of physical design might ask for novel bandits that adopt the notion of heterogeneous arms (indices, views, or statistics), or hierarchical models where individual choices at lower levels (\eg the choice of indices or materialised views) influence decisions at a higher level (\eg index merging due to memory constraints). 
\section{Conclusions}
\label{sec:conclusion}

This paper develops a multi-armed bandit learning fra\-mework for online index selection. 
Benefits include eschewing the DBA and the (error-prone) query optimiser by learning the benefits of indices through strategic exploration and observation. 
We justify our choice of MAB over general reinforcement learning for online index tuning, comparing MAB against DDQN, a popular RL algorithm based on deep neural networks, demonstrating significantly faster convergence of the MAB. 
Furthermore, our extensive experimental evaluation demonstrates advantages of MAB over an existing commercial physical design tool (up to 75\% speed up, and 23\% on average), and exemplifies robustness to data skew, unpredictable ad-hoc workloads \malinga{and complex HTAP environments}.
\appendix
\makeorange
\section{Theoretical Regret Analysis}
\label{sec:theory}

As first claimed by Qin \etal~\cite{c2ucb}, C$^2$UCB enjoys an $\alpha$-regret bound of $\Tilde{O}(\sqrt{n})$ as described by the following theorem.
\begin{theorem}[Theorem 4.1 from \cite{c2ucb}]
\label{theoremc2ucb}
    Without loss of generality, assuming that $||\bm{\theta}_\star||_2\leq S,\, ||\mathbf{x}_t(i)||_2 \leq 1,\, r_t(i)\in[0,1] \forall t\geq 0,\,i \in \{1,2,\cdots,k\},$ and assuming that we set $\alpha_t = \sqrt{d\log{\left(\frac{1+tk/\lambda}{\delta}\right)}} + \lambda^{1/2}S$ for $0<\delta<1,$ then C$^2$UCB has $\alpha$-regret bounded as 
    \begin{align*}
      \sum_{t=1}^T R&eg_t^{\alpha}\leq C\sqrt{64Td\log{(1+Tk/d\lambda})}\\
      &\cdot\left(\sqrt{\lambda}S + \sqrt{2\log{(1/\delta)+d\log{(1+Tk/(\lambda d))}}}\right)  
    \end{align*}
    for any $n\geq 0$ with probability at least $1-\delta$.
\end{theorem}
The original bound leverages the following result.

\begin{lemma}[Lemma 4.2 from \cite{c2ucb}]
\label{lemmac2ucb}
    Let $\mathbf{V} \in \mathbb{R}^{d\times d}$ be a positive definite matrix, $s_t \subseteq \{1,\cdots,k\}$ where $|s_t|\leq \ell$ for $t=1,2,\dots$, and $\mathbf{V}_T=\mathbf{V}+\sum_{t=1}^T\sum_{i\in s_t}\mathbf{x}_t(i)\mathbf{x}_t(i)'$. Then, if $\forall t,i \, \,\,\lambda\geq\ell$ and $||\mathbf{x}_t(i)||_2\leq 1$, we have
    \begin{align*}
        \sum_{t=1}^T\sum_{i\in s_t}\|\mathbf{x}_t(i)\|^2_{\mathbf{V}_{t-1}^{-1}} &\leq 2 \log \det \mathbf{V}_T - 2 \log \det \mathbf{V}\\
        &\leq 2d \log ((\Tr(\mathbf{V})+T\ell)/d) - \\
        &\qquad2\log \det \mathbf{V}.
    \end{align*}
\end{lemma}
Unfortunately the original proof of this lemma suffers from an error when the following incorrect claim is made.
\begin{claim}
    \label{claim:incorrect}
    Let $k, T$ be natural numbers, $\mathbf{V}$ be a $d \times d$ real and positive definite matrix, and $s_t \subseteq \{1,\cdots, k\}$ so that $|s_t|\leq k$ for $t\in\{1,\cdots,T\}$. Let $\mathbf{x}_t(i)\in\mathbb{R}^d$ be vectors for $t\in\{1,\cdots,T\}, i\in\{1,\cdots,k\}$, and define the matrix containing the sum of the outer products of the contexts to be $\mathbf{V}_T = \mathbf{V} + \sum_{t=1}^T\sum_{i\in s_t}\mathbf{x}_t(i)\mathbf{x}_t(i)'$. If we define $\|\mathbf{a}\|_{\mathbf{M}}=\sqrt{\mathbf{a}^\prime\mathbf{M}\mathbf{a}}$, then
    $\det(\mathbf{V}_T) = \det(\mathbf{V})\prod_{t=1}^T\left(1+\sum_{i\in s_t}\|\mathbf{x}_t(i)\|^2_{\mathbf{V}_{t-1}^{-1}}\right)$.
\end{claim}
Counterexamples to this claim are found by inserting arbitrary values of variables when $T=2$. We relax the claim's relation, as proven in~\cite{C2UCBProofNote}, by instead proving an inequality: $$\det(\mathbf{V}_T) \geq \det(\mathbf{V})\prod_{t=1}^T\left(1+\sum_{i\in s_t}\|\mathbf{x}_t(i)\|^2_{\mathbf{V}_{t-1}^{-1}}\right).$$ Fortunately, this still achieves a proof of Lemma~\ref{lemmac2ucb} as originally stated with non-trivial modification to the proof but with no further changes needed in proving Theorem~\ref{theoremc2ucb}.

Denote by $\lambda_i(\mathbf{A})$ the eigenvalues of the $n \times n$ matrix $\mathbf{A}$, where, without loss of generality, $\lambda_1(\mathbf{A}) \leq \lambda_2(\mathbf{A}) \leq \dots \leq \lambda_n(\mathbf{A})$.
    We likewise order the super arm at round $t$ as $s_t = \{ s_{(1,t)}, s_{(2,t)}, \dots, s_{(|s_t|,t)}\},$ where $s_{(1,t)} < s_{(2,t)} < \dots < s_{(|s_t|,t)}$. In the proof, we will make use of the Generalised Matrix Determinant Lemma: Let $\mathbf{A}$ be an invertible $n \times n$ matrix, and $\mathbf{B}, \mathbf{C}$ be $n \times m$ matrices, then
$\det(\mathbf{A}+\mathbf{B}\mathbf{C}^T) = \det(\mathbf{I}_m+\mathbf{C}^T\mathbf{A}^{-1}\mathbf{B})\det(\mathbf{A})$.

We begin our proof by collecting the contexts of the super arm $\mathbf{X}_{T} = \begin{bmatrix}
    \mathbf{x}_{T}(s_{(1,T)}) &\dots &\mathbf{x}_{T}(s_{(|S_{T}|,T)})
\end{bmatrix}$. Then,
\begin{align*}
    \det(\mathbf{V}_T)
    &= \det\left(\mathbf{V} + \sum_{t=1}^T\sum_{i\in s_t}\mathbf{x}_t(i)\mathbf{x}_t(i)'\right)\\
    &= \det\left(\mathbf{V} + \sum_{t=1}^{T-1}\sum_{i\in s_t}\mathbf{x}_t(i)\mathbf{x}_t(i)' +\right.\\
    &\quad\quad\quad\quad\left.\sum_{i\in s_T} \mathbf{x}_T(i)\mathbf{x}_T(i)'\right)\\
    &= \det\left(\mathbf{V}_{T-1} + \mathbf{X}_{T}\mathbf{X}_{T}'\right)\\
    &= \det(\mathbf{V}_{T-1})\det\left(\mathbf{I}_{|s_{T}|}+\mathbf{X}_{T}'\mathbf{V}_{T-1}^{-1}\mathbf{X}_{T}\right)\\
    &= \det(\mathbf{V}_{T-1})\left[ \prod_{i=1}^{|s_{T}|}\lambda_i\left(\mathbf{I}_{|s_{T}|}+\mathbf{X}_{T}'\mathbf{V}_{T-1}^{-1}\mathbf{X}_{T}\right)\right]\\
    &= \det(\mathbf{V}_{T-1})\left[ \prod_{i=1}^{|s_{T}|}\left(1+\lambda_i\left(\mathbf{X}_{T}'\mathbf{V}_{T-1}^{-1}\mathbf{X}_{T}\right)\right)\right] \;,
\end{align*}
where the fourth and final equalities follow from the Generalised Matrix Determinant Lemma and the fact that adding the identity to a square matrix increases eigenvalues by one. 
Now, the final line's product term can be expanded as
\begin{align}
    1 + &\sum_{i=1}^{|s_{T}|} \lambda_i(\mathbf{X}_{T}'\mathbf{V}_{T-1}^{-1}\mathbf{X}_{T})+\nonumber\\
    &\sum_{1 \leq i_1 < i_2 \leq |s_{T}|} \lambda_{i_1}(\mathbf{X}_{T}'\mathbf{V}_{T-1}^{-1}\mathbf{X}_{T})\lambda_{i_2}(\mathbf{X}_{T}'\mathbf{V}_{T-1}^{-1}\mathbf{X}_{T})+\nonumber\\
    &\cdots +\prod_{i=1}^{|s_{T}|}\lambda_i(\mathbf{X}_{T}'\mathbf{V}_{T-1}^{-1}\mathbf{X}_{T})\;. \label{eq:prodexpanse}
\end{align}

Since $\mathbf{V}$ is positive definite and $\mathbf{x}_t(i)\mathbf{x}_t(i)'$ is positive semi-definite (with one eigenvalue being $\mathbf{x}_t(i)' \mathbf{x}_t(i)$ and the remainder all zero) for all $t$ and $i$, we have that $\mathbf{V}_{T-1} = \mathbf{V} + \sum_{t=1}^{T-1}\sum_{i\in s_t}\mathbf{x}_t(i)\mathbf{x}_t(i)'$ is positive definite. %
Therefore, we conclude that $\mathbf{V}_{T-1}^{-1}$ is also positive definite, hence it has a symmetric square root matrix $\mathbf{V}_{T-1}^{-1/2}$. It also follows that $\mathbf{X}_{T}'\mathbf{V}_{T-1}^{-1}\mathbf{X}_{T}$ is positive semi-definite. 
\ifdefined\longpap
This is because  for any $|S_n| \times 1$ vector $\mathbf{a}$, where $\mathbf{a}\neq \mathbf{0}$,
\begin{align*}
    \mathbf{a}^T(\mathbf{X}_n^T\mathbf{V}_{n-1}^{-1}\mathbf{X}_n)\mathbf{a} &= \mathbf{a}^T\mathbf{X}_n^T\mathbf{V}_{n-1}^{-1/2}\mathbf{V}_{n-1}^{-1/2}\mathbf{X}_n\mathbf{a} \\
    &= \mathbf{a}^T\mathbf{X}_n^T\mathbf{V}_{n-1}^{-T/2}\mathbf{V}_{n-1}^{-1/2}\mathbf{X}_n\mathbf{a}\\
    &= (\mathbf{V}_{n-1}^{-1/2}\mathbf{X}_n\mathbf{a})^T(\mathbf{V}_{n-1}^{-1/2}\mathbf{X}_n\mathbf{a})\\
    &= \|\mathbf{V}_{n-1}^{-1/2}\mathbf{X}_n\mathbf{a}\|^2\\
    &\geq 0,
\end{align*}
which shows that all the eigenvalues of $\mathbf{X}_n^T\mathbf{V}_{n-1}^{-1}\mathbf{X}_n$ are non-negative.
\fi
Therefore, the terms starting from the third term in the expansion \eqref{eq:prodexpanse} are all non-negative because they are products of the eigenvalues of $\mathbf{X}_{T}'\mathbf{V}_{T-1}^{-1}\mathbf{X}_{T}$. Thus we have,
\begin{align*}
    \det(\mathbf{V}_{T})
    &= \det(\mathbf{V}_{T-1})\left[ \prod_{i=1}^{|s_{T}|}\left(1+\lambda_i\left(\mathbf{X}_{T}'\mathbf{V}_{T-1}^{-1}\mathbf{X}_{T}\right)\right)\right]\\
    &\geq \det(\mathbf{V}_{T-1})\left(1 + \sum_{i=1}^{|s_{T}|} \lambda_i(\mathbf{X}_{T}'\mathbf{V}_{T-1}^{-1}\mathbf{X}_{T})\right)\\
    &= \det(\mathbf{V}_{T-1})\left( 1 + \Tr(\mathbf{X}_{T}'\mathbf{V}_{T-1}^{-1}\mathbf{X}_{T})\right) \\
     &= \det(\mathbf{V}_{T-1})\left( 1 + \sum_{i\in s_{T}} \mathbf{x}_{T}(i)\mathbf{V}_{T-1}^{-1}\mathbf{x}_{T}(i)\right)\\
     &= \det(\mathbf{V}_{T-1})\left( 1 + \sum_{i\in s_{T}} \|\mathbf{x}_{T}(i)\|^2_{\mathbf{V}_{T-1}^{-1}}\right)\; .
\end{align*}
Applying our recurrence relation on $\mathbf{V}_t$ for $1\leq t \leq T$, we can telescope to arrive at
\ifdefined\longpap
\begin{align*}
    \det(\mathbf{V}_{n})
    &\geq \det(\mathbf{V}_{n-1})\left( 1 + \sum_{i\in S_{n}} \|\mathbf{x}_{n}(i)\|^2_{\mathbf{V}_{n-1}^{-1}}\right)\\
    &\geq \det(\mathbf{V}_{n-2})\left( 1 + \sum_{i\in S_{n-1}} \|\mathbf{x}_{n-1}(i)\|^2_{\mathbf{V}_{n-2}^{-1}}\right)\left( 1 + \sum_{i\in S_{n}} \|\mathbf{x}_{n}(i)\|^2_{\mathbf{V}_{n-1}^{-1}}\right)\\
    &\geq \dots\\
    & \geq \det(\mathbf{V})\prod_{t=1}^n\left(1+\sum_{i\in S_t}\|\mathbf{x}_t(i)\|^2_{\mathbf{V}_{t-1}^{-1}}\right),
\end{align*}
which proves 
\fi
the wanted inequality.

\color{black}

\bibliographystyle{spmpsci}
\bibliography{bibliography_medium}

\end{document}